\documentclass[12pt]{article}
\pdfoutput=1
\usepackage{jcappub}

\usepackage{amsthm,graphicx,multirow}
\usepackage{epsfig}
\usepackage{latexsym, amssymb} \usepackage{soul}
\usepackage{amsmath}
\usepackage{epstopdf}
\usepackage[perpage]{footmisc}
\usepackage{hyperref}
\usepackage{xcolor}

\usepackage{multirow}

\usepackage{mhchem}

\newcommand{\dd}{\mathrm{d}}

\begin{document}
\title{Cosmological constraints on the gravitational constant}
\author[1,2,3,4]{Mario Ballardini,}
\author[2,3]{Fabio Finelli,}
\author[5]{Domenico Sapone}

\affiliation[1]{Dipartimento di Fisica e Astronomia, Alma Mater Studiorum Universit\`a di Bologna, via Gobetti 93/2, I-40129 Bologna, Italy}
\affiliation[2]{INAF/OAS Bologna, via Piero Gobetti 101, I-40129 Bologna, Italy}
\affiliation[3]{INFN, Sezione di Bologna, via Irnerio 46, I-40126 Bologna, Italy}
\affiliation[4]{Department of Physics \& Astronomy, University of the Western Cape, Cape Town 7535, South Africa}
\affiliation[5]{Cosmology and Theoretical Astrophysics group, Departamento de Física, FCFM, Universidad de
Chile, Blanco Encalada 2008, Santiago, Chile}

\emailAdd{mario.ballardini@inaf.it}
\emailAdd{fabio.finelli@inaf.it}
\emailAdd{domenico.sapone@uchile.cl}

\abstract{\small
We study the variation of the gravitational constant on cosmological scales in scalar-tensor theories 
of gravity. We focus on the simplest models of scalar-tensor theories with a coupling to the Ricci 
scalar of the form $F(\sigma) = N_{\rm pl}^2 + \xi\sigma^2$, such as extended Jordan-Brans-Dicke 
($N_{\rm pl}=0$), or a non-minimally coupled scalar field with $N_{\rm pl}=M_{\rm pl}$, which permits 
the gravitational constant to vary self-consistently in time and space. In addition, we allow the 
effective gravitational constant on cosmological scales to differ from the Newton's measured constant 
$G$, i.e. $G_{\rm eff}(z=0) = G\left(1+\Delta\right)^2$. 
We study the impact of this imbalance $\Delta$ jointly with the coupling $\xi$ into anisotropies of the 
cosmic microwave background and matter power spectrum at low-redshift. 
Combining the information from {\em Planck} 2018 CMB temperature, polarization and lensing, together with a compilation of BAO measurements from the release DR12 of the Baryon Oscillation Spectroscopic 
Survey (BOSS), we constrain the imbalance to $\Delta = -0.022 \pm 0.023$ (68\% CL) and the coupling parameter to $10^3\, \xi < 0.82$ (95\% CL) for Jordan-Brans-Dicke and for a non-minimally coupled scalar 
field with $F(\sigma) = M^2_{\rm pl} + \xi\sigma^2$ we constrain the imbalance to $\Delta > -0.018$ ($< 0.021$) and the coupling parameter to $\xi < 0.089$ ($\xi > - 0.041$) both at 95\% CL. With current data, we observe that the degeneracy between $\Delta$, the coupling $\xi$ to the Ricci scalar, 
and $H_0$ allows for a larger value of the Hubble constant increasing the consistency  between the 
distance-ladder measurement of the Hubble constant from supernovae type Ia by the SH0ES team and its 
value inferred by CMB data. 
We also study how future cosmological observations can constrain the gravitational Newton's constant. 
Future data such as the combination of CMB anisotropies from LiteBIRD and CMB-S4, and large-scale 
structures galaxy clustering from DESI and galaxy shear from LSST reduce the uncertainty in $\Delta$ 
to $\sigma(\Delta) \simeq 0.004$.}

\maketitle

\section{Introduction}
The Universe is a unique and peculiar laboratory to test fundamental physical laws and possible clues 
for physics beyond the current standard understanding. In particular, the exciting possibility that 
fundamental constants \cite{Uzan:2002vq} could vary in time, which has long been proposed by Dirac 
\cite{Dirac:1937ti}, can be tested through cosmological observations at higher and higher precision.
Cosmology can indeed uniquely probe lengths and/or timescales otherwise inaccessible on ground and 
from Solar System experiments.

Among the different fundamental constants, Newton's constant remains the one with the largest relative 
uncertainty from laboratory measurements. 
According to CODATA\footnote{\href{https://codata.org/initiatives/strategic-programme/fundamental-physical-constants/}{https://codata.org/initiatives/strategic-programme/fundamental-physical-constants/}}, 
the value of Newton's constant G is $6.67430(15) \times 10^{-8}$ cm$^3$ g$^{-1}$ s$^{-2}$ with a 
relative uncertainty of $2.2 \times 10^{-5}$. 

General Relativity (GR) has been tested exquisitely in the Solar System and on extra-galactic systems 
\cite{Will:2005va} setting observational challenges for the theories alternative to Einstein gravity.
As Solar System tests on parameterized post Newtonian (PPN) parameters we quote the time dilation due 
to the effect of the Sun's gravitational field measured very accurately using the signal from Cassini 
satellite giving a constraint $\gamma_{\rm PN} = \left(2.1 \pm 2.3\right) \times 10^{-5}$ 
\cite{Bertotti:2003rm} and the perihelion shift of Mercury 
$\beta_{\rm PN} - 1 = \left(-4.1 \pm 7.8\right) \times 10^{-5}$ \cite{Will:2014kxa}, assuming the 
Cassini bound. 
Lunar laser ranging data provide a tight constraint on the time variation of the gravitational constant 
$\dot{G}/G = \left(2 \pm 7\right) \times 10^{-13}$ yr$^{-1}$ \cite{Muller:2007zzb}.
Other extra-galactic constraints on the time-variation of $G$ come from stellar 
\cite{GarciaBerro:2011wc,Mould:2014iga,Bellinger:2019lnl} and pulsar timing \cite{Zhu:2018etc}.  

As previously stated, cosmological observations can constrain the gravitational constant at totally 
different scale and redshift.
The primordial abundances of light elements allow to constrain gravitational constant during Big Bang 
Nucleosynthesis (BBN) based on the modified ratio of the expansion rate and a given prior value/range 
of the baryon density $\omega_{\rm b}$ 
\cite{Casas:1990fz,Casas:1991ui,Santiago:1997mu,Clifton:2005xr,Bambi:2005fi,Copi:2003xd,Alvey:2019ctk}. 
Current measurements of the primordial abundances of helium and deuterium constrain 
$0.04 < (\delta G/G) < -0.08 $ at 95\% CL \cite{Alvey:2019ctk} at nucleosynthesis by assuming 
$\omega_{\rm b} = 0.02236 \pm 0.00030$ at 68\% CL \cite{Aghanim:2018eyx}.
A different value of the gravitational constant in the Einstein equations can be interpreted as a 
change in the background expansion history and distance measurements 
\cite{Umezu:2005ee,Chan:2007fe,Perenon:2019dpc,Hanimeli:2019wrt,Sapone:2020wwz,Sakr:2021nja}, and in 
recombination for cosmic microwave background (CMB) physics \cite{Zahn:2002rr,Galli:2009pr,Martins:2010gu}.

Scalar-tensor theories of gravity, with a scalar field non-minimally coupled to gravity, modify GR by 
dynamically determining the value of the gravitational constant and can accommodate self-consistently 
space and time variation of the gravitational constant.
The most recent {\em Planck} 2018 and baryon acoustic oscillation (BAO) data constrain the variation of 
the gravitational constant with respect to the radiation era to be smaller than 3\% at 95\% CL 
\cite{Ballardini:2020iws} by assuming adiabatic initial condition for scalar fluctuations 
\cite{Paoletti:2018xet} and the effective gravitational constant at present corresponding to 
the Newton's measured constant (see 
\cite{Umezu:2005ee,Chan:2007fe,Umilta:2015cta,Ballardini:2016cvy,Rossi:2019lgt} for previous 
constraints).
In these scalar-tensor theories of gravity, the cosmological constraints on the time variation of the 
gravitational constant can be tighter than those from the Lunar Laser ranging.

This paper wants to go further and study in detail the impact of an imbalance $\Delta$ between the 
effective gravitational constant at present and measured value of the Newton's constant, defined as 
$G_{\rm eff}(z=0) = G (1+\Delta)^2$, jointly with the coupling to the curvature in simple scalar-tensor 
theories where the gravitational constant can vary self consistently in space and time. 
We will explore the impact of $\Delta \ne 0$ on the CMB anisotropies and matter power spectrum at low 
redshift calculating the joint constraints on  $\Delta$ and on the coupling to the Ricci scalar with 
publicly available data and future cosmological observations.

It is important to stress that these minimal scalar-tensor theories are not only workhorse models to 
study how cosmology can constrain gravity on large scales, but are also of great current interest since 
alleviate the existing tension \cite{Schoneberg:2021qvd,DiValentino:2020zio} between distance-ladder 
measurement of the Hubble constant from supernovae type Ia by the SH0ES team and its value inferred by 
CMB data. Moreover the mismatch between different values of the Hubble constant can be reduced assuming 
a late-time transition of the effective gravitational constant \cite{Marra:2021fvf,Alestas:2021luu}.

The paper is organized as follows. 
After this introduction, we describe the implementation of the variation of the gravitational constant 
in the context of scalar-tensor theories in Section~\ref{sec:G}. In Section~\ref{sec:data} we describe 
the datasets and prior considered and in Section~\ref{sec:results} we discuss our results in light of 
CMB and BAO data. We present the Fisher methodology for CMB and LSS for our science forecasts and the 
results in Section~\ref{sec:forecast}. In Section~\ref{sec:conclusion} we draw our conclusions.
In App.~\ref{sec:appendix1}-\ref{sec:appendix2}, we collect all the tables and triangle plots with the 
constraints on the cosmological parameters obtained with our MCMC analysis.

\section{Varying $G$ within minimal scalar-tensor theory} \label{sec:G}
In this paper, we use a scalar field  non-minimally coupled (NMC) to the Ricci scalar as the simplest 
scalar-tensor theory of gravity \cite{Jordan:1949zz,Brans:1961sx} with which we test deviations from 
GR and constrain the variation of the effective gravitational constant $G_{\rm eff}$ from cosmology, 
as previously done in \cite{Chen:1999qh,Nagata:2003qn,Acquaviva:2004ti,Li:2013nwa,Avilez:2013dxa,Umilta:2015cta,Ballardini:2016cvy,Ooba:2016slp,Ooba:2017gyn,Rossi:2019lgt,SolaPeracaula:2019zsl,Ballesteros:2020sik,Braglia:2020iik,Ballardini:2020iws,Braglia:2020auw,Cheng:2021yvh,Joudaki:2020shz}. 

By using NMC scalar fields, we have a self-consistent way of modifying both the background dynamics of 
the universe and that of the perturbations accurately for both early- and late-time probes.
In order to do so, we consider the NMC theory described by 
\begin{equation} \label{eqn:action}
    S = \int \dd^{4}x \sqrt{-g} \left[ \frac{F(\sigma)}{2}R 
    - \frac{g^{\mu\nu}}{2} \partial_\mu \sigma \partial_\nu \sigma - V(\sigma) + {\cal L}_m \right] \,,
\end{equation}
where $\sigma$ is a scalar field, $F(\sigma) = N_{\rm pl}^2+\xi\sigma^2$, $R$ is the Ricci scalar, and 
${\cal L}_m$ is the Lagrangian density for matter fields. We restrict ourselves to a potential of the 
type $V(\sigma) \propto F^2(\sigma)$ \cite{Amendola:1999qq,Finelli:2007wb} in which the scalar field 
is effectively massless and the effective gravitational constant $G_{\rm eff}$ between two test masses 
is \cite{Boisseau:2000pr}
\begin{equation}
    G_{\rm eff}(z=0)  = \frac{1}{8 \pi F_0}\frac{2F_0+4F_{0,\sigma}^2}{2F_0+3F_{0,\sigma}^2} \,.
\end{equation}
We define also the gravitational constant entering in the NMC background equations 
$G_{\rm N} = (8\pi F)^{-1}$. For the full set of background equations and linear cosmological 
perturbations in NMC scalar-tensor models, we refer the interested reader to 
Refs.~\cite{Boisseau:2000pr,Umilta:2015cta,Rossi:2019lgt}.
In this paper, we allow an imbalance $\Delta$ between $G_{\rm eff}(z=0)$ and $G$
\begin{equation} \label{eqn:boundary}
    G_{\rm eff}(z=0) = G\left(1+\Delta\right)^2 \,,
\end{equation}
which was fixed to zero in many previous studies 
\cite{Umilta:2015cta,Ballardini:2016cvy,Rossi:2019lgt,Ballardini:2020iws}\footnote{Note that in 
Refs.~\cite{Avilez:2013dxa,Joudaki:2020shz} the so-called {\em unrestricted} evolution corresponds to 
$\Delta \ne 0$ for Jordan-Brans-Dicke model, which is equivalent to the IG case studied here by a 
redefinition of the scalar field. However, we have here different theoretical priors on the effective 
coupling to the curvature and for the imbalance.}.

We consider the following three cases of the model: induced gravity (IG) described by 
$N_{\rm pl} = 0$ and $\xi > 0$, a conformally coupled scalar field (CC) described by 
$N_{\rm pl} = M_{\rm pl}$ and $\xi = -1/6$, and a NMC for which $N_{\rm pl} = M_{\rm pl}$ and the 
free parameter is $\xi \ne 0$. 
For all models, the effective value of the Newton's gravitational constant $G_{\rm eff}$ decreases 
with time but for $\Delta < 0$ we can find a late-time period of weaker gravitational strength compared 
to the GR case, i.e. $G_{\rm eff} < G$, and vice versa for $\Delta > 0$.

Note that we have previously studied NMC scalar fields by considering primary extra parameters 
$(N_{\rm pl}, \xi)$ with $\Delta=0$ \cite{Rossi:2019lgt,Ballardini:2020iws} or $(\sigma_i, \xi)$ with 
$\sigma_i$ as the initial value of the scalar field and $N_{\rm pl} = M_{\rm pl}$ 
\cite{Braglia:2020iik,Braglia:2020auw}. 
Here instead we would like to promote the imbalance $\Delta$ to a primary parameter by fixing 
$N_{\rm pl} = M_{\rm pl}$ in order to avoid introducing parameters which would be hardly constrained 
by data.

\begin{figure}
\centering
\includegraphics[width=0.98\textwidth]{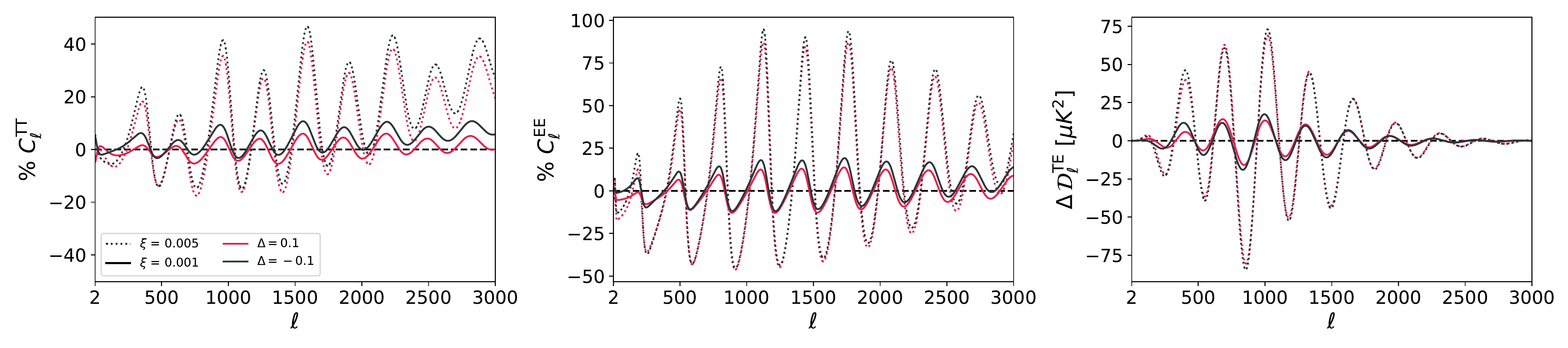}
\includegraphics[width=0.98\textwidth]{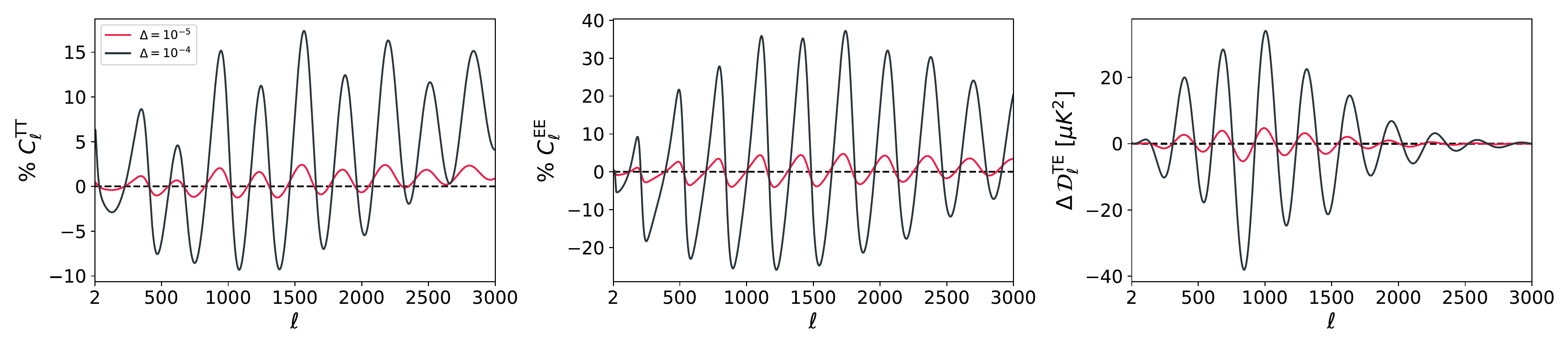}
\caption{Differences with respect to the $\Lambda$CDM CMB lensed angular power spectra with IG (top 
panels) for $\xi = 0.001,\,0.005$ (solid, dotted) and $\Delta = -0.1,\,0.1$ (black, red), and CC 
(bottom panels) for $\Delta = 0.00001,\,0.0001$ (red, black). 
${\cal D}_\ell \equiv  \ell(\ell+1)C_\ell/(2\pi)$ are the band-power angular power spectra.}
\label{fig:ig_cc_cl}
\end{figure}

We show the effect of allowing $\Delta$ to vary on CMB anisotropies temperature (TT), E-mode 
polarization (EE), and temperature-E-mode correlation (TE) angular power spectra in 
Figs.~\ref{fig:ig_cc_cl}-\ref{fig:nmc_cl}; the CMB lensing potential and the linear matter power 
spectra at $z=0$ are shown in Figs.~\ref{fig:ig_cc_mpk}-\ref{fig:nmc_mpk}.

\subsection{Induced gravity}
For IG, i.e. $N_{\rm pl} = 0$ and $\xi > 0$, Eq.~\eqref{eqn:boundary} leads to
\begin{equation}
    \left(\frac{\sigma_0}{M_{\rm pl}}\right)^2 = \frac{1}{\xi}\frac{1+8\xi}{1+6\xi}
    \frac{1}{\left(1+\Delta\right)^2} =
    \left(\frac{\overline{\sigma}_0}{M_{\rm pl}}\right)^2\frac{1}{\left(1+\Delta\right)^2} \,.
\end{equation}
where $\overline{\sigma}_0 \equiv \sigma_0\left(\Delta=0\right)$. In this case, we consider $\xi > 0$, 
and both positive and negative values for $\Delta$. 

\begin{figure}
\centering
\includegraphics[width=0.98\textwidth]{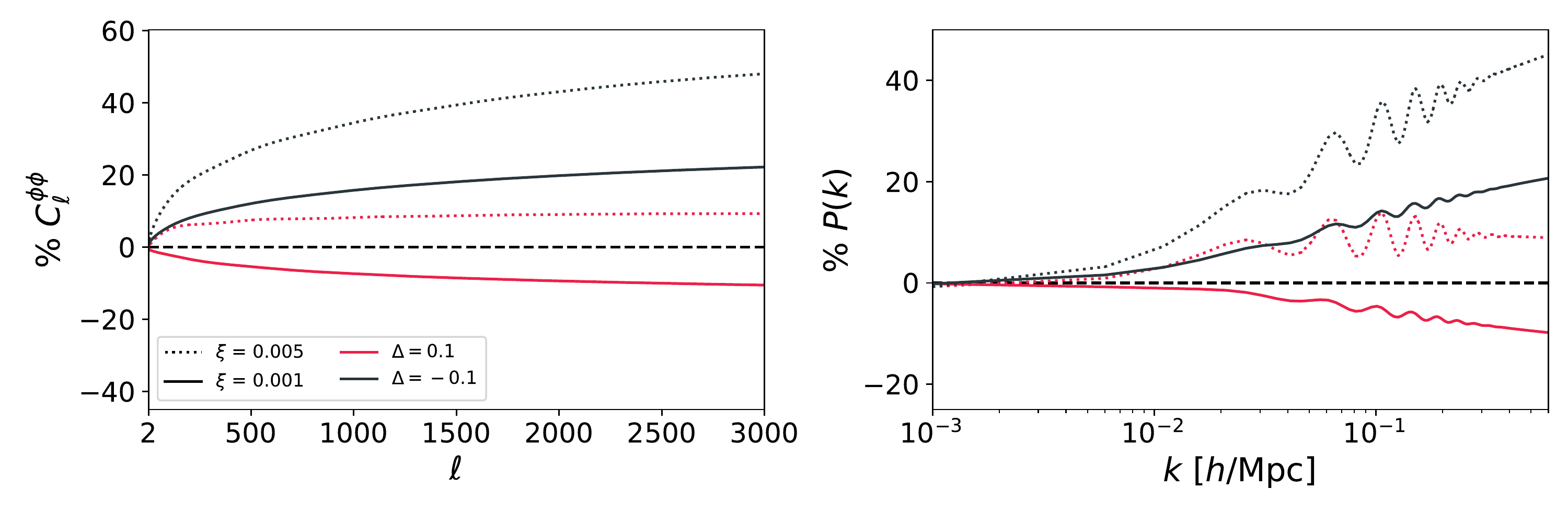}
\includegraphics[width=0.98\textwidth]{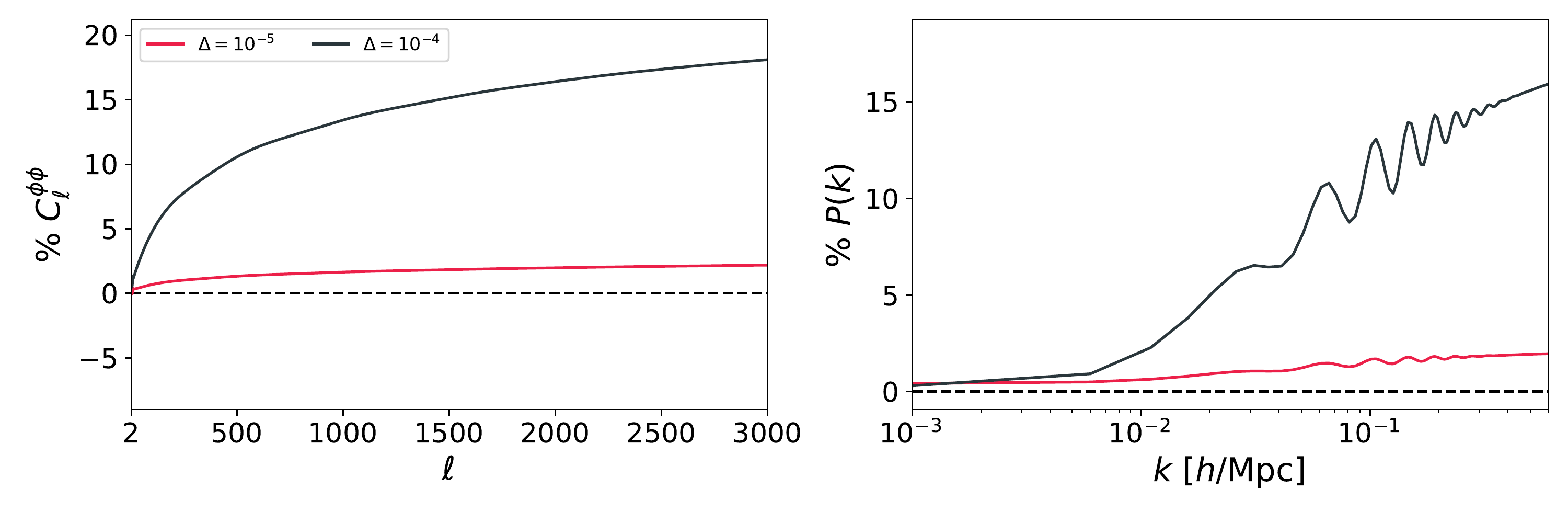}
\caption{Relative differences of the CMB lensing potential angular power spectrum (left panels) 
and linear matter power spectrum at $z=0$ (right panels) for IG (top panels) with 
$\xi = 0.001,\,0.005$ (solid, dotted) and $\Delta = -0.1,\,0.1$ (black, red), and CC (bottom panels) 
for $\Delta = 0.00001,\,0.0001$ (red, black) with respect to the $\Lambda$CDM.}
\label{fig:ig_cc_mpk}
\end{figure}

\begin{figure}
\centering
\includegraphics[width=\textwidth]{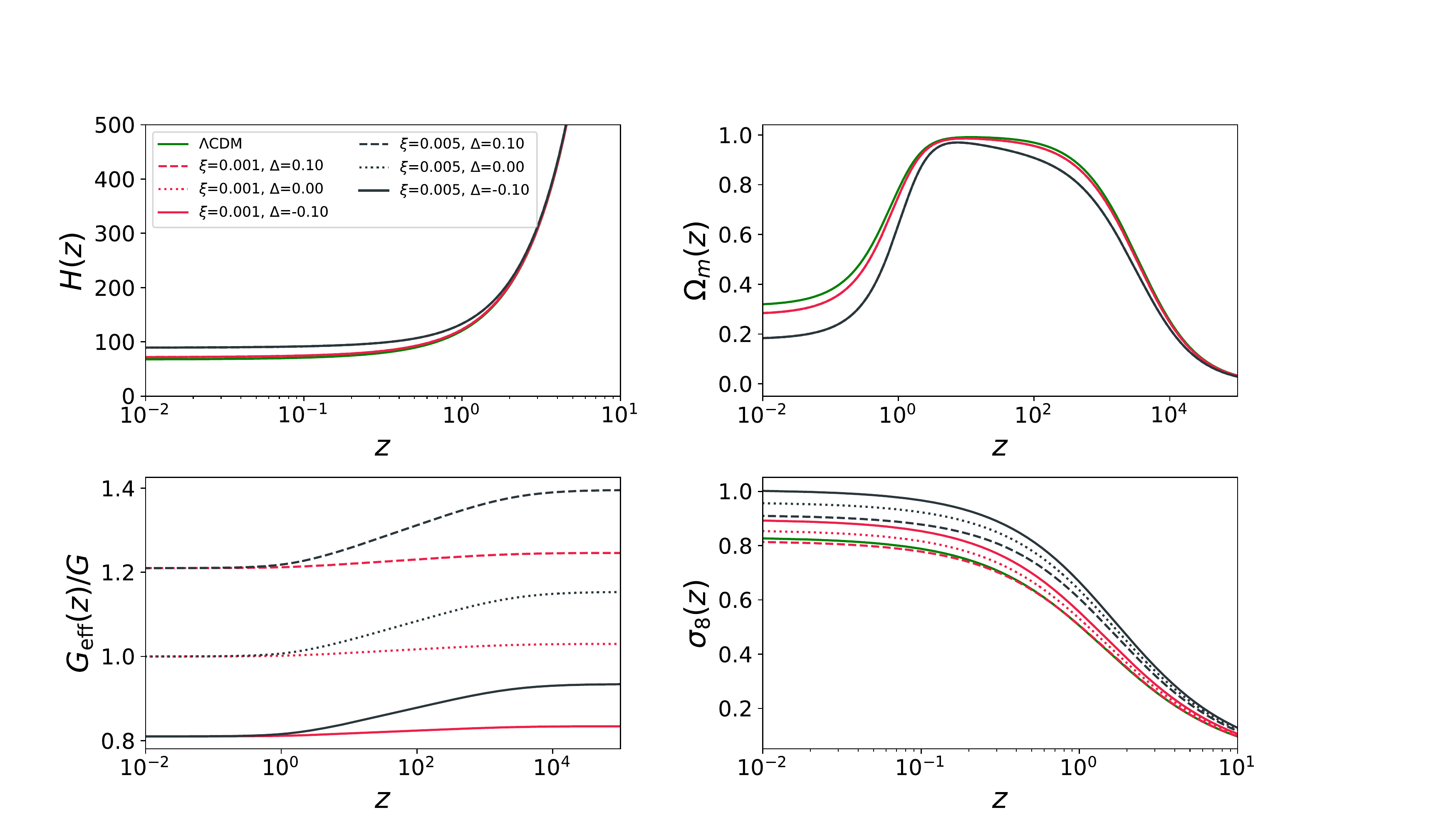}
\caption{Evolution of $H(z)$ (top-left panel), $\Omega_{\rm m} = \rho_{\rm m}/(3H_0^2)$ (top-right 
panel), $G_{\rm eff}(z)/G$ (bottom-left panel), and $\sigma_8(z)$ (bottom-right panel) as function of 
$z$ for $\xi = 10^{-3},\ 5\times 10^{-3}$ and different choices of $\Delta$ from $\Delta = -0.1$ to 
$\Delta = 0.1$.}
\label{fig:ig_bkg}
\end{figure}

Fig.~\ref{fig:ig_cc_mpk} shows that for $\Delta >0$ the CMB lensing potential spectrum 
$C_\ell^{\phi\phi}$ and the matter power spectrum at $z=0$ $P(k)$ can be smaller than in $\Lambda$CDM 
for small scales, a behavior which does not occur for $\Delta < 0$ or for $\Delta=0$ 
\cite{Umilta:2015cta}.
In Fig.~\ref{fig:ig_bkg}, we show the evolution of the background quantities $H(z)$, 
$\Omega_{\rm m}(z)$, $G_{\rm eff}/G(z)$, and $\sigma_8(z)$ for IG. We see that for $\xi = 0.001$ and 
$\Delta = 0.1$ the amplitude of matter perturbation today $\sigma_8$ is slightly smaller than the one 
in $\Lambda$CDM.

\subsection{A non-minimally coupled scalar field}
We study $F(\sigma) = M_{\rm pl}^2 + \xi \sigma^2$. In this case we find two different branches of 
the parameter space $\{\xi > 0,\,\Delta < 0\}$ (NMC+) and $\{\xi < 0,\,\Delta > 0\}$ (NMC$-$) in order 
to satisfy the positiveness of $\sigma$ and $F(\sigma)$ for all times, with 
\begin{align} \label{eqn:boundary_nmc}
    \left(\frac{\sigma_0}{M_{\rm pl}}\right)^2 = &\frac{1}{2(1+\Delta)^2 \xi  (1+6\xi)}\Bigg[-1 +2\xi -2 \Delta(2+\Delta)(1+3 \xi) \notag\\
    &+ \sqrt{1+4 \xi  \left(-1+\xi+\Delta  \left(2+\Delta\right) \left(-5+3\xi\left(-2+3 \Delta  \left(2+\Delta\right)\right)\right)\right)} \Bigg] \,.
\end{align}

\begin{figure}
\centering
\includegraphics[width=0.98\textwidth]{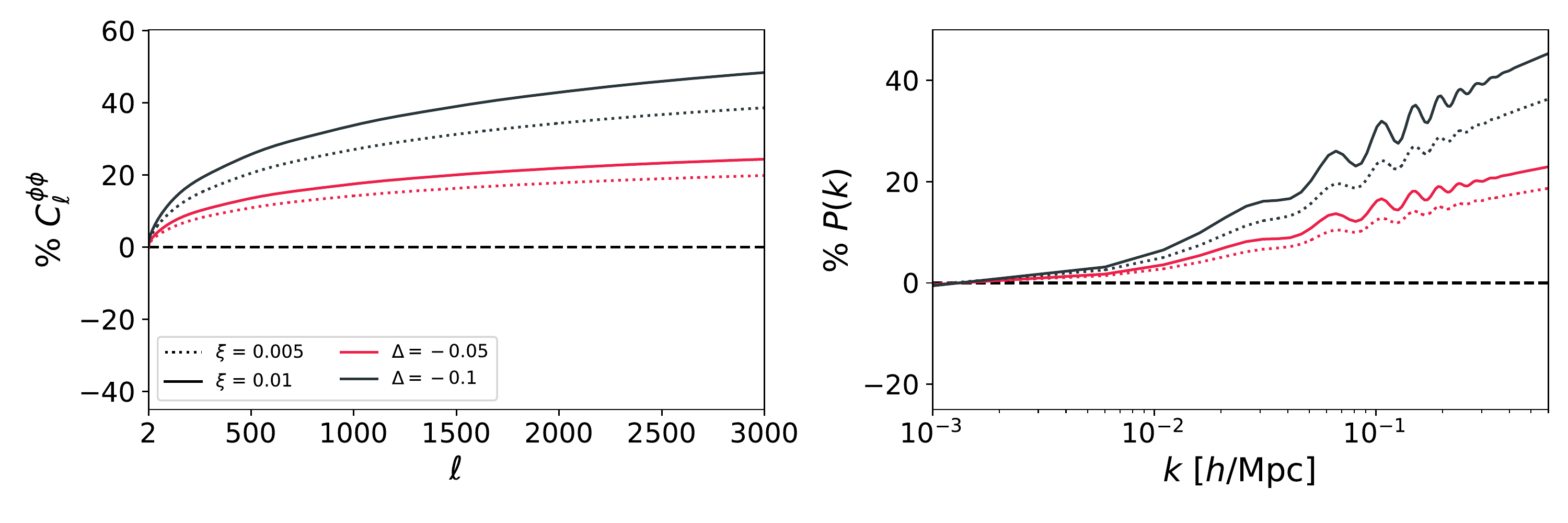}
\includegraphics[width=0.98\textwidth]{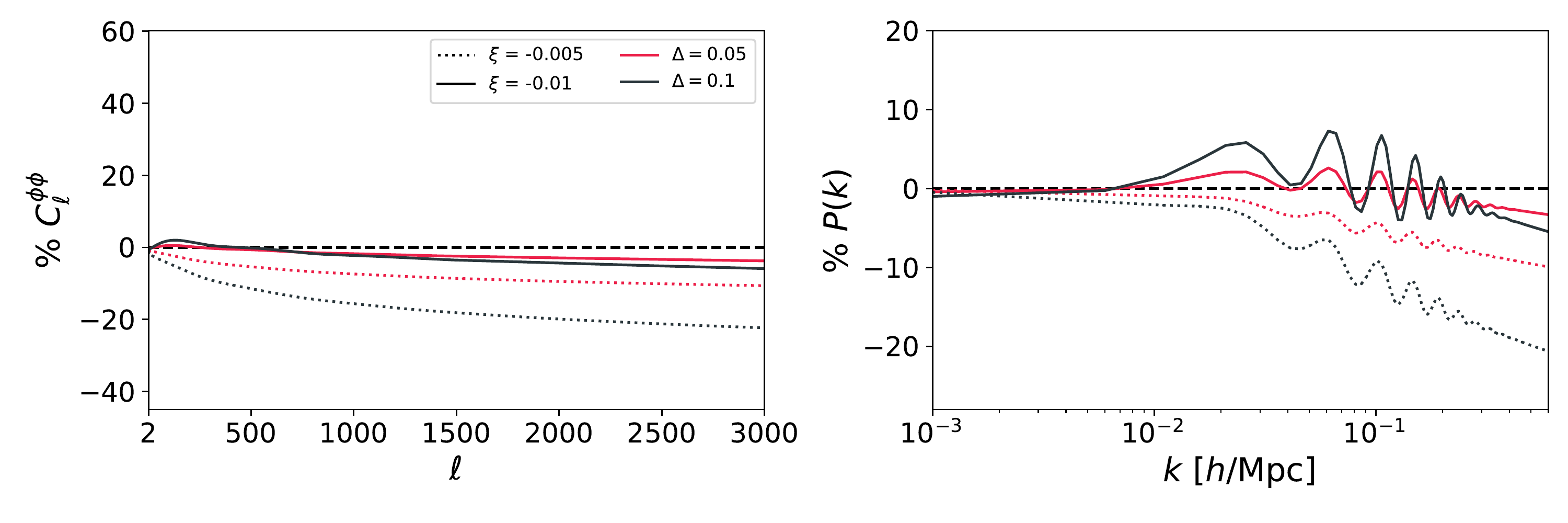}
\caption{Relative differences of the CMB lensing potential angular power spectrum (left panels) and 
linear matter power spectrum at $z=0$ (right panels) for NMC+ (top panels) with $\xi = 0.005,\,0.01$ 
(dotted, solid) and $\Delta = -0.05,\,-0.1$ (red, black), and for NMC- (bottom panels) with 
$\xi = -0.005,\,-0.01$ (dotted, solid) and $\Delta = 0.05,\,0.1$ (red, black) with respect to the 
$\Lambda$CDM.}
\label{fig:nmc_mpk}
\end{figure}

The CC case $\xi=-1/6$ has to be treated separately  
\begin{equation}
    \left(\frac{\sigma_0}{M_{\rm pl}}\right)^2 = 18 \frac{\left(1+\Delta\right)^2-1}{1+3\left(1+\Delta\right)^2} \,.
\end{equation}
In the NMC case $\Delta$ can not be both negative and positive for the same branch of $\xi$ implying 
that for a given sign of $\xi$ the effective gravitational constant can be only larger or only smaller 
than the value of the gravitational constant.
This is due to the assumption $N_{\rm pl} = M_{\rm pl}$ and the condition $F(\sigma) > 0$ to avoid 
negative kinetic energy states in the tensor sector \cite{Gannouji:2006jm}. 
Fig.~\ref{fig:nmc_mpk} shows that for $\Delta >0$ the CMB lensing potential spectrum $C_\ell^{\phi\phi}$ 
and the matter power spectrum at $z=0$ $P(k)$ can be smaller than in $\Lambda$CDM for small scales, a 
trend which does not occur for $\Delta < 0$ or for $\Delta=0$ and $N_{\rm pl} \ne M_{\rm pl}$  \cite{Rossi:2019lgt}.

\begin{figure}
\centering
\includegraphics[width=0.98\textwidth]{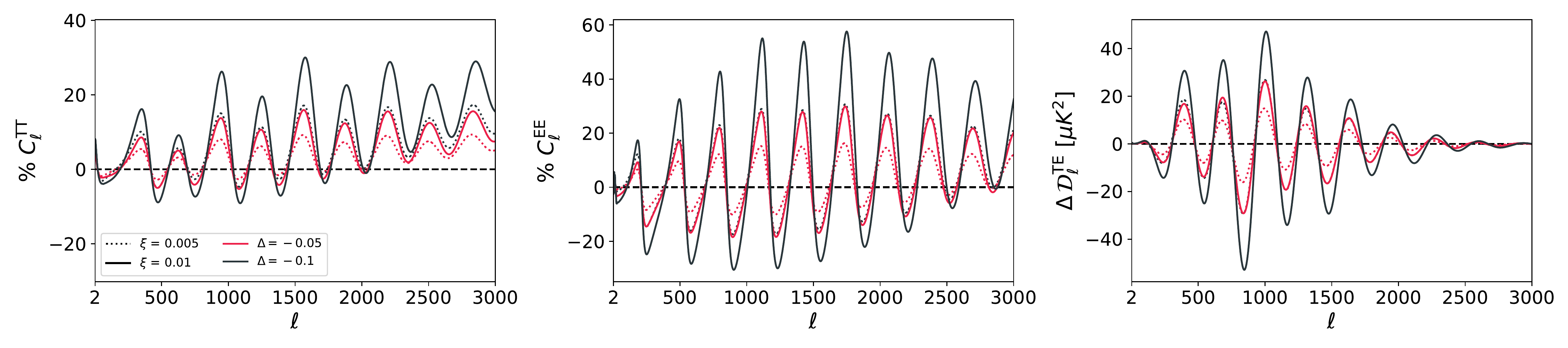}
\includegraphics[width=0.98\textwidth]{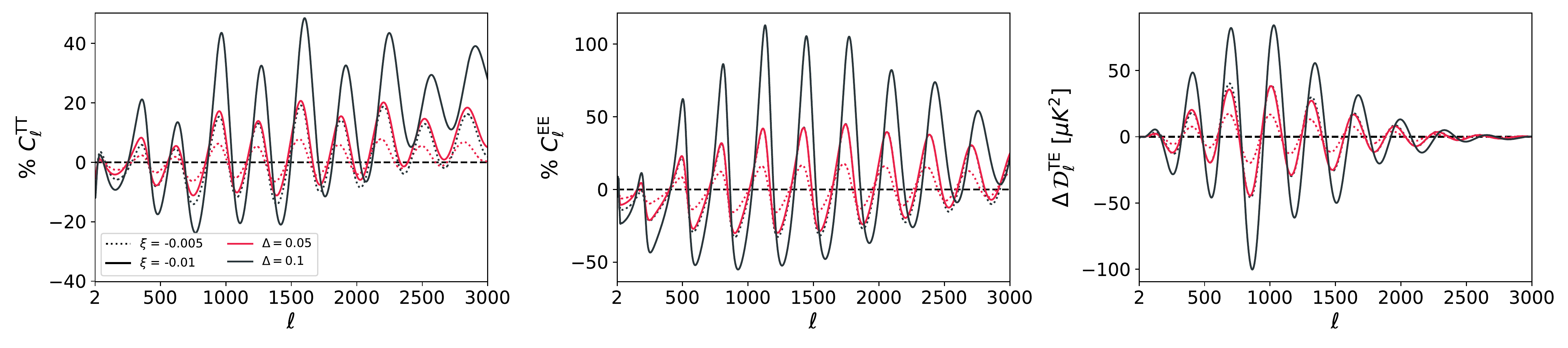}
\caption{Differences with respect to the $\Lambda$CDM CMB lensed angular power spectra for NMC+ with 
$\xi = 0.005,\,0.01$ (dotted, solid) and $\Delta = -0.05,\,-0.1$ (red, black), and for 
NMC- (bottom panels) with $\xi = -0.005,\,-0.01$ (dotted, solid) and $\Delta = 0.05,\,0.1$ (red, 
black). ${\cal D}_\ell \equiv \ell(\ell+1)C_\ell/(2\pi)$ are the band-power angular power spectra.}
\label{fig:nmc_cl}
\end{figure}

\section{Methodology and datasets} \label{sec:data}
In order to derive the constraints on the cosmological parameters we perform a Markov Chain Monte Carlo 
(MCMC) analysis by using the publicly available code 
{\tt MontePython}\footnote{\href{https://github.com/brinckmann/montepython\_public}{https://github.com/brinckmann/montepython\_public}} 
\cite{Audren:2012wb,Brinckmann:2018cvx} connected to our modified version of the code 
{\tt CLASS}\footnote{\href{https://github.com/lesgourg/class\_public}{https://github.com/lesgourg/class\_public}} 
\cite{Lesgourgues:2011re,Blas:2011rf}, i.e. {\tt CLASSig} \cite{Umilta:2015cta}. 
Mean values and uncertainties on the reported parameters, as well as the plotted contours, have been 
obtained using 
{\tt GetDist}\footnote{\href{https://getdist.readthedocs.io/en/latest}{https://getdist.readthedocs.io/en/latest}} \cite{Lewis:2019xzd}. We use adiabatic initial conditions for the scalar field perturbations 
\cite{Paoletti:2018xet,Rossi:2019lgt}.

We vary the six cosmological parameters for a flat $\Lambda$CDM concordance model, i.e. 
$\omega_{\rm b}$, $\omega_{\rm c}$, $H_0$, $\tau$, $\ln\left(10^{10}A_{\rm s}\right)$, $n_{\rm s}$, 
plus the extra parameters related to the coupling to the Ricci curvature. For IG 
$(N_{\rm pl}=0,\,\xi>0)$, we sample on the quantity $\zeta_{\rm IG} \equiv \ln\left(1 + 4\xi\right)$, 
according to \cite{Umilta:2015cta,Ballardini:2016cvy,Ballardini:2020iws} in the prior range 
$[0,\,0.039]$ and $\Delta \in [-0.3,\, 0.3]$\footnote{As in our previous works, we use linear priors 
on $\zeta_{\rm IG}$ which are essentially linear priors on the coupling to the curvature $\xi$ or on 
the deviation of the post-Newtonian parameter $\gamma_{\rm PN}$ for $\xi \ll 1$, which turns out to 
be the range allowed from observations. 
We caution the interested reader in bearing in mind different priors when comparing the constraints on 
the $\xi$ obtained here with those obtained in Refs.~\cite{Avilez:2013dxa,Joudaki:2020shz} where 
priors on $\omega_{\rm BD}$ or $\ln \omega_{\rm BD}$ are considered.}. 
For CC $(N_{\rm pl}=M_{\rm pl},\,\xi=-1/6)$, we sample on $\Delta \in [0,\, 0.1]$. For NMC 
$(N_{\rm pl}=M_{\rm pl},\,\xi\ne0)$, we sample separately on the positive branch with $\xi \in [0,0.3]$ 
and $\Delta \in [-0.1,\, 0]$, and on the negative branch with $\xi \in [-0.3,0]$ and 
$\Delta \in [0,\, 0.1]$. We assume 2 massless neutrino with $N_{\rm eff} = 2.0328$, 
and a massive one with fixed minimum mass $m_\nu = 0.06$ eV. 
We fix the primordial \ce{^{4}He} mass fraction $Y_{\rm p}$ according to the 
prediction from {\tt PArthENoPE} \cite{Pisanti:2007hk,Consiglio:2017pot}, by taking into account 
the relation with the baryon fraction $\omega_{\rm b}$ and the varying gravitational constant which 
enters in the Friedman equation during nucleosynthesis. We indeed consider the varying gravitational 
constant as an additional contribution to the effective relativistic species 
$\Delta N_{\rm eff} = \left[3.046 + \frac{8}{7}\left(\frac{11}{4}\right)^\frac{4}{3}\right]\left(\frac{G_{\rm N}}{G} - 1\right)$ in $Y_{\rm BBN}(\omega_b,\,N_{\rm eff})$ at $z_{\rm BBN}$ \cite{Hamann:2007sb}.

We constrain the cosmological parameters using the CMB anisotropies measurements from the {\em Planck} 
2018 legacy release (hereafter P18), in combination with BAO measurements from galaxy redshift surveys, 
and a Gaussian likelihood based on the determination of the Hubble constant from Hubble Space 
Telescope (HST) observations (hereafter R19), i.e. $H_0 = (73.4 \pm 1.4)$ km s$^{-1}$Mpc$^{-1}$ 
\cite{Reid:2019tiq}.
Our CMB measurements combine temperature, polarization, and weak lensing CMB anisotropies angular power 
spectra \cite{Aghanim:2019ame,Aghanim:2018oex}. The high-multipoles likelihood is based on {\tt Plik} 
likelihood, the low-$\ell$ likelihood is based on the {\tt Commander} likelihood (temperature-only) plus 
the {\tt SimAll} EE-only likelihood, the CMB lensing likelihood is considered on the {\em conservative} 
multipoles range, i.e. $8 \leq \ell \leq 400$. 
We marginalize over foreground and calibration nuisance parameters of the {\em Planck} likelihoods which 
are also varied together with the cosmological ones.
We use BAO data from Baryon Spectroscopic Survey (BOSS) DR12 \cite{Alam:2016hwk} {\em consensus} results 
in three redshift slices with effective redshifts $z_{\rm eff} = 0.38,\,0.51,\,0.61$ in combination with 
measure from 6dF \cite{Beutler:2011hx} at $z_{\rm eff} = 0.106$ and the one from SDSS DR7 
\cite{Ross:2014qpa} at $z_{\rm eff} = 0.15$. In some of our analysis, we also include a Gaussian prior 
on $S_8\equiv \sigma_8\sqrt{\Omega_{\rm m}/0.3}$, $p(S_8)$, based on the inverse-variance weighted 
combination of the weak lensing measurements of DES \cite{DES:2017myr}, KV-450 
\cite{Hildebrandt:2016iqg,Hildebrandt:2018yau}, and HSC \cite{HSC:2018mrq}, i.e. 
$S_8 = 0.770 \pm 0.017$.

\section{Results} \label{sec:results}
In this section we present the results for IG, CC, and NMC based on the combination 
of {\em Planck} 2018 data (P18) \cite{Aghanim:2019ame,Aghanim:2018oex}, BOSS DR12 BAO 
{\em consensus} data \cite{Alam:2016hwk}, and a Gaussian prior on $H_0$ from \cite{Reid:2019tiq}.
We collect tables with the full constraints on the cosmological parameters obtained with our 
MCMC analysis in App.~\ref{sec:appendix1} and the full triangle plots with the standard 
cosmological parameters in App.~\ref{sec:appendix2}.

For {\bf IG}, we obtain the following joint constraints at 68\% CL correspond to 
$\Delta = -0.032^{+0.029}_{-0.025}$, $10^3\, \xi < 2.1$  (P18), 
$\Delta = -0.022\pm 0.023$, $10^3\, \xi < 0.82$ (P18 + BAO), 
and $\Delta = -0.026\pm 0.024$, $10^3\, \xi = 0.74^{+0.52}_{-0.54}$  (P18 + BAO + R19). 
Constraints on the ratio of the effective gravitational constant correspond to 
$G_{\rm eff}/G = 0.938^{+0.056}_{-0.049}$  (P18), 
$G_{\rm eff}/G = 0.957\pm 0.045$ (P18 + BAO), and 
$G_{\rm eff}/G = 0.949\pm 0.048$ (P18 + BAO + R19) at 68\% CL. 

\begin{figure}
\centering
\includegraphics[width=0.98\textwidth]{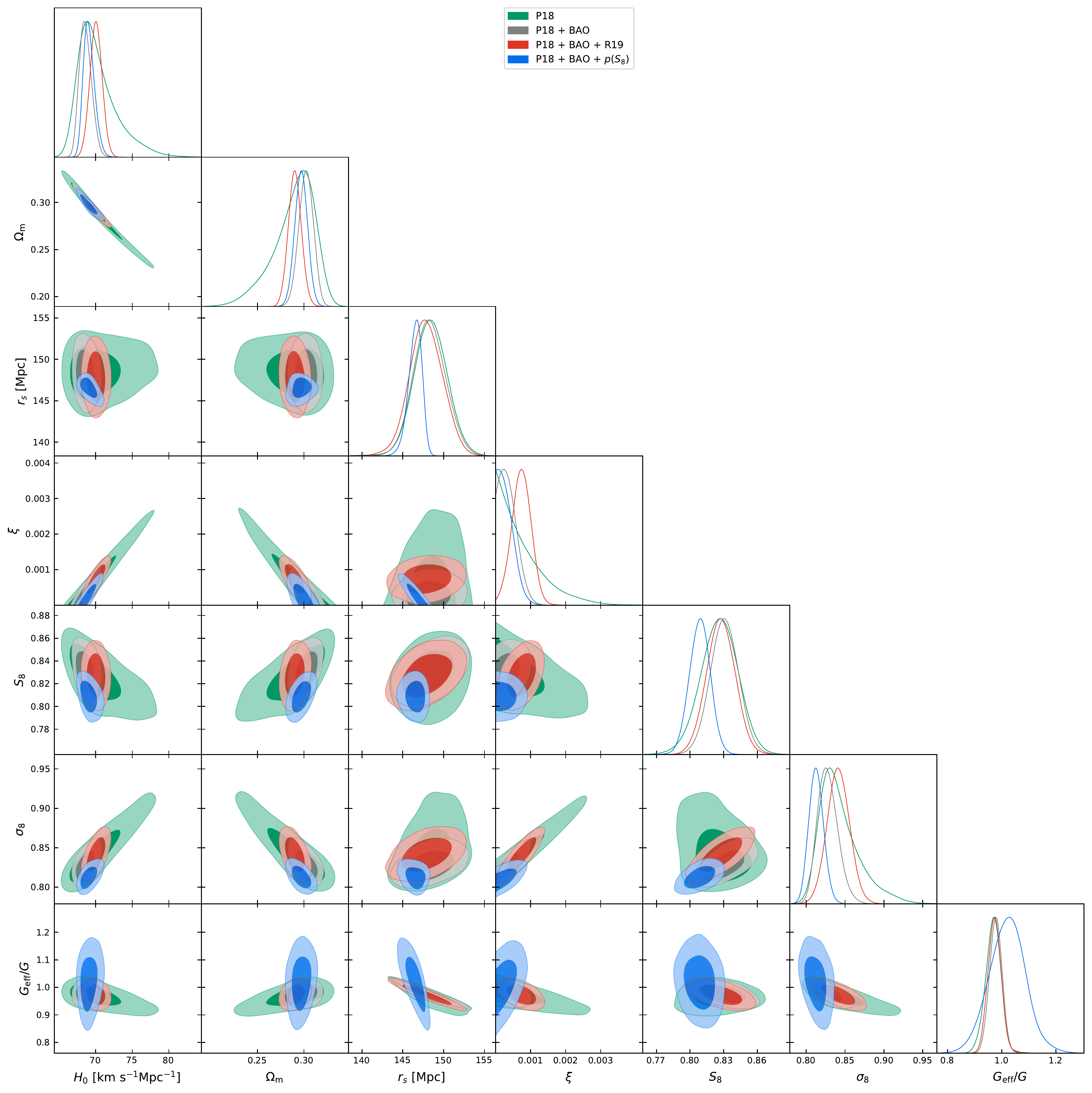}
\caption{Marginalized joint 68\% and 95\% CL regions 2D parameter space using the $Planck$ legacy data 
(green), its combination with BAO DR12, i.e. P18 + BAO (gray), and P18 + BAO + R19 (red), for the IG 
model. The blue contours include a Gaussian prior on $S_8$.}
\label{fig:dig}
\end{figure}

\begin{figure}
\centering
\includegraphics[height=0.3\textwidth]{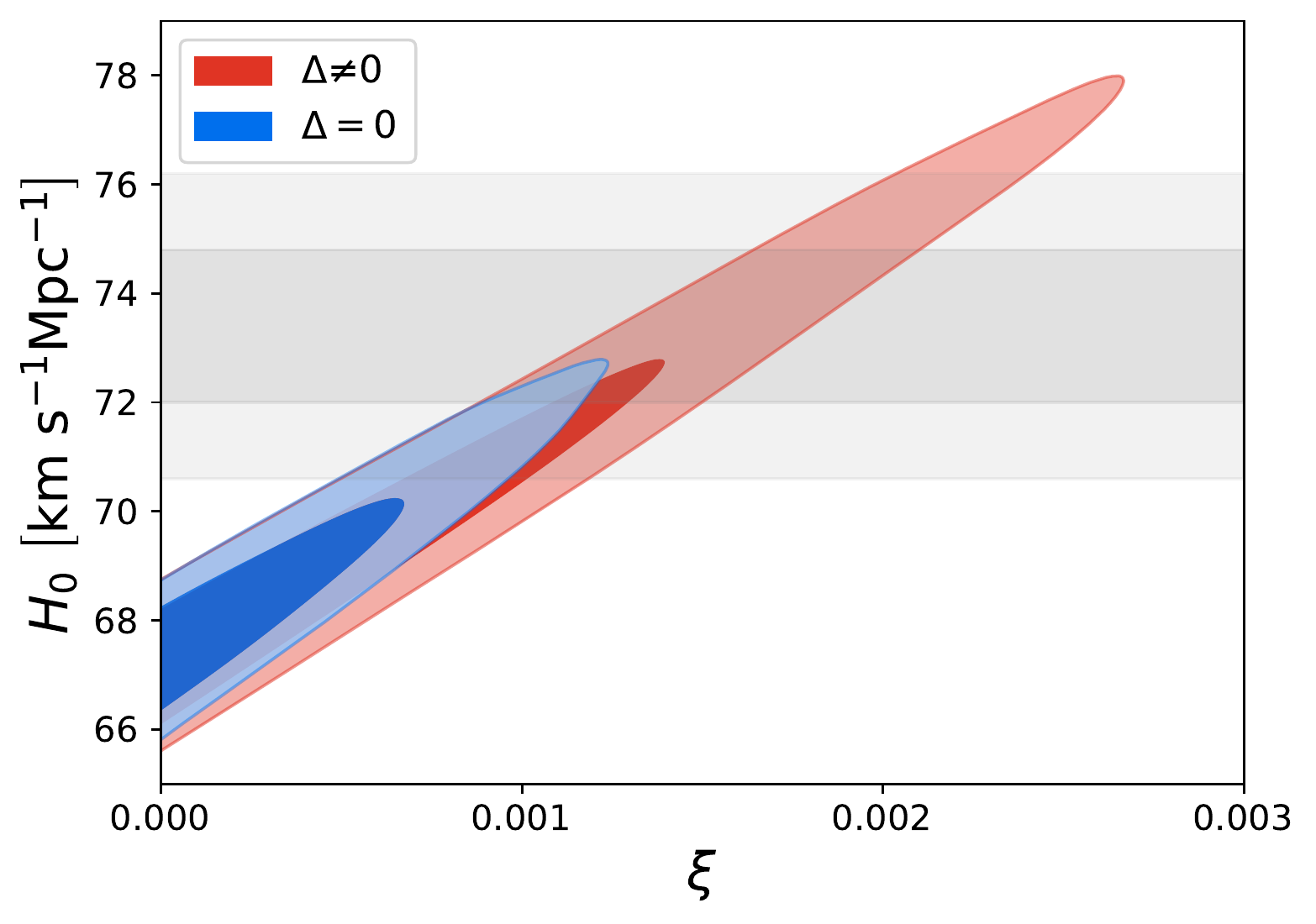}
\includegraphics[height=0.3\textwidth]{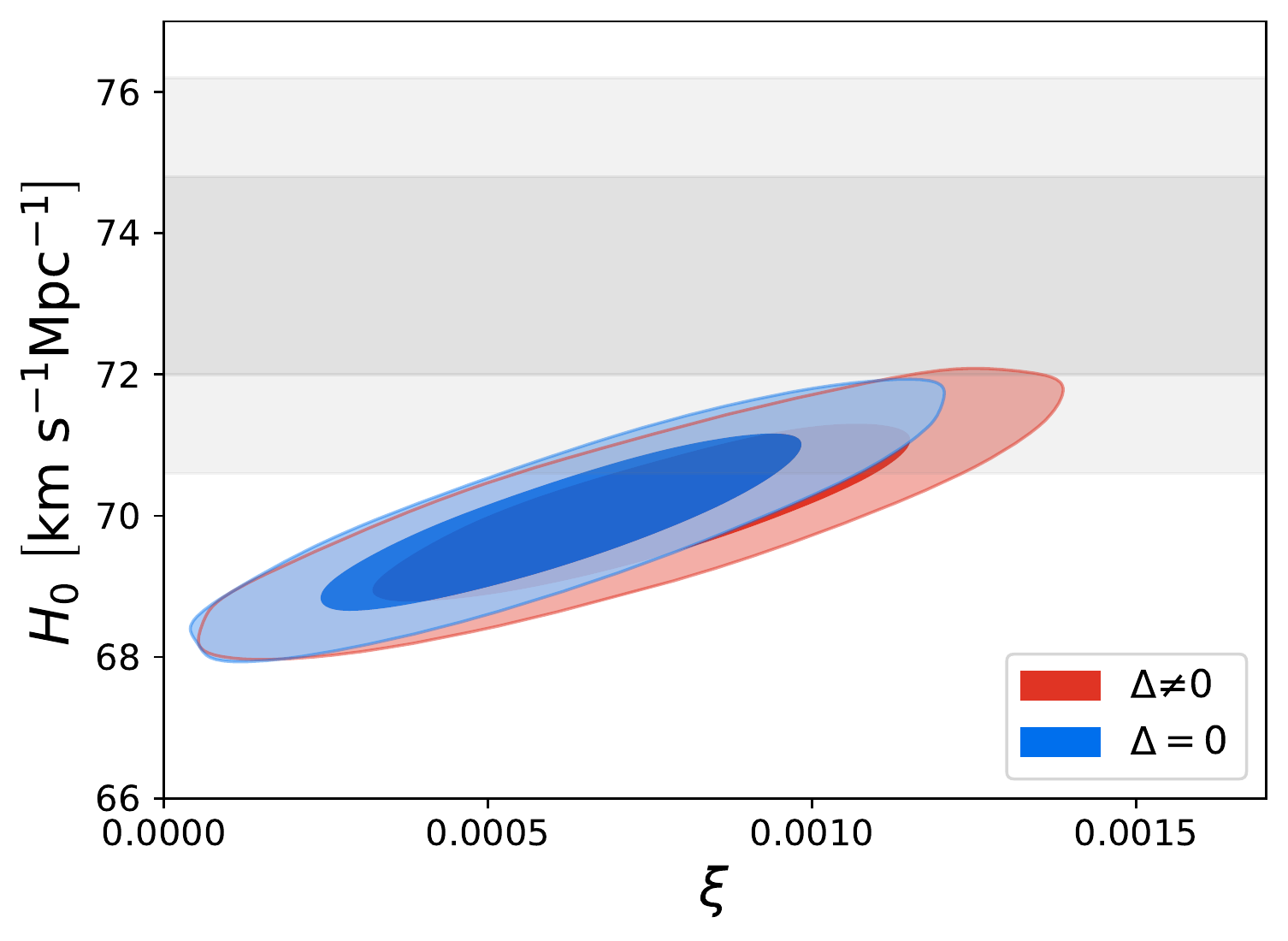}
\includegraphics[height=0.3\textwidth]{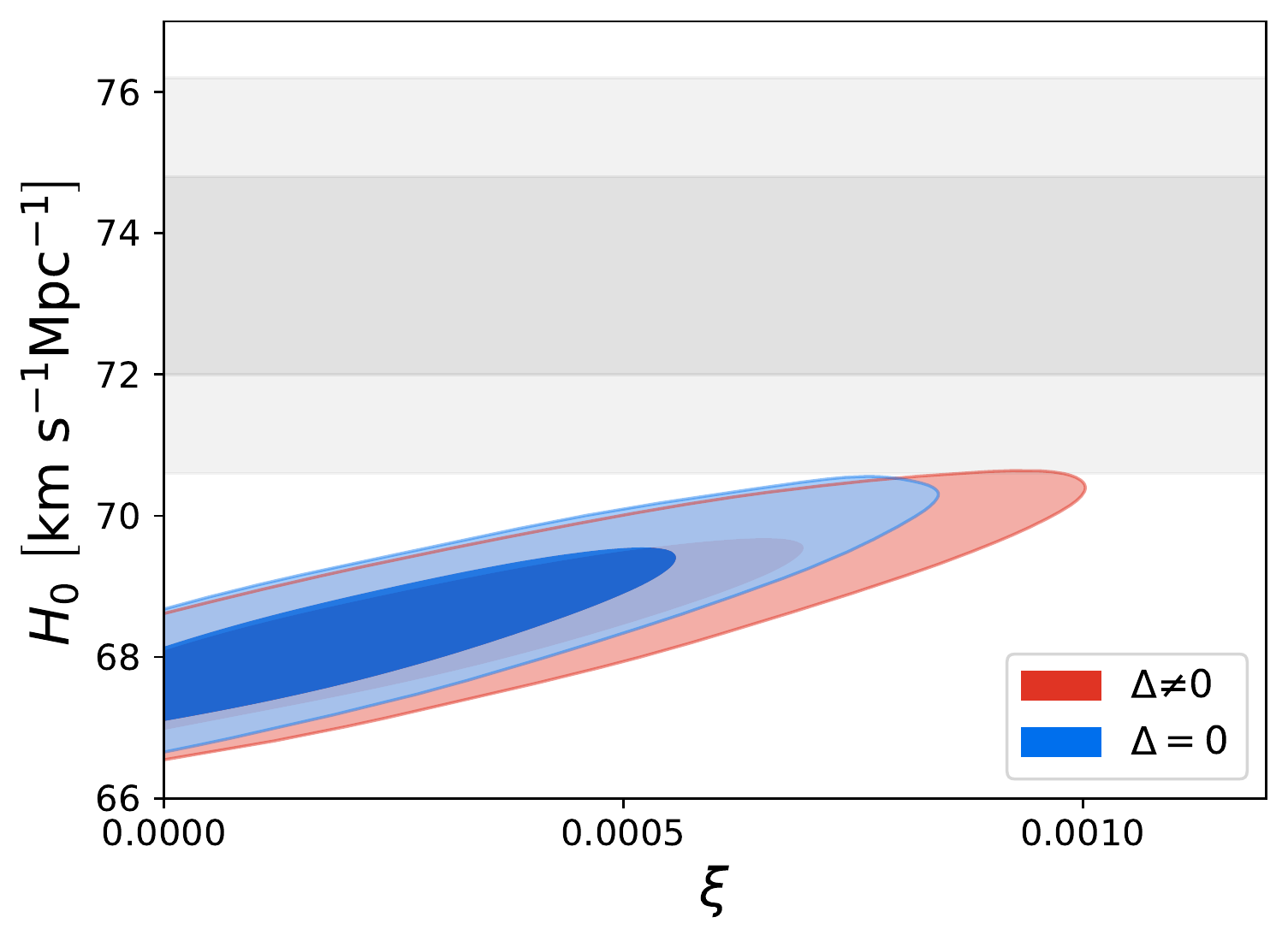}
\includegraphics[height=0.3\textwidth]{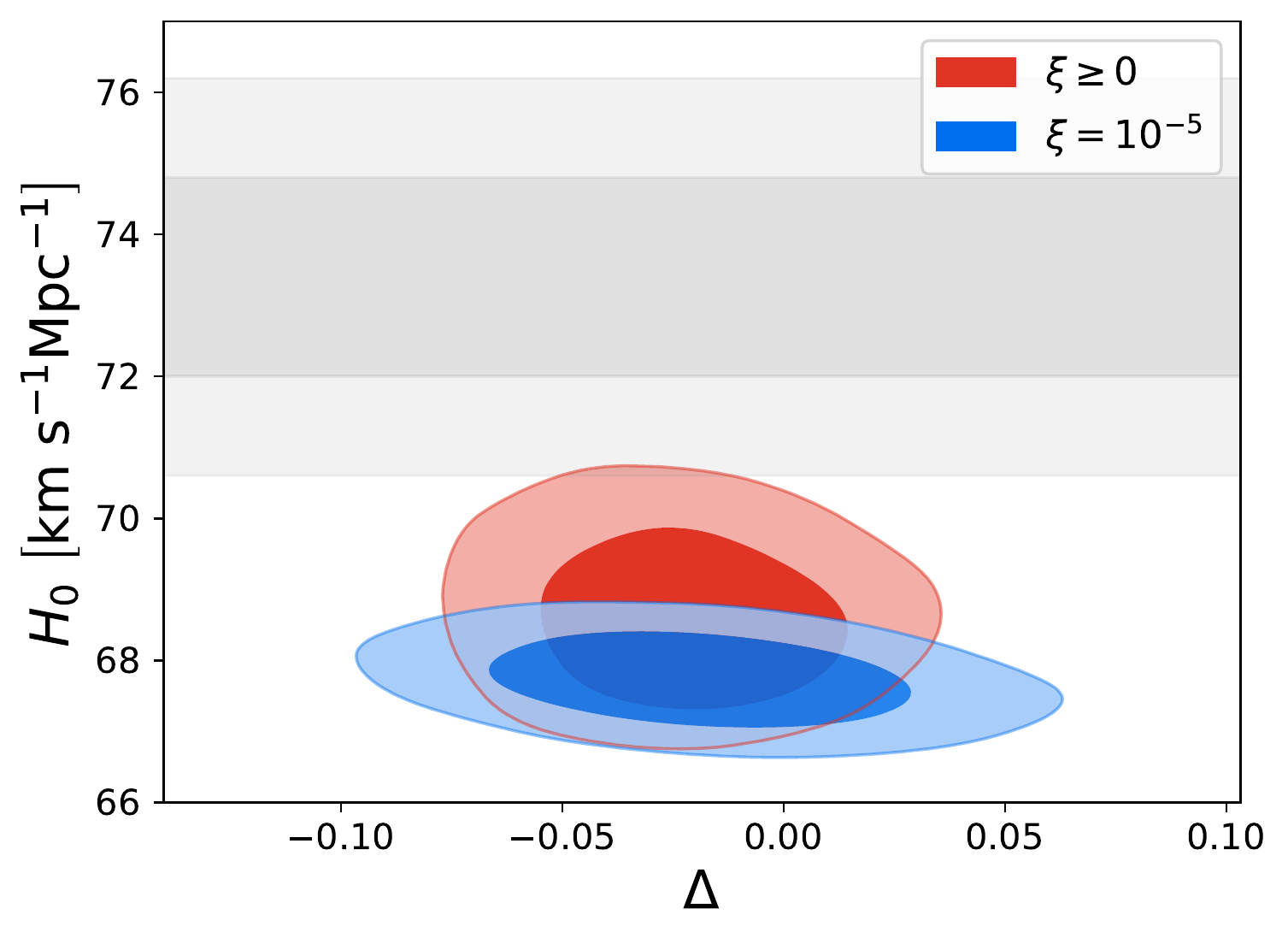}
\caption{Marginalized joint 68\% and 95\% CL regions 2D parameter space using the $Planck$ legacy data 
(upper left panel), its combination with BAO DR12, i.e. P18 + BAO (bottom left panel), and 
P18 + BAO + R19 (upper right panel) for the IG model with $\Delta \ne 0$ ($\Delta = 0$) in red (blue). 
We show also the combination P18 + BAO on the parameter space $H_0$-$\Delta$ where we fix 
$\xi = 10^{-5}$ (bottom right panel).}
\label{fig:dig_comparison}
\end{figure}

In Fig.~\ref{fig:dig_comparison}, we compare the results with the case $\Delta = 0$ over the parameter space $H_0$-$\xi$. We see that relaxing the condition on $G_{\rm eff}$, i.e. Eq.~\eqref{eqn:boundary}, 
the constraints on the coupling become larger: $10^3\, \xi < 2.1,\, < 0.82,\, = 0.74^{+0.52}_{-0.54}$ 
at 95\% CL with $\Delta \ne 0$ compared to $10^3\, \xi < 0.96,\,< 0.68,\, = 0.62^{+0.48}_{-0.46}$ with 
$\Delta = 0$ for P18, P18 + BAO, P18 + BAO + R19 respectively. In particular, varying $\Delta$ the 
uncertainties on $\xi$ become two times larger using CMB data alone, $\sim 21\%$ larger for P18 + BAO, 
and $\sim 13\%$ larger for P18 + BAO+ R19. Once we include R19, the larger value of $H_0$ comes with a 
$2.5\sigma$ detection of the coupling $\xi = 0.00074^{+0.00052}_{-0.00054}$ at 95\% CL.

\begin{figure}
\centering
\includegraphics[width=0.98\textwidth]{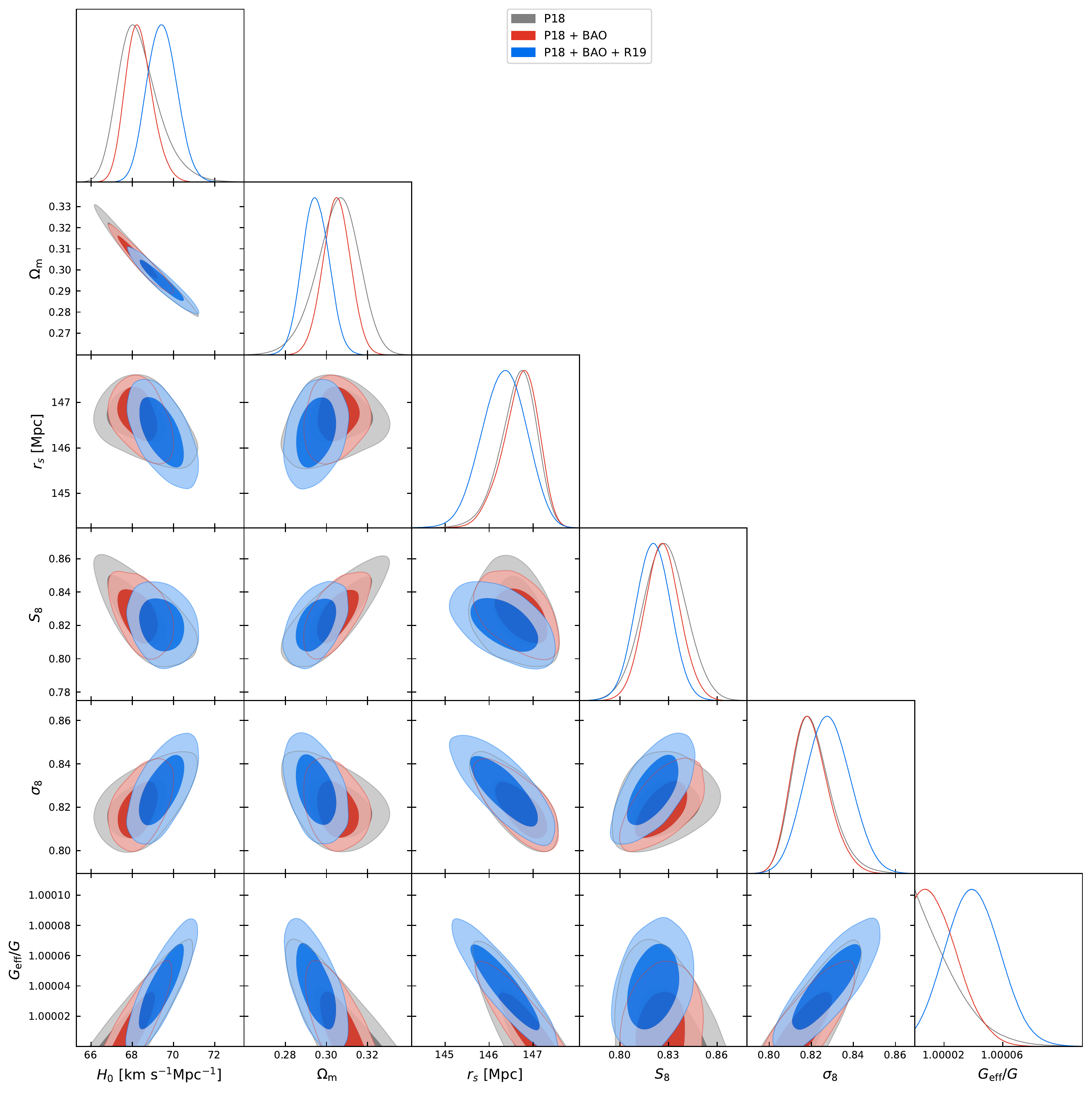}
\caption{Marginalized joint 68\% and 95\% CL regions 2D parameter space using the $Planck$ legacy data 
(gray), its combination with BAO DR12, i.e. P18 + BAO (red), and P18 + BAO + R19 (blue) for the CC model.}
\label{fig:dcc}
\end{figure}

\begin{figure}
\centering
\includegraphics[height=0.23\textwidth]{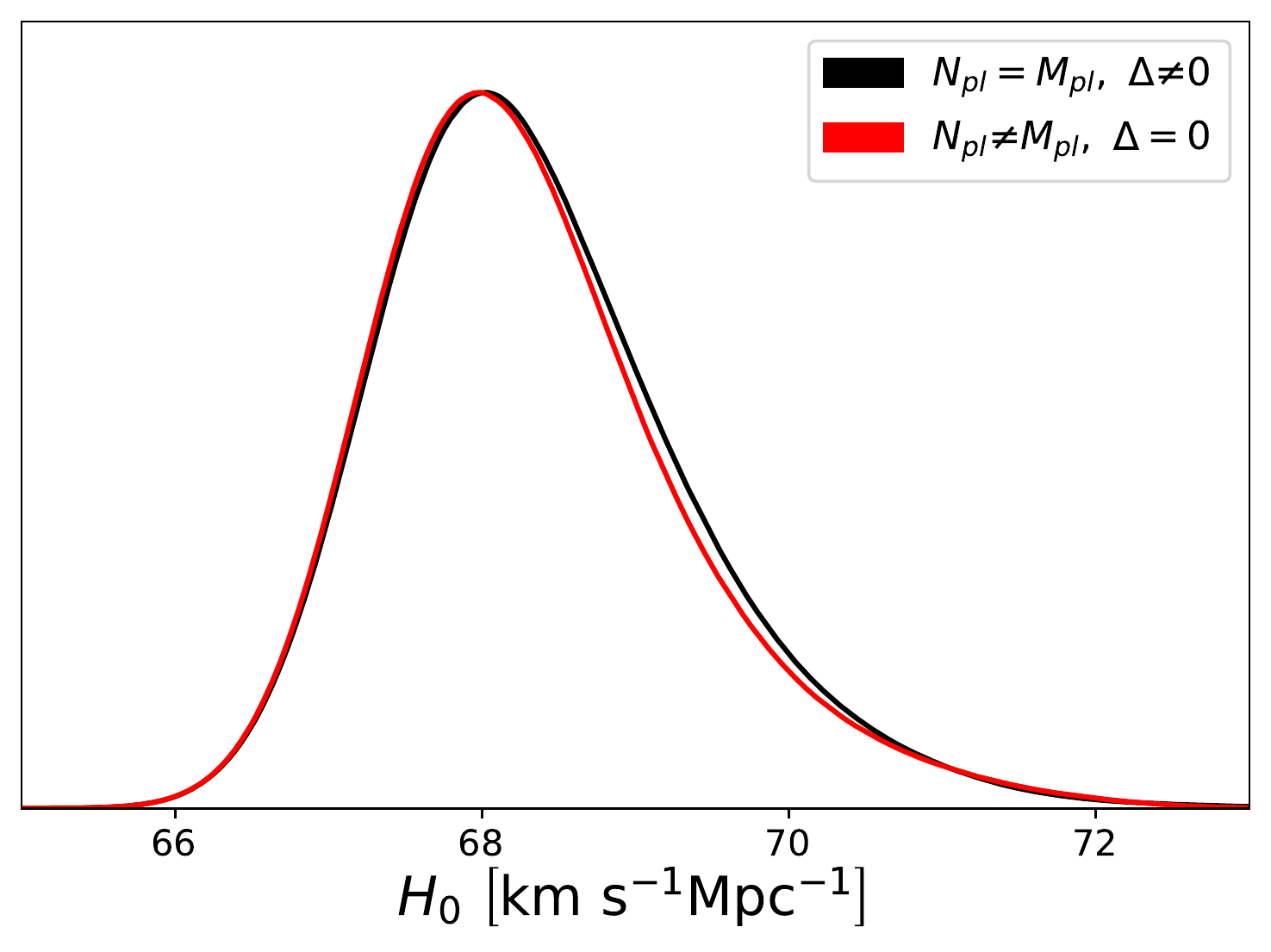}
\includegraphics[height=0.23\textwidth]{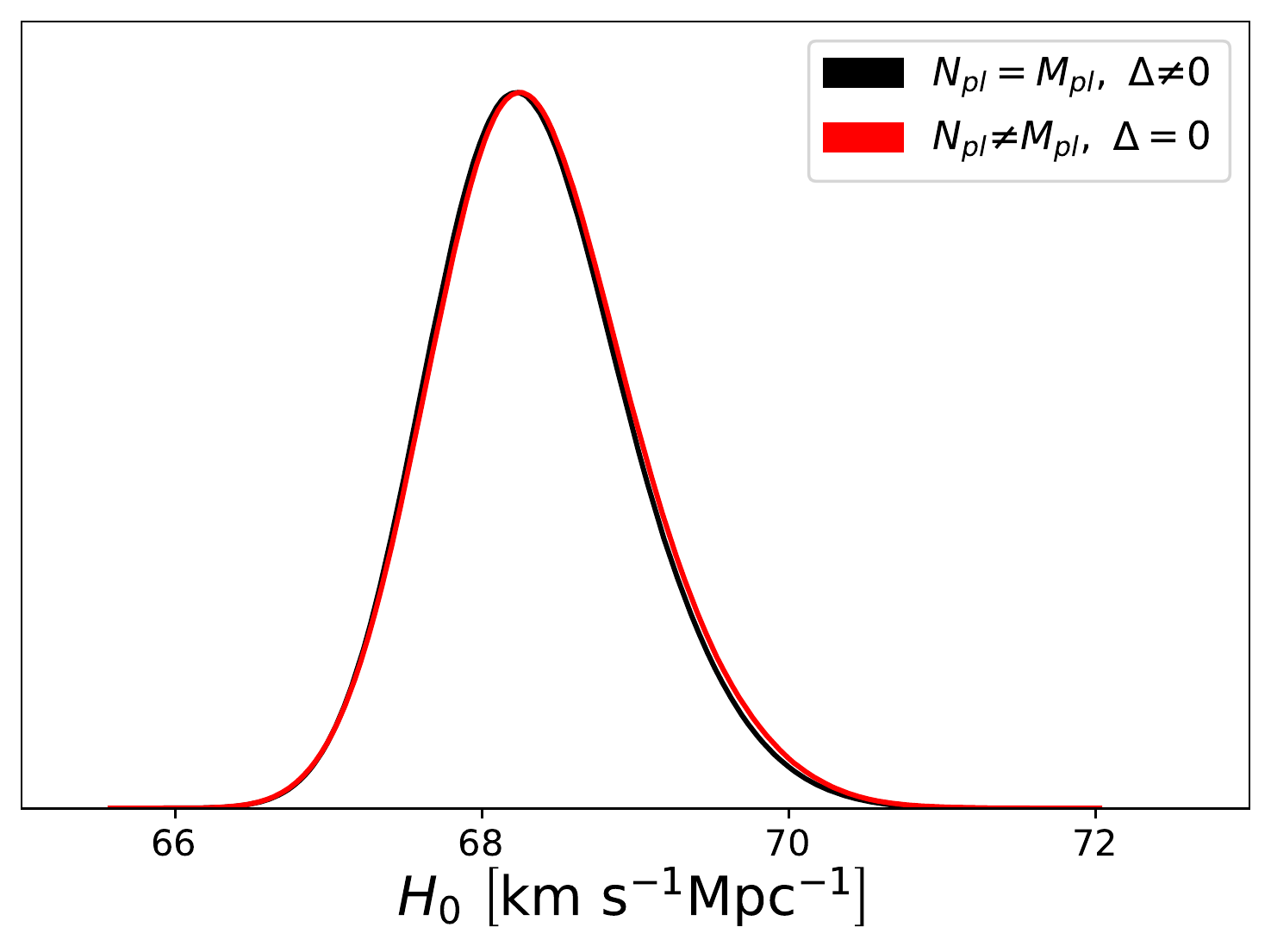}
\includegraphics[height=0.23\textwidth]{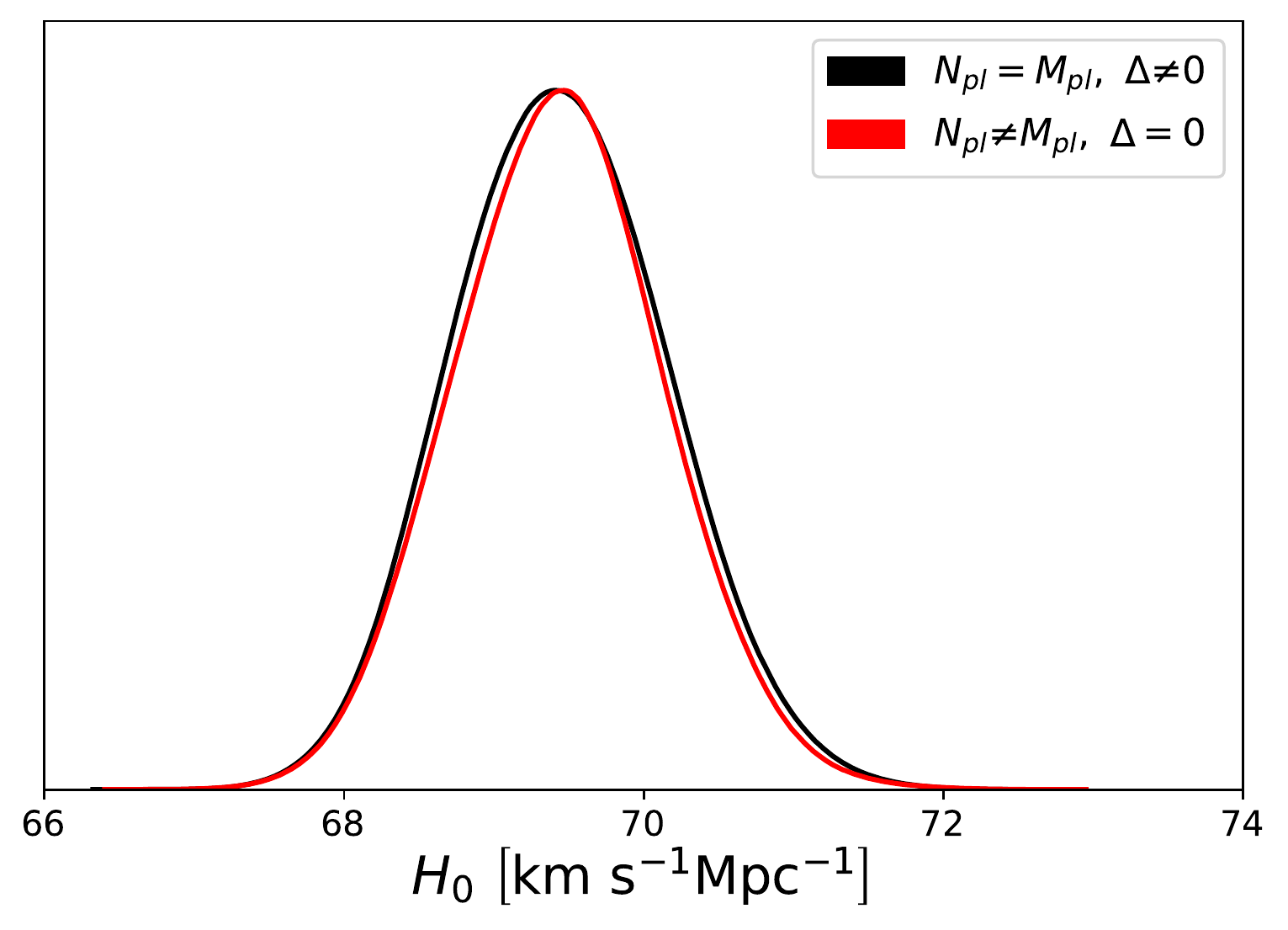}
\caption{Marginalized joint 68\% and 95\% CL regions 2D parameter space using the $Planck$ legacy data 
(left panel) in combination with BAO DR12, i.e. P18 + BAO (central panel), and P18 + BAO + R19 (right 
panel) for the CC model with $\Delta \ne 0$ ($\Delta = 0$) in red (black).}
\label{fig:dcc_comparison}
\end{figure}

The {\bf CC} case remains tightly constrained, as shown in Fig.~\ref{fig:dcc}, see also 
Refs.~\cite{Rossi:2019lgt,Ballardini:2020iws,Braglia:2020auw}. The large value of $\xi = -1/6$ 
requires a small value of the scalar field $\sigma (z)/M_{\rm pl}$ at early time in order to satisfy 
the CMB constraints on $F(\sigma)$. This is reflected on a tight constraint 
$\Delta < 2.3 \times 10^{-5}$ at 95\% CL for P18 + BAO (analogously to the constraints 
$N_{\rm pl} < 1.000023\ $M$_{\rm pl}$ \cite{Ballardini:2020iws} at 95\% CL for P18 + BAO).
We show in Fig.~\ref{fig:dcc_comparison} how negligible is the dependence on different priors for the 
CC case either sampling linearly on $\Delta$ with $N_{\rm pl}=M_{\rm pl}$ as done here or on 
$N_{\rm pl} > M_{\rm pl}$ as in Refs.~\cite{Rossi:2019lgt,Ballardini:2020iws}, the posterior probability 
for cosmological parameters are the same.

For {\bf NMC+} ({\bf NMC$-$}) we sample $\Delta$ over range $[-0.1,\, 0]$ ($[0,\,0.1]$) so that the 
value of the effective gravitational constant today is always smaller (larger) than $G$. Constraints 
on $\Delta$ for NMC+ correspond to $\Delta > -0.018$ (P18 + BAO) at 95\% CL and to 
$\Delta = -0.0072^{+0.0053}_{-0.0020}$ (P18 + BAO + R19) at 68\% CL. Constraints on the ratio of the 
effective gravitational constant correspond to $G_{\rm eff}/G > 0.964$ (P18 + BAO) at 95\% CL and 
$G_{\rm eff}/G = 0.986^{+0.011}_{-0.0041}$ (P18 + BAO + R19) at 68\% CL. Analogously for NMC$-$, we 
obtain $\Delta < 0.021$ (P18 + BAO) and $\Delta < 0.030$ (P18 + BAO + R19) both at 95\% CL. 
Constraints at 95\% CL on the ratio of the effective gravitational constant correspond to 
$G_{\rm eff}/G < 1.04$ (P18 + BAO) and $G_{\rm eff}/G < 1.06$ (P18 + BAO + R19). 

As previously observed in Refs.~\cite{Rossi:2019lgt,Ballardini:2020iws,Braglia:2020iik}, there is a strong 
degeneracy between the coupling parameters for the form $F(\sigma) = N_{\rm pl}^2 + \xi \sigma^2$ also 
in this case with $N_{\rm pl} = M_{\rm pl}$ opening to $\Delta \ne 0$.
Since our data constrains the deviations ${\cal O}(\xi\sigma^2/M_{\rm pl}^2)$ from $M_{\rm pl}^2$, 
we loose constraining power on $\xi$ for small values of $\Delta$ corresponding to the limit for 
$\sigma_0 \to 0$ (see Eq.~\eqref{eqn:boundary_nmc}).

\begin{figure}
\centering
\includegraphics[width=0.98\textwidth]{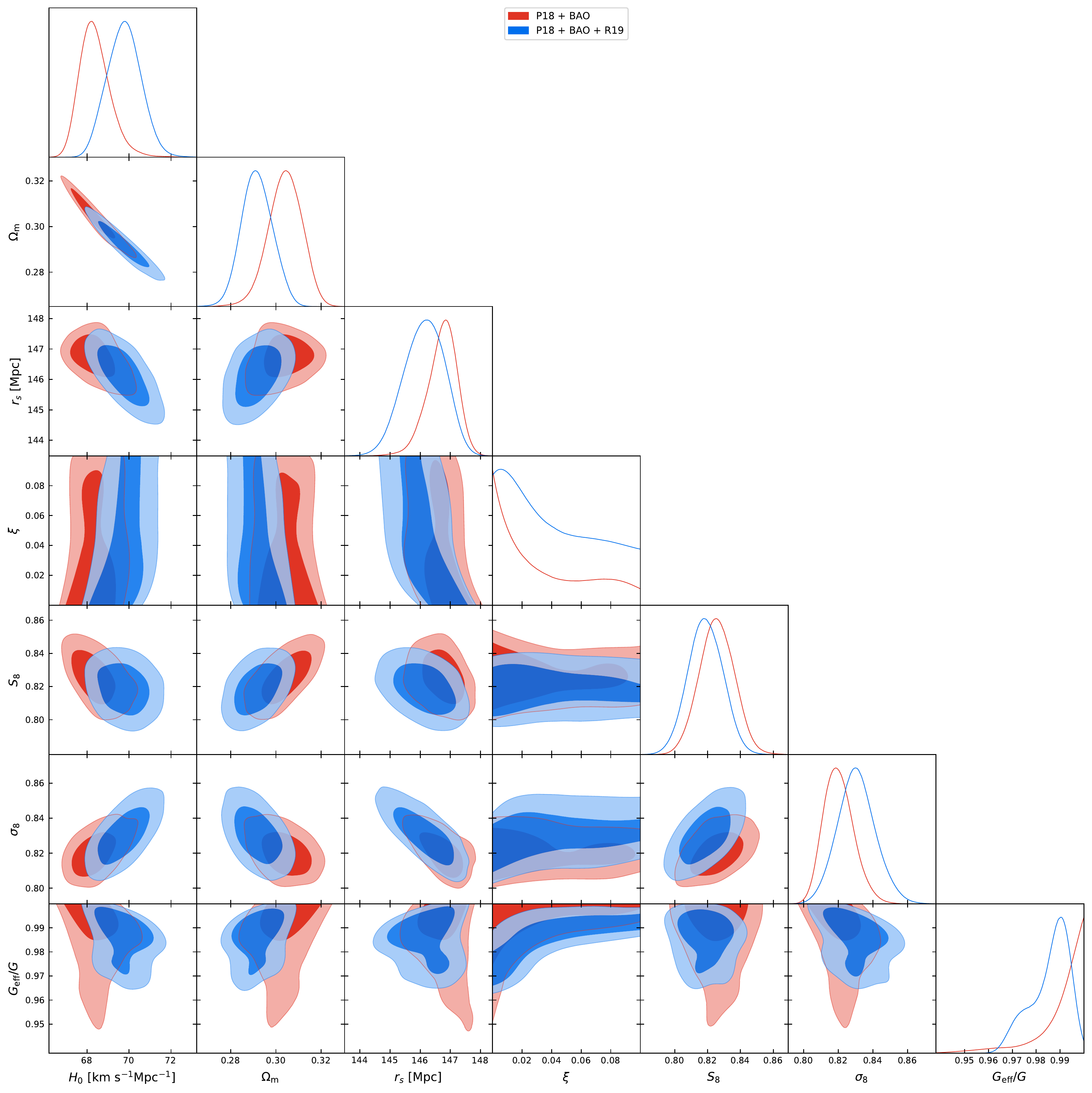}
\caption{Marginalized joint 68\% and 95\% CL regions 2D parameter space using the $Planck$ legacy data 
in combination with BAO DR12, i.e. P18 + BAO (red) and P18 + BAO + R19 (blue) for the NMC+ model.}
\label{fig:dnmc_xp}
\end{figure}

\begin{figure}
\centering
\includegraphics[width=0.98\textwidth]{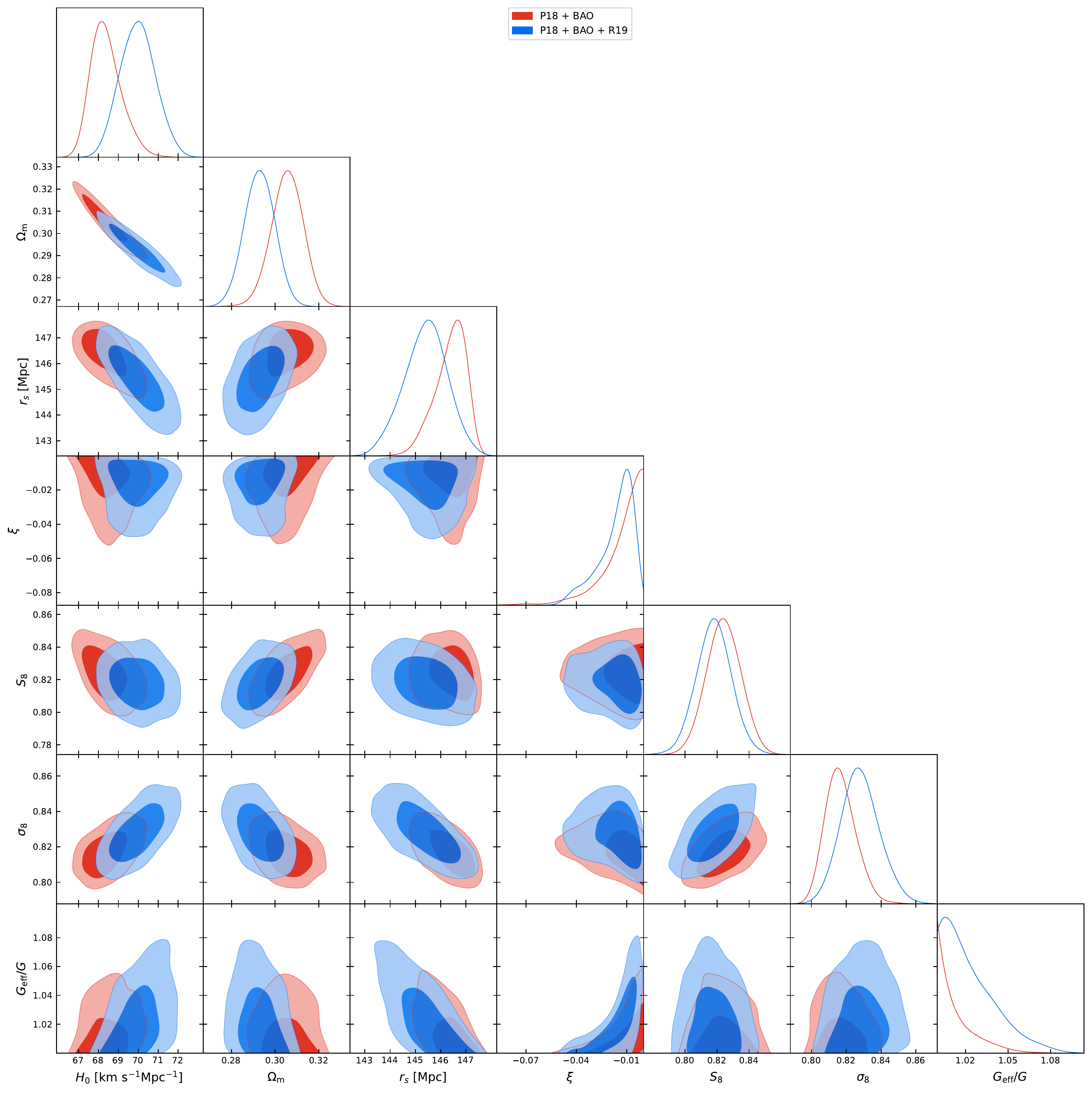}
\caption{Marginalized joint 68\% and 95\% CL regions 2D parameter space using the $Planck$ legacy data 
in combination with BAO DR12, i.e. P18 + BAO (red) and P18 + BAO + R19 (blue) for the NMC- model.}
\label{fig:dnmc_xm}
\end{figure}

\subsection{Implications for the $H_0$ and $S_8$ tensions}
As already pointed out in previous studies (see Refs.~\cite{Umilta:2015cta,Ballardini:2020iws}), these 
models alleviate the $H_0$ {\em tension} compared to the $\Lambda$CDM concordance model thanks to the 
early-time contribution to the radiation density budget in the radiation-dominated epoch and to the 
modification of the background expansion history.
Extending the models to $\Delta \ne 0$, we find that the constraints on $H_0$ from CMB alone are larger 
while they are slightly affected once BAO are included, see 
Figs.~\ref{fig:dig_comparison}-\ref{fig:dcc_comparison}.
We find for IG $H_0 = \left(70.2^{+1.2}_{-3.1}\right)$, $\left(68.61^{+0.72}_{-0.94}\right)$, 
$\left(70.04\pm 0.83\right)$ km s$^{-1}$ Mpc$^{-1}$ with $\Delta \ne 0$ at 68\% CL, compared to 
$H_0 = \left(68.82^{+0.8}_{-1.7}\right)$, $\left(68.57^{+0.62}_{-0.90}\right)$, 
$\left(69.93\pm 0.81\right)$ km s$^{-1}$ Mpc$^{-1}$ with $\Delta = 0$ for P18, P18 + BAO, 
P18 + BAO + R19 respectively. Indeed, the larger value of $H_0$ inferred is mainly driven by $\xi$ once 
BAO are included. In Fig.~\ref{fig:dig_comparison} (bottom-right panel), we show that the marginalized 
posterior distribution of $H_0$ shrinks towards smaller values when fixing $\xi = 10^{-5}$. The differences 
are smaller for the CC model.

NMC scalar-tensor models usually lead to a larger value of $\sigma_8$, in particular a positive 
correlation between the NMC parameters and both $H_0$ and $\sigma_8$ alleviating the $H_0$ tension while 
exacerbating the discrepancy between the value of $S_8$ inferred and the one observed by galaxy shear 
experiments.
These theories predict a decreasing effective gravitational constant which is always larger than the 
Newton's measured constant if we impose Eq.~\eqref{eqn:boundary} \cite{Umilta:2015cta,Rossi:2019lgt}. 
Relaxing the boundary condition on the present value of the effective gravitational constant, it is 
possible to generate a regime of weaker or stronger gravity at low redshift connected with an higher 
or lower value of $\sigma_8$ compared to the $\Lambda$CDM prediction; in particular it is possible to 
reduce the value of $\sigma_8$ for values $\Delta > 0$ in IG and NMC$-$ keeping a larger value of $H_0$, 
as shown in Figs.~\ref{fig:ig_cc_mpk}-\ref{fig:nmc_mpk} and in 
Figs.~\ref{fig:dig}-\ref{fig:dcc}-\ref{fig:dnmc_xp}-\ref{fig:dnmc_xm}, in contrast to what happens in 
NMC with $\Delta = 0$ or in EDE models.

\begin{figure}
\centering
\includegraphics[width=0.6\textwidth]{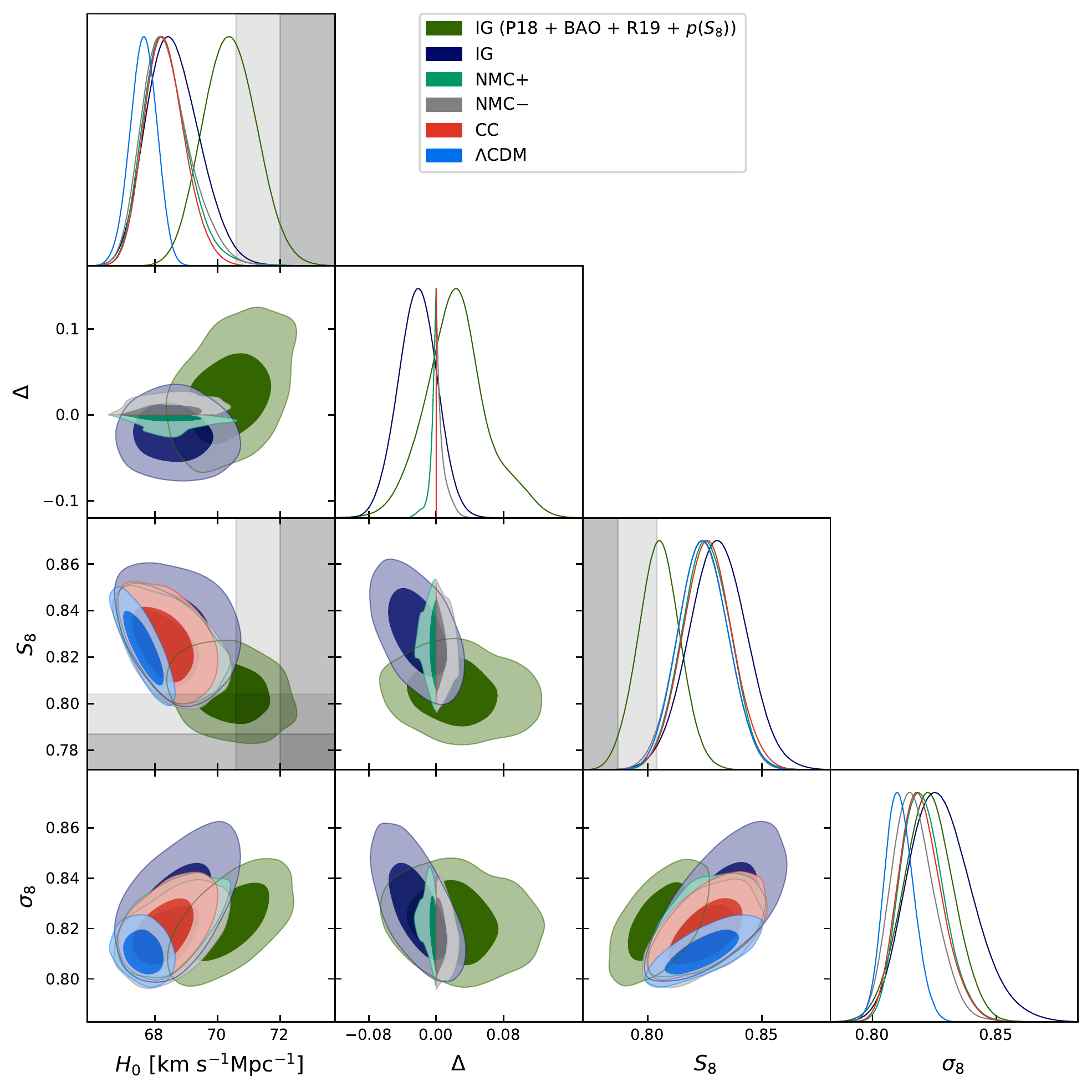}
\caption{Marginalized joint 68\% and 95\% CL regions 2D parameter space using the $Planck$ legacy data 
in combination with DR12 BAO data (P18 + BAO) for all the models analysed (if not otherwise stated). 
The gray bands denote the local Hubble parameter measurement from R19, i.e. $H_0 = (73.4 \pm 1.4)$ 
km s$^{-1}$Mpc$^{-1}$ \cite{Reid:2019tiq}, and the inverse-variance weighted combination of the 
DES/KV-450/HSC weak lensing measurements, i.e. $S_8 = 0.770 \pm 0.017$ 
\cite{DES:2017myr,Hildebrandt:2016iqg,Hildebrandt:2018yau,HSC:2018mrq}.}
\label{fig:H0S8}
\end{figure}

Finally we compare the theoretical predictions of different models on the parameters $S_8$ and $H_0$ 
for the combination of CMB and BAO data, see Fig.~\ref{fig:H0S8}. As it can be seen for IG in 
Fig.~\ref{fig:ig_bkg}, while $\xi$ affects all the quantities plotted and can be connected to both the 
$H_0$ and $S_8$ tension, $\Delta$ plays a role mainly for the $S_8$ tension and does not affect 
significantly the value of the Hubble constant, see also Ref.~\cite{Joudaki:2020shz}. In order to put 
to test a possible connection between $\Delta$ and $S_8$, we test the addition of a Gaussian prior 
$p(S_8)$ on $S_8$ to our fit. In this way for IG we obtain an higher value for 
$\Delta = 0.001^{+0.035}_{-0.029}$ at 68\% CL and approximately a 2$\sigma$ decrease in the value of 
$S_8$, compared to P18 + BAO results in Table~\ref{tab:dig}.
We leave the purpose of a full weak lensing analysis to a future work.

\section{Forecasts} \label{sec:forecast}
We perform the Fisher forecast analysis for the 6 standard cosmological parameters $\omega_c$, 
$\omega_b$, $H_0$, $\tau$, $n_s$, $\ln(10^{10}A_s)$, and the extra parameters $\xi$, $\Delta$.
For the CMB we consider also the optical depth at reionization $\tau$ and then we marginalize over it 
before combining it with the CMB Fisher matrix with the LSS ones. For the standard parameters, we assume 
as fiducial model a flat cosmology with best-fit parameters corresponding to $\omega_{\rm c}=0.12$, 
$\omega_{\rm b}=0.02237$, $H_0=67.36$ km s$^{-1}$ Mpc$^{-1}$, $\tau=0.0544$, $n_{\rm s}=0.9649$, 
$\ln\left(10^{10}\ A_{\rm s}\right)=3.044$, and one massive neutrino with $m_\nu=0.06$ eV consistent 
with the results of $Planck$ DR3 \cite{Aghanim:2018eyx}.
As fiducial value for the extra parameters for IG, we choose $\xi=10^{-5}$ and $\Delta = 10^{-3}$.

\subsection{Cosmic microwave background anisotropies}
As specifications for future CMB measurements, we consider the combination of the Lite (Light) satellite 
for the study of B-mode polarization and Inflation from cosmic background Radiation Detection (LiteBIRD) 
\cite{LiteBIRD:2020khw}, selected by the Japan Aerospace Exploration Agency (JAXA) as a strategic large 
class mission, and CMB-S4 \cite{Abazajian:2019eic} as representative of current and future CMB 
experiments from ground, which also include SPT-3G \cite{SPT-3G:2014dbx}, Simons Observatory 
\cite{SimonsObservatory:2018koc}. For LiteBIRD, we consider the multipole range between 
$2 \leq \ell < 30$. For CMB-S4, following \cite{Ballardini:2019tho,Abazajian:2019eic}, we assume a 
sensitivity $\sigma_{\rm T}^{1/2} = \sigma_{\rm P}^{1/2}/\sqrt{2} = 1\ \mu$K-arcmin with a beam 
resolution of $\theta_{\rm FWHM} = 3$ arcmin over 40\% of the sky, with $\ell_{\rm min}= 30$ and a 
different cut at small scales of $\ell^{\rm T}_{\rm max} = 3000$ in temperature and 
$\ell^{\rm P}_{\rm max}= 5000$ in polarization motivated by the excess of foreground contamination 
expected on the small scales in temperature. We use the CMB lensing information in the range 
$30 \leq \ell \leq 3000$, assuming the minimum variance quadratic estimator for the lensing 
reconstruction, combining the TT, EE, BB, TE, TB, and EB  estimators, calculated according to 
Ref.~\cite{Hu:2001kj} and applying iterative lensing reconstruction (see 
Ref.~\cite{Hirata:2003ka,Smith:2010gu}).

\subsection{Spectroscopic galaxy clustering} \label{sec:gc_forecast}
To describe the main galaxy clustering observable, we follow \cite{Ballardini:2019tho} and we modify 
the observed galaxy power spectrum in order to account for the non-linear effects according to 
\cite{Seo:2003pu,Song:2008qt,Wang:2012bx,Euclid:2019clj}. The full anisotropic non-linear observed 
galaxy power spectrum is given by
\begin{eqnarray}
P_\text{obs}(k,\mu;\,z) &=& 
\frac{D^2_{{\rm A},r}(z) H(z)}{D^2_{\rm A}(z) H_{r}(z)}
\left\{\frac{\left[b\sigma_8(z)+f\sigma_8(k,z)\mu^2\right]^2}{1+k^2\mu^2f^2(k,z)\sigma_{\rm p}^2(z)}\right\}\times\nonumber \\ 
&\times&\frac{P_\text{dw}(k,\mu;z)}{\sigma^2_8(z)}
F_z(k,\mu;z) +P_\text{s}(z) \,,
\label{eq:GC:pkobs}
\end{eqnarray}
where the subscript $r$ refers to the reference (or fiducial) cosmology. $P_{\rm s}(z)$ is a  scale-independent nuisance parameter due to imperfect removal of shot-noise. The Alcock-Paczynski (AP) 
effect takes into account the incorrect cosmological models from the fiducial one and it is 
parameterised through the rescaling of the angular diameter distance $D_{\rm A}(z)$ and the Hubble 
parameter $H(z)$. The AP effect enters as a multiplicative factor to the galaxy power spectrum, and in 
both $k$ and $\mu$, see \cite{Alcock:1979mp,Euclid:2019clj} for more details. 
The term in the curly brackets in Eq.~\eqref{eq:GC:pkobs} is the redshift space distortion (RSD) 
\cite{Kaiser:1987qv} which is corrected for the non-linear finger-of-God (FoG) effect. The RSD is 
parameterised through the galaxy bias $b(z)$ and  the growth rate $f(k,z)$, both multiplied by the root 
mean square of matter density fluctuation $\sigma_8(z)$, whereas $\mu$ is the cosine of the angle of 
the wave mode with respect to the line of sight pointing into the direction $\hat{r}$, and $k$ is the 
scale of the perturbation. 

$P_{\rm dw}(k,\mu;\,z)$ is the de-wiggled power spectrum which models the smearing of the BAO signal 
due to non-linearities, and it is defined as
\begin{equation} \label{eq:GC:pk_dw}
    P_\text{dw}(k,\mu;z) = P_{\rm m}(k;z)\,\text{e}^{-g_\mu k^2} + P_\text{nw}(k;z)\left(1-\text{e}^{-g_\mu k^2}\right) \,,
\end{equation}
where $P_{\rm nw}(k;z) $ is the ‘no-wiggle’ power spectrum obtained directly from the matter power 
spectrum but without BAO features. Modified gravity models usually predict a scale dependent growth 
rate $f(z,k)$ which needs to be taken into account when evaluating the Fisher matrix. For the model 
used in this work, we found that the deviation from a constant value is at most $0.1\%$ over the whole 
range of $k$ used in the analysis. Hence, we can safely assume the growth rate to be independent of 
scale in the non-linear terms appearing in $P_{\rm dw}$. The pairwise velocity dispersion, 
$\sigma_{\rm p}(z)$ and the velocity dispersion, $\sigma_{\rm v}(z)$ are then equal and the function 
$g_\mu$ in Eq.~\eqref{eq:GC:pk_dw}
\begin{eqnarray}
    \sigma^2_{\rm v}(z) =\sigma^2_{\rm p}(z)  &=& \frac{1}{6\pi^2}\int{\rm d} k\, P_{\rm m}(k,z)\,,\label{eq:sigmas}\\
    g_{\mu}(z, \mu) &\simeq& \sigma^2_{\rm v}(z)\left[1 - \mu^2 + \mu^2\left[1+f(\bar{k},z)\right]^2\right]\,.\label{eq:gmu}
\end{eqnarray}
where we choose the mean value of the $\bar{k} = 0.05\,h/{\rm Mpc}$.   
Finally, the total galaxy power spectrum in Eq.~\eqref{eq:GC:pkobs} includes the errors on redshift 
through the factor 
\begin{equation}
    F_z(k, \mu;z) = \text{e}^{-k^2\mu^2\sigma_{r}^2(z)}\,,
\end{equation}
where $\sigma_{r}^2(z) = c(1+z)\sigma_{0,z}/H(z)$ and $\sigma_{0,z}$ is the error on the measured 
redshifts. 

The no-wiggle matter power spectrum $P_{\rm nw}(k;z)$ entering Eq.~\eqref{eq:GC:pk_dw} has been obtained 
using a Savitzky-Golay filter to the matter power spectrum $P_{\rm m}(k;z)$. The Savitzky-Golay filter 
is usually applied to noisy data in order to smooth their behavior. In practice, we treat the BAO 
wiggles in the matter power spectrum as if they were noise in the overall shape of the matter power 
spectrum; by smoothing the noise, we recover exactly the same shape and amplitude of the matter power 
spectrum without the BAO wiggles, see \cite{Boyle:2017lzt}. 

The final Fisher matrix for the galaxy clustering observable for one redshift bin $z_i$ is 
\begin{equation} \label{eq:fisher-gc}
    F_{\alpha\beta}(z_i) = \frac{1}{8\pi^2}\int_{-1}^{1}{\rm d}\mu\int_{k_{\rm min}}^{k_{\rm max}} V_{\rm eff}(z_i,k)\cdot\frac{\partial \ln P_{\rm obs}(k,\mu;\,z_i)}{\partial p_\alpha}\frac{\partial \ln P_{\rm obs}(k,\mu;\,z_i)}{\partial p_\beta}\cdot k^2{\rm d}k
\end{equation}
where the derivatives are evaluated at the parameter values of the fiducial model and $V_{\rm eff}$ is 
the effetive volume of the survey, given by
\begin{equation}
    V_{\rm eff}(k,\mu;\,z) = V_{\rm s}\left[\frac{n(z)P_{\rm obs}(k,\mu;\,z)}{1+n(z)P_{\rm obs}(k,\mu;\,z)}\right]^2
\end{equation}
being $V_{\rm s}$ the volume of the survey and $n(z)$ the number of galaxies in a redshift bin. 

In our analysis we used 8 cosmological parameters constant for all redshifts and 2 redsfhit dependent 
parameters
\begin{equation} 
    p_\alpha = \left\{\Omega_{\rm m},\,\Omega_{\rm b},\,h,\,n_{\rm s},\,\xi,\,\Delta,\,\sigma_8,\,\ln b\sigma_8(z),\,P_{\rm s}(z)\right\}
\end{equation}
where $\Omega_X$ corresponds to the density parameter at current time and 
$\sigma_8 \equiv \sigma_8(z=0)$. The total Fisher matrix for GC is constructed summing up directly the 
elements of the Fisher matrices at each bin for the redsfhit independent parameters, whereas the 
z-dependent parameters will be added in sequence to the final Fisher matrix. In the specific, the final 
Fisher matrix has dimensions $7+n_{\rm bin}\times 2$, where $n_{\rm bin}$ is the number of redshift 
bins. Both $\ln b\sigma_8(z)$ and $P_{\rm s}(z)$ are nuisance parameters and they are marginalized over. 

We present GC results for three different range of wavenumbers: quasi-linear scales with 
$k_{\rm max}^{\rm GC} = 0.15\ h/$Mpc and two case including non-linear scales with 
$k_{\rm max}^{\rm GC} = 0.25,\, 0.30\ h/$Mpc.

We forecast the GC constraints for the ground-based Dark Energy Spectroscopic Instrument (DESI) 
\cite{DESI:2016fyo}. Following \cite{Ballardini:2016hpi}, we consider a unified effective sample 
combining the populations of LRGs, ELGs and QSOs, covering thirteen redshift bins between z = 0.6 and 
1.9 with width of $\Delta z = 0.1$, with a volume of the survey of $14,000\,{\rm deg}^2$. 
Finally, we complete the analysis by include low-redshift spectroscopic information from BOSS 
\cite{BOSS:2016wmc}, the volume of the survey is $9,329\,{\rm deg}^2$, in a redshift range 
$z \in [0.2,\,0.75]$ divided in two bins with $\Delta z = [0.3,\,0.25]$.

\subsection{Weak gravitational lensing} \label{sec:wl_forecast}
Here we report the main equations of the weak lensing tomographic signal and we refer to the literature 
for further details \cite{Euclid:2019clj,Amendola:2007rr, Majerotto:2015bra, Palma:2017wxu}. The weak 
lensing convergence power spectrum is a linear function of the matter power spectrum convoluted with 
the lensing properties of the survey. In the $\Lambda$CDM cosmology, we can write it as
\begin{equation} \label{eq:cell_wl}
    P_{i\,j}(\ell) = H^4_0 \int_{0}^{\infty} \frac{{\rm d}z}{H^2(z)}W_i(z)W_j(z)P_{\rm nl}\left(k = \frac{\ell\,H_0}{r(z)}\,,z\right) \,,
\end{equation}
where $\ell$ is the multipole number, $r(z)$ is the comoving distance between lens and objects, $H(z)$ 
is the Hubble parameter, and the subscripts $i\,,j$ refer to the redshift bins around the redshifts 
$z_i$ and $z_j$. The window function, $W_i(z)$, which takes into account the lensing properties of 
space, is defined as 
\begin{equation}
    W_i(z) = \frac32 \left(\frac{H_0}{c}\right)^2 \Omega_m(1+z)r(z) \int_z^{z_{\rm max}}{\rm d}z' D(x) \left[1-\frac{r(z)}{r(z')}\right] 
\end{equation}
being $z_{\rm max}$ the maximum redshift of the $i$-th bin and  the radial distribution function of 
galaxies is 
\begin{equation}
D(z) = z^\alpha \exp\left[-\left(z/z_0\right)^\beta\right]
\end{equation}
where $\alpha\,,\beta\,,z_0$ are constants that depend on the survey strategy. 
For the Vera C. Rubin Observatory’s Legacy Survey of Space and Time (LSST),  we assume $\alpha = 1.27$, 
$\beta = 1.02$ and $z_0 = 0.5$ \cite{Schaan:2016ois, LSSTDarkEnergyScience:2018jkl}. Moreover, we 
consider a survey up to $z_{\rm max} = 3$ divided into 10 bins each containing the same number of 
galaxies. 

The Fisher matrix for the weak lensing signal is 
\begin{equation}
F_{\alpha \beta} = f_{\rm sky} \sum_{\ell} \frac{(2\ell+1)\Delta \ell}{2}\frac{\partial P_{ij}}{\partial p_\alpha}C^{-1}_{jk}\frac{\partial P^{-1}_{km}}{\partial p_\beta}C^{-1}_{mi}
\end{equation}
where $\Delta\ell$ is the step in multipoles, to which we chose 100 step in logarithm scale, and 
$f_{\rm sky} = 0.4363$; whereas $p_\alpha$ are the cosmological parameters. The covariances are defined 
as
\begin{equation}
    C_{jk} = P_{jk}+\delta_{jk}\langle\gamma^2_{\rm int}\rangle n^{-1}_j
\end{equation}
where $\gamma_{\rm int}$ is the rms intrinsic shear, which we assume $\langle\gamma^2_{\rm int}\rangle = 0.26$. 

The number of galaxies per steradians in each bin is defined as:
\begin{equation}
    n_j = 3600 d\left(\frac{180}{\pi}\right)^2\hat{n}_j
\end{equation}
where the number density is $d = 26$ galaxy per arcmin and $\hat{n}_j$ is the fraction of sources that belongs 
to the $j$-th bin.

The final set of parameter for WL is
\begin{equation} \label{eqn:LSSparams}
    p_\alpha = \left\{\Omega_{\rm m},\,\Omega_{\rm b},\,h,\,n_{\rm s},\,\xi,\,\Delta,\,\sigma_8\right\}\,.
\end{equation}
We present WL results for two different range of multipoles: a conservative case with 
$\ell_{\rm max}^{\rm WL} = 1500$ and a case including all the information down to 
$\ell_{\rm max}^{\rm WL} = 5000$.

\begin{figure}
\centering
\includegraphics[height=0.21\textwidth]{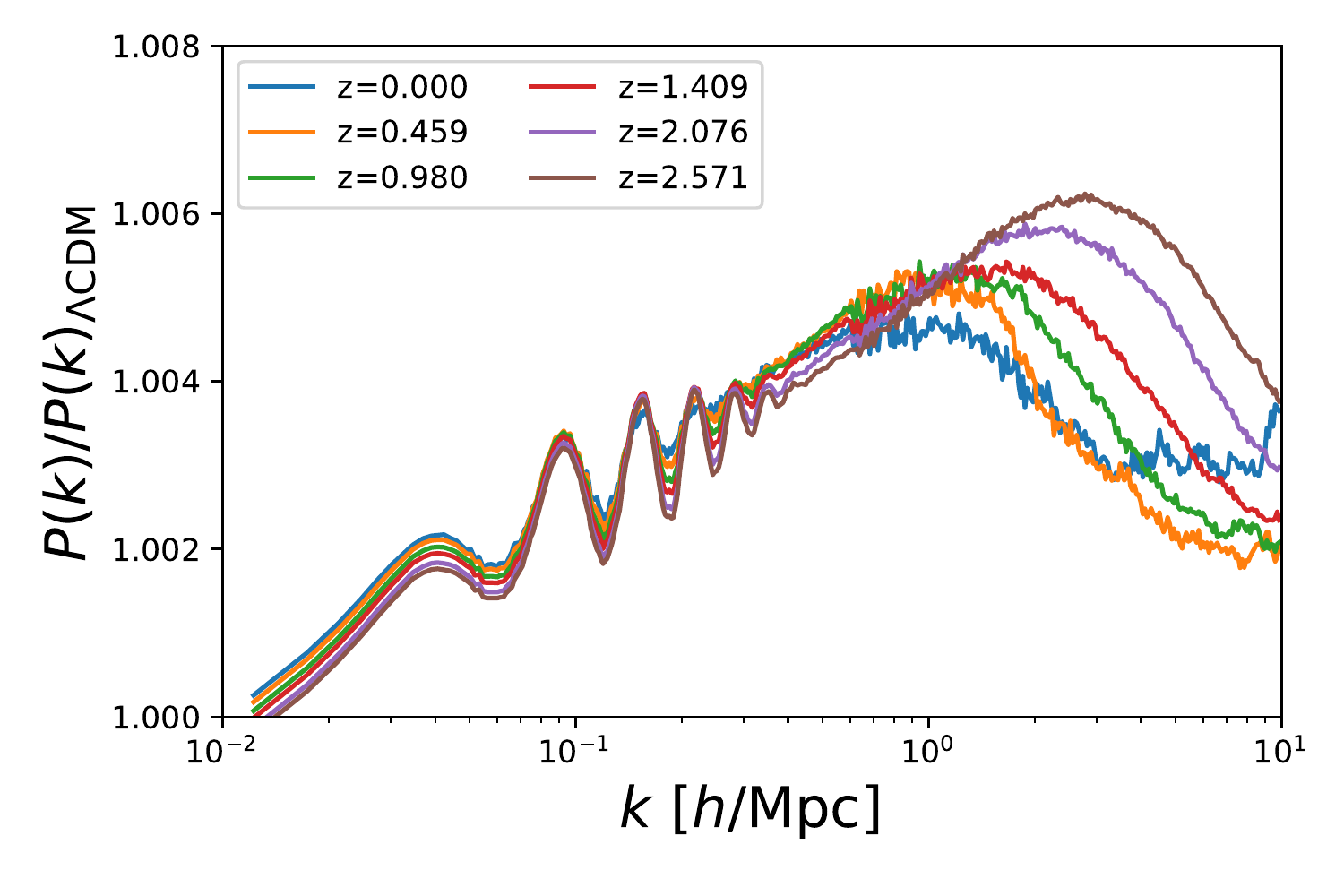}
\includegraphics[height=0.21\textwidth]{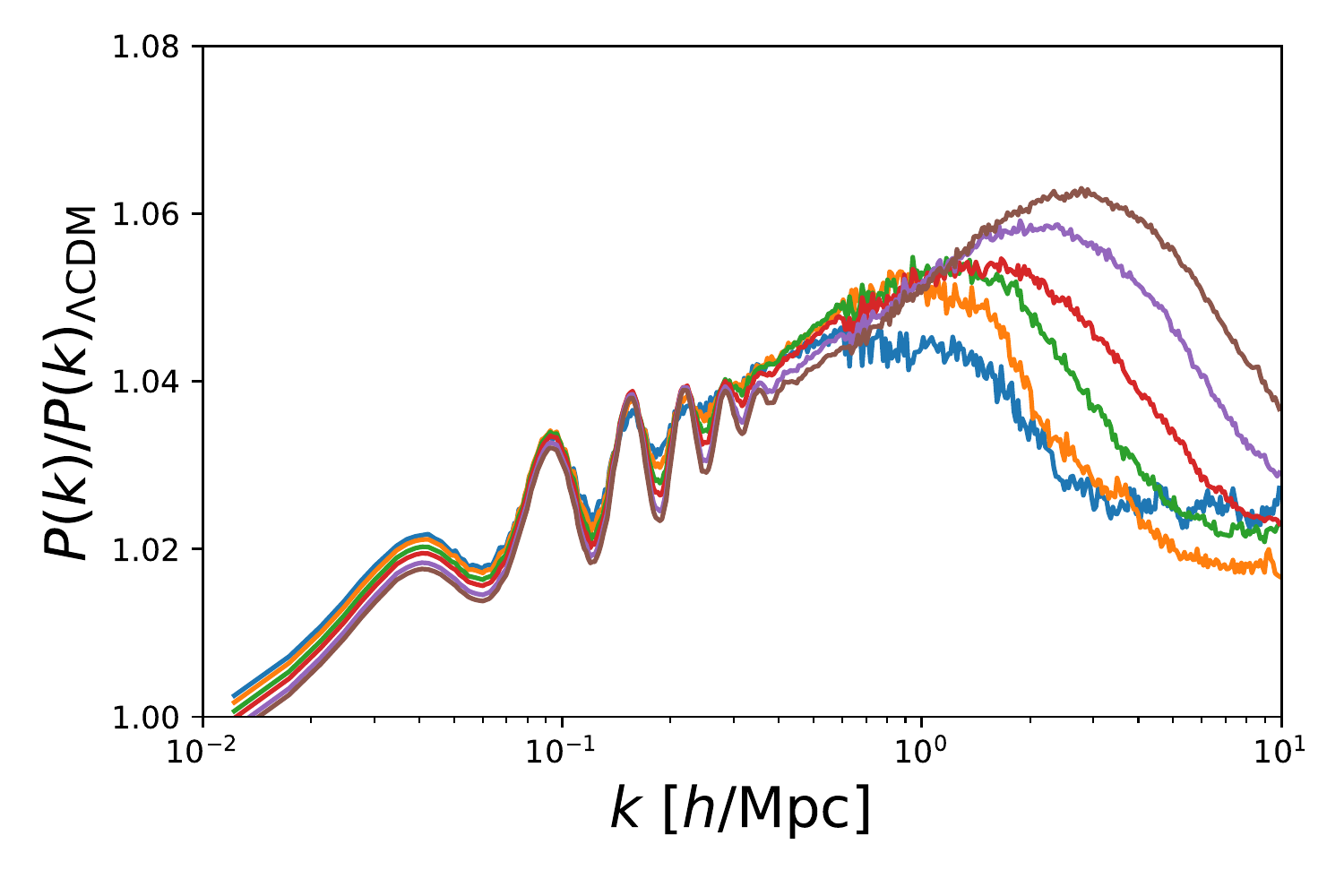}
\includegraphics[height=0.21\textwidth]{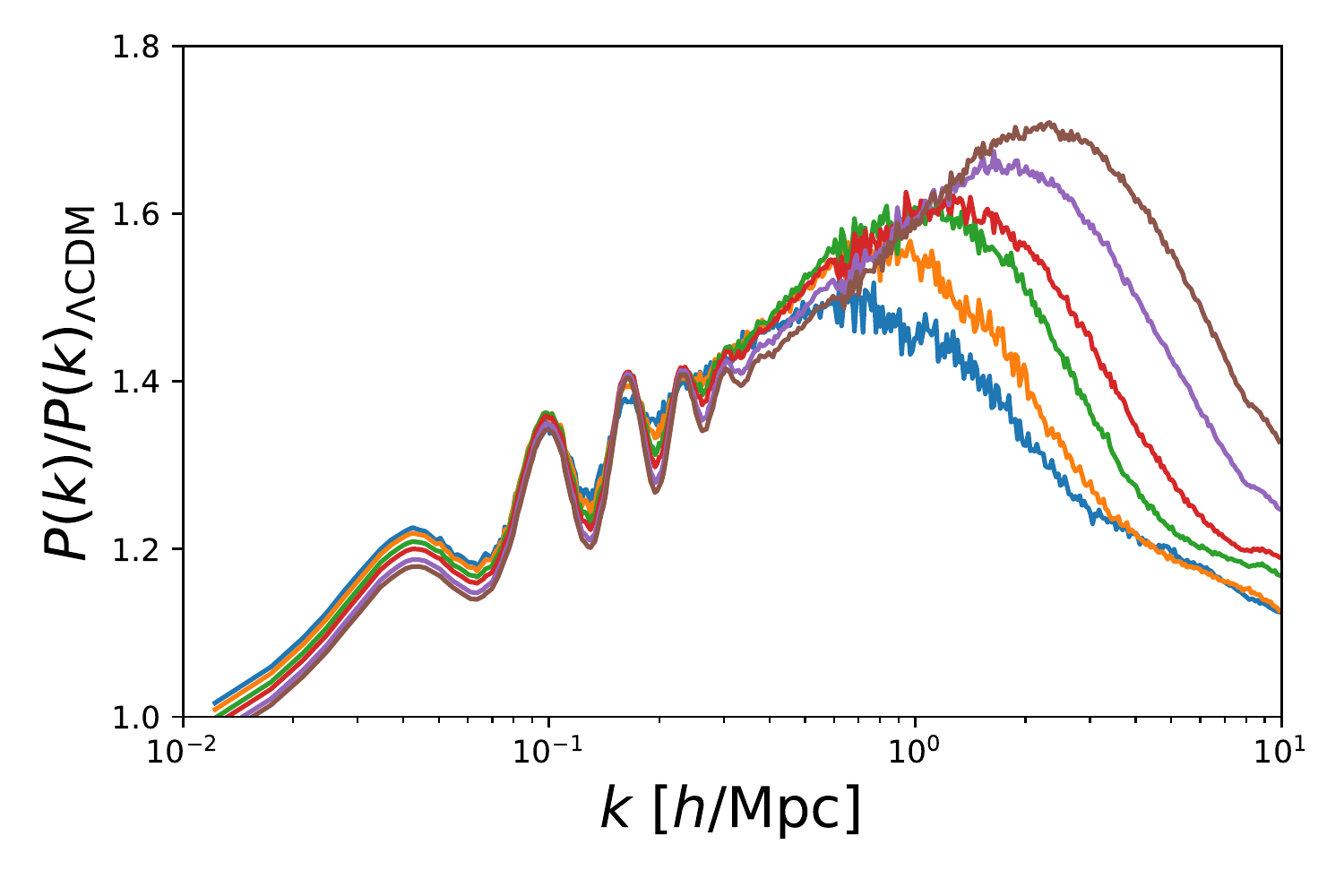}
\caption{Relative difference of the matter power spectrum with respect to $\Lambda$CDM from the 
simulations of IG for $\xi = 0.0001, 0.001, 0.01$ from left to right.}
\label{fig:nonlinear}
\end{figure}

We study the non-linear evolution of the matter power spectrum $P_{\rm nl}(k,z)$ with a modified version 
of the COmoving Lagrangian Accelerator (COLA) 
code\footnote{\href{https://github.com/HAWinther/MG-PICOLA-PUBLIC}{https://github.com/HAWinther/MG-PICOLA-PUBLIC}} \cite{Tassev:2013pn,Winther:2017jof}.
We generate simulations with $N = 1024^3$ particles in a box size $L = 1024\ {\rm Mpc}/h$ to cover large 
scales and $N = 512^3$ particles in a box size $L = 200\ {\rm Mpc}/h$ to cover the small scales, for the 
fiducial parameters used in the forecast analysis and $\xi = 0.0001, 0.001, 0.01$. The variation of 
$\Delta$ is propagated through the linear matter power spectrum \cite{Joudaki:2020shz}.
Fig.~\ref{fig:nonlinear} shows the relative differences to $\Lambda$CDM.

\subsection{Results}
We carry out a Fisher matrix analysis using $\xi$ and $\Delta$ in addition to the standard $\Lambda$CDM 
cosmological parameters. We marginalize the CMB Fisher matrix over the optical depth parameter $\tau$ 
and we project it over the LSS parameters in Eq.~\eqref{eqn:LSSparams}.
Uncertainties on the cosmological parameters are calculated as the square root of the diagonal elements 
of the inverse of the Fisher matrix $\sqrt{\left(F^{-1}\right)_{\alpha\alpha}}$. 

\begin{table}[!h]
\centering
\begin{tabular}{|c|c|c|}
\hline
\rule[-2mm]{0mm}{.6cm}
 &  $10^3\sigma(\xi)$  &  $10^2\sigma(\Delta)$ \\
\hline
\rule[-2mm]{0mm}{.6cm}
S4+LiteBIRD & 0.14 & 0.7 \\
\hline
\rule[-2mm]{0mm}{.6cm}
BOSS+DESI ($k_{\rm max}^{\rm GC} = 0.15,\, 0.25,\, 0.30$ $h/$Mpc) & 1.6/0.86/0.76 & 18/11/9.9 \\
\hline
\rule[-2mm]{0mm}{.6cm}
LSST ($\ell_{\rm max}^{\rm WL} = 1500/5000$) & 2.1/1.2 & 8.9/5.4 \\
\hline
\rule[-2mm]{0mm}{.6cm}
S4+LiteBIRD & \multirow{2}{*}{0.061/0.04/0.035} & \multirow{2}{*}{0.62/0.61/0.61} \\
+ BOSS+DESI ($k_{\rm max}^{\rm GC} = 0.15,\, 0.25,\, 0.30$ $h/$Mpc)   & &   \\
\hline
\rule[-2mm]{0mm}{.6cm}
S4+LiteBIRD & \multirow{2}{*}{0.031/0.020} & \multirow{2}{*}{0.54/0.44} \\
+ LSST ($\ell_{\rm max}^{\rm WL} = 1500/5000$)   & &   \\
\hline
\rule[-2mm]{0mm}{.6cm}
BOSS+DESI ($k_{\rm max}^{\rm GC} = 0.15,\, 0.25,\, 0.30$ $h/$Mpc) & \multirow{2}{*}{0.49/0.43/0.40} & \multirow{2}{*}{7.0/6.6/6.2} \\
+ LSST ($\ell_{\rm max}^{\rm WL} = 1500$)   & &   \\
\hline
\rule[-2mm]{0mm}{.6cm}
BOSS+DESI ($k_{\rm max}^{\rm GC} = 0.15,\, 0.25,\, 0.30$ $h/$Mpc) & \multirow{2}{*}{0.43/0.34/0.33} & \multirow{2}{*}{4.1/3.7/3.5} \\
+ LSST ($\ell_{\rm max}^{\rm WL} = 5000$)   & &   \\
\hline
\rule[-2mm]{0mm}{.6cm}
S4+LiteBIRD & \multirow{3}{*}{0.029/0.024/0.023} & \multirow{3}{*}{0.51/0.49/0.49} \\
+ BOSS+DESI ($k_{\rm max}^{\rm GC} = 0.15,\, 0.25,\, 0.30$ $h/$Mpc)   & &   \\
+ LSST ($\ell_{\rm max}^{\rm WL} = 1500$)   & &   \\
\hline
\rule[-2mm]{0mm}{.6cm}
S4+LiteBIRD & \multirow{3}{*}{0.019/0.018/0.017} & \multirow{3}{*}{0.43/0.41/0.41} \\
+ BOSS+DESI ($k_{\rm max}^{\rm GC} = 0.15,\, 0.25,\, 0.30$ $h/$Mpc)   & &   \\
+ LSST ($\ell_{\rm max}^{\rm WL} = 5000$)   & &   \\
\hline
\end{tabular}
\caption{Marginalized uncertainties (68\% CL) for $\xi = 10^{-5}$ and $\Delta = 10^{-3}$.}
\label{tab:errors}
\end{table}

We collect the uncertainties for the single probes and the various combinations of those in 
Tab.~\ref{tab:errors}. The uncertainties from single surveys are dominated by the CMB information. 
By the expected future measurements of CMB anisotropies from the combination of LiteBIRD and CMB-S4, 
which constrain both distance measurements as well as the growth rate at early times, we obtain 68\% CL 
uncertainties $10^4\, \sigma(\xi) \simeq 1.4$ and $10^2\, \sigma(\Delta) \simeq 0.7$. These results 
improve the current {\em Planck} constraints by an order of magnitude on $\xi$ and by a factor of 4 on 
$\Delta$. Uncertainties on the coupling $\xi$ are consistent with the results obtained in 
Ref.~\cite{Ballardini:2019tho} for IG with $\Delta = 0$ showing no appreciable widening of the 
constraints for such future experiments.

\begin{figure}
\centering
\includegraphics[width=0.48\textwidth]{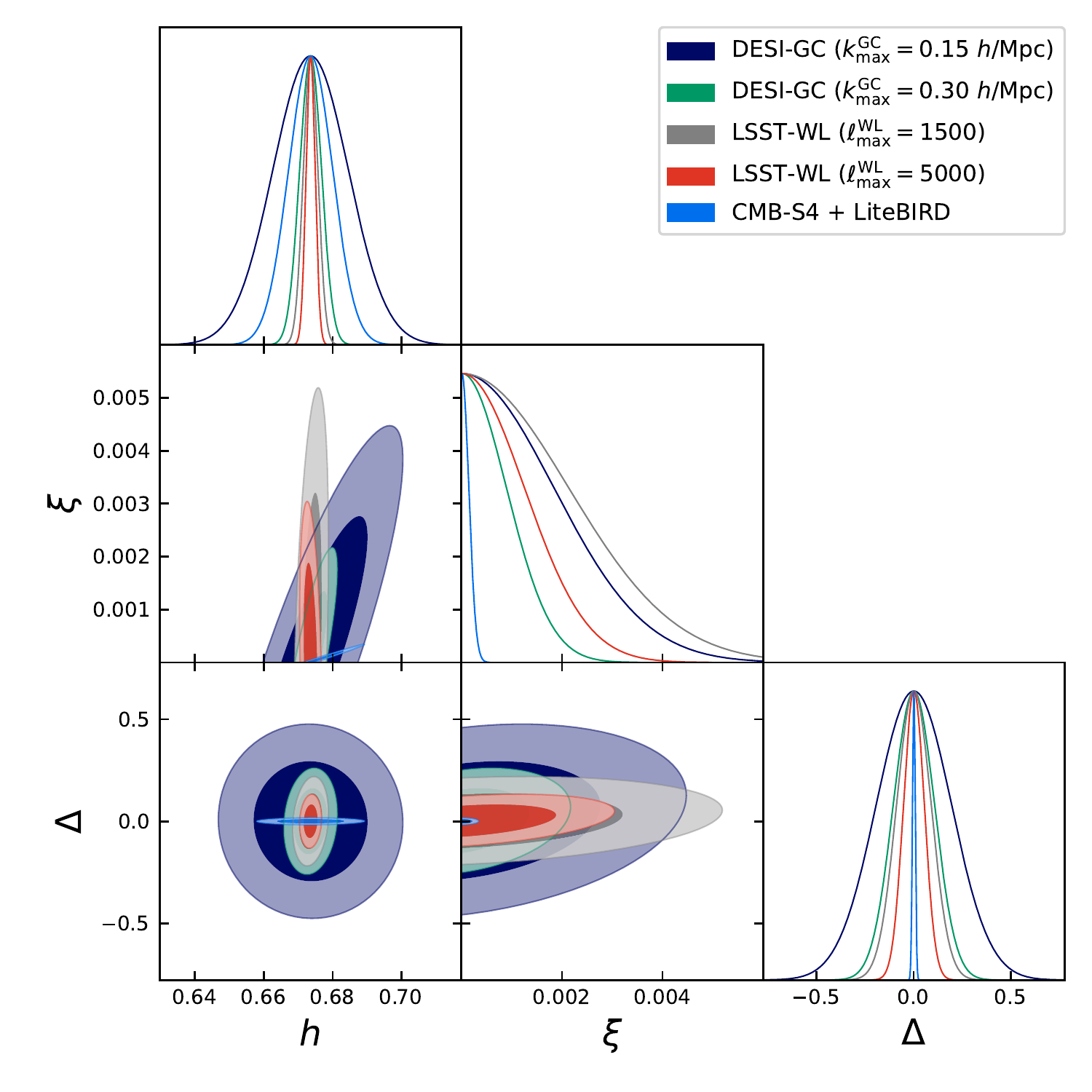}
\includegraphics[width=0.48\textwidth]{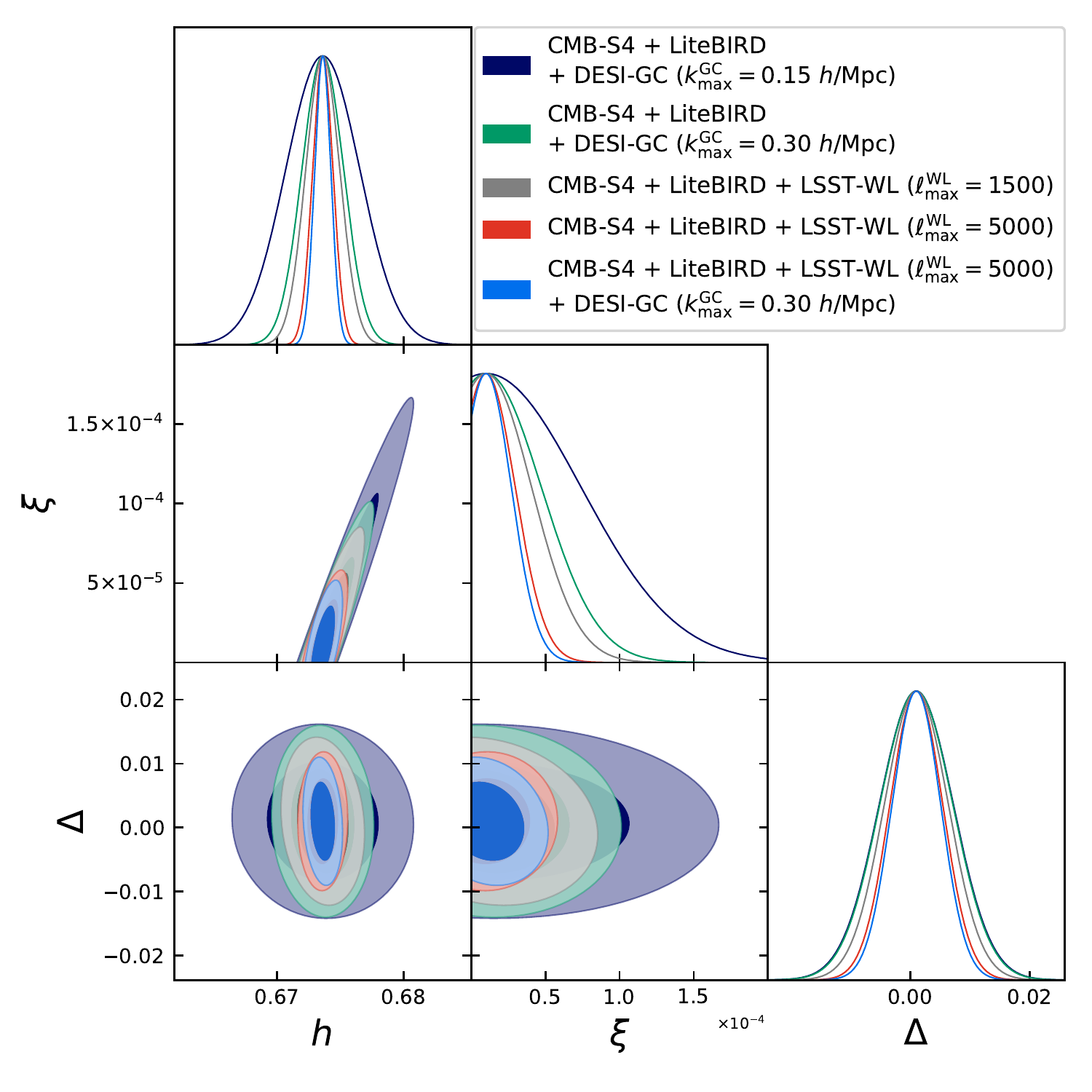}
\caption{Left: marginalized 68\% and 95\% CL constraints on the parameter space $h$-$\xi$-$\Delta$ 
from single probes alone. Right: marginalized 68\% and 95\% CL constraints on the parameter space 
$h$-$\xi$-$\Delta$ from different combination of probes.}
\label{fig:forecasts}
\end{figure}

In Fig.~\ref{fig:forecasts}, we can see the important role that the complementarity between early-time 
and late-time cosmological probes plays. It is well established \cite{Alonso:2016suf,Ballardini:2019tho} 
that the CMB and late-time measurements will combine to supply powerful constraints mitigating the 
degeneracies between $H_0$ and the coupling $\xi$ parameters.
While GC data alone constrain better the coupling $\xi$, WL measurements are more sensitive to $\Delta$.

Finally, we derive tighter constraints from the combination of our three cosmological 
probes, which are $10^5\, \sigma(\xi) \simeq 1.7$ and $10^3\, 
\sigma(\Delta) \simeq 4.1$ for $k_{\rm max}^{\rm GC} = 0.30\ h/$Mpc and 
$\ell_{\rm max}^{\rm WL} = 5000$. It is interesting to note that these constraints are weakly affected 
($< 5\%$) by reducing the GC nonlinear information to $k_{\rm max}^{\rm GC} = 0.15\ h/$Mpc.

\section{Conclusions} \label{sec:conclusion}
We have studied general constraints on the gravitational constant from cosmology.
We have used the framework of a scalar field non-minimally coupled to the Ricci scalar, i.e. the 
simplest scalar-tensor theory of gravity, to study self-consistently variations in time and space of 
the gravitational constant. Going beyond what previously done, in this paper we have investigated in 
a direct way the effect of an imbalance $\Delta$ between the effective gravitational constant and G, 
i.e. $G_{\rm eff}(z=0) = G (1+\Delta)^2$.

We computed the effects of the imbalance $\Delta$ on cosmological observables and computed how current 
cosmological data can constrain it by also allowing the coupling to the Ricci scalar and the rest of 
cosmology to vary. {\em Planck} 2018 data in combination with BAO from BOSS DR12 data constrain the 
imbalance to $\Delta = -0.022 \pm 0.023$ at 68\% CL and the coupling parameter to $10^3\, \xi < 0.82$ 
at 95\% CL for $F(\sigma) = \xi\sigma^2$ and for a non-minimally coupled scalar field with 
$F(\sigma) = M^2_{\rm pl} + \xi\sigma^2$ constrain the imbalance to $\Delta > -0.018$ ($< 0.021$) and 
the coupling parameter to $\xi < 0.089$ ($\xi > - 0.041$) both at 95\% CL.
These bounds correspond to constrain $G_{\rm eff}(z=0)/G$ to about 4-15\% at 95\% CL. By allowing 
$\Delta$ to vary the constraints on the coupling to the Ricci scalar $\xi$ degrade with respect to 
$\Delta=0$ \cite{Ballardini:2020iws}.

We also explored the limit that could be achieved by future experiments. We forecasted 
$\sigma(G_{\rm eff}(z=0)/G) \simeq 0.014$ at 68\% CL by the combination of CMB anisotropy measurements 
from LiteBIRD and CMB-S4. Combining the CMB information with galaxy clustering from BOSS + DESI and 
galaxy shear from LSST we found $\sigma(G_{\rm eff}(z=0)/G) \simeq 0.008$ at 68\% CL.

We note that the extended phenomenology studied here is not only relevant for testing gravity on large 
scales, but can also help in interpreting the current tensions in the estimates of cosmological 
parameters from different observations. Since the first {\em Planck} data release, we noticed how one 
of the simplest scalar-tensor gravity model such as induced gravity (equivalent to Jordan-Brans-Dicke) 
with a quartic potential could accommodate for a larger value of $H_0$ compared the one in $\Lambda$CDM 
\cite{Umilta:2015cta} due to a degeneracy between the coupling to the Ricci scalar and $H_0$.
Subsequent studies generalized this result for different simple potentials and couplings 
\cite{Ballardini:2016cvy,Rossi:2019lgt,Ballardini:2020iws}.
While this model is able to accommodate for a smaller values of $\sigma_8$ for large value of 
$\Delta > 0$, we do not see any reduction in term of the parameter $S_8$ by fitting {\em Planck} 2018 
CMB data and BOSS DR12 measurements. Work in this direction is in progress.

\section*{Acknowledgments}
We would like to thank Matteo Braglia for contributing to the early stage of the work.
MB and FF acknowledges financial support from the contract ASI/ INAF for the Euclid mission 
n.2018-23-HH.0. FF acknowledges financial support from the contract by the agreement n. 2020-9-HH.0 
ASI-UniRM2 "Partecipazione italiana alla fase A della missione LiteBIRD". DS acknowledges financial 
support from the Fondecyt Regular project number 1200171. 
This research was also partially supported by the Munich Institute for Astro- and Particle Physics 
(MIAPP) which is funded by the Deutsche Forschungsgemeinschaft (DFG, German Research Foundation) 
under Germany`s Excellence Strategy - EXC-2094 - 390783311.

    \newpage
\appendix
\section{Tables} \label{sec:appendix1}

\begin{table*}[h!]
{\small
\centering
\begin{tabular}{l|ccc}
\hline
\hline
                                         & P18 & P18 + BAO & P18 + BAO + R19  \\
\hline
$\omega_{\rm b}$                         & $0.02218\pm 0.00024$        & $0.02221\pm 0.00024$     & $0.02223\pm 0.00025$  \\
$\omega_{\rm c}$                         & $0.1198\pm 0.0013$          & $0.1200\pm 0.0011$       & $0.1201\pm 0.0011$  \\
$H_0$ [km s$^{-1}$Mpc$^{-1}$]            & $70.2^{+1.2}_{-3.1}$        & $68.61^{+0.72}_{-0.94}$  & $70.04\pm 0.83$ \\
$\tau$                                   & $0.0551\pm 0.0075$          & $0.0543\pm 0.0072$       & $0.0555^{+0.0067}_{-0.0079}$  \\
$\ln \left(  10^{10} A_{\rm s} \right)$  & $3.041\pm 0.017$            & $3.040\pm 0.016$         & $3.043\pm 0.017$  \\
$n_{\rm s}$                              & $0.9608\pm 0.0077$          & $0.9604\pm 0.0074$       & $0.9617\pm 0.0077$  \\
$\zeta_{\rm IG}$                         & $< 0.0084$ (95\% CL)        & $< 0.0033$ (95\% CL)     & $0.0029\pm 0.0011$  \\
$\Delta$                                 & $-0.032^{+0.029}_{-0.025}$  & $-0.022\pm 0.023$        & $-0.026\pm 0.024$  \\
\hline
$\xi$                                                     & $< 0.0021$ (95\% CL)       & $< 0.00082$ (95\% CL)     & $0.00074^{+0.00052}_{-0.00054}$ (95\% CL)  \\
$\gamma_{\rm PN}$                                             & $> 0.9917$ (95\% CL)       & $> 0.9968$ (95\% CL)      & $0.9971\pm 0.0011$  \\
$\delta G_\mathrm{N}/G_\mathrm{N}$ (z=0)                  & $> -0.060$ (95\% CL)       & $> -0.0240$ (95\% CL) & $-0.0216\pm 0.0079$  \\
$10^{13} \dot{G}_\mathrm{N}/G_{\rm N}$ (z=0) [yr$^{-1}$]  & $> -2.34$ (95\% CL)        & $> -0.98$ (95\% CL)       & $-0.88\pm 0.32$  \\
$G_\mathrm{N}/G$ (z=0)                                    & $0.937^{+0.057}_{-0.050}$  & $0.956\pm 0.045$          & $0.948\pm 0.048$  \\
$G_\mathrm{eff}/G$ (z=0)                                  & $0.938^{+0.056}_{-0.049}$  & $0.957\pm 0.045$          & $0.949\pm 0.048$  \\
$\sigma_i$ [Mpl]    &   $< 80$ (95\% CL)  &   $73^{+10}_{-40}$  &   $< 70$ (95\% CL)  \\
\hline
$\Omega_{\rm m}$                        & $0.290^{+0.027}_{-0.013}$  & $0.3023\pm 0.0078$         & $0.2903\pm 0.0071$  \\
$\sigma_8$                              & $0.842^{+0.013}_{-0.030}$  & $0.828^{+0.011}_{-0.015}$  & $0.841\pm 0.014$  \\
$S_8$                                   & $0.826\pm 0.016$           & $0.831\pm 0.013$           & $0.827\pm 0.013$  \\
$r_s$ [Mpc]                             & $148.4\pm 2.0$   & $148.3^{+1.8}_{-2.1}$      & $147.9\pm 2.0$  \\
%\hline
%$\Delta \chi^2$                         & $-6.0$ & $-2.7$  & $-16.8$  \\
\hline
\hline
\end{tabular}}
\caption{\label{tab:dig} 
Constraints on the main and derived parameters (at 68\% CL if not otherwise stated) considering 
P18 in combination with BAO and BAO + R19 for the IG model.}
\end{table*}

    \newpage
\begin{table*}[h!]
{\small
\centering
\begin{tabular}{l|cc}
\hline
\hline
                                         & P18 + BAO + $p(S_8)$ & P18 + BAO + R19 + $p(S_8)$ \\
\hline
$\omega_{\rm b}$                         & $0.02254\pm 0.00021$     & $0.02266^{+0.019}_{-0.025}$  \\
$\omega_{\rm c}$                         & $0.1192^{+0.0012}_{-0.0014}$       & $0.1198^{+0.0015}_{-0.0016}$  \\
$H_0$ [km s$^{-1}$Mpc$^{-1}$]            & $69.10^{+0.60}_{-0.89}$  & $70.41\pm 0.85$ \\
$\tau$                                   & $0.0537\pm 0.0074$       & $0.0551\pm 0.0073$  \\
$\ln \left(  10^{10} A_{\rm s} \right)$  & $3.044\pm 0.017$         & $3.054^{+0.016}_{-0.019}$  \\
$n_{\rm s}$                              & $0.9697^{+0.0042}_{-0.0047}$       & $0.9740^{+0.0041}_{-0.0050}$  \\
$\zeta_{\rm IG}$                         & $< 0.0029$ (95\% CL)     & $0.0026\pm 0.0011$  \\
$\Delta$                                 & $0.010^{+0.035}_{-0.029}$        & $0.024^{+0.029}_{-0.039}$  \\
\hline
$\xi$                                                     & $< 0.00073$ (95\% CL)     & $0.00066^{+0.00053}_{-0.00056}$ (95\% CL)  \\
$\gamma_{\rm PN}$                                             & $> 0.9971$ (95\% CL)      & $0.9974\pm 0.0011$  \\
$\delta G_\mathrm{N}/G_\mathrm{N}$ (z=0)                  & $> -0.021$ (95\% CL)     & $-0.0192\pm 0.0081$  \\
$10^{13} \dot{G}_\mathrm{N}/G_{\rm N}$ (z=0) [yr$^{-1}$]  & $> -0.88$ (95\% CL)       & $-0.79\pm 0.34$  \\
$G_\mathrm{N}/G$ (z=0)                                    & $1.020^{+0.068}_{-0.061}$   & $1.049^{+0.056}_{-0.081}$  \\
$G_\mathrm{eff}/G$ (z=0)                                  & $1.021^{+0.068}_{-0.061}$   & $1.050^{+0.056}_{-0.082}$  \\
$\sigma_i$ [Mpl]                                          & $83^{+20}_{-60}$          & $44^{+10}_{-20}$  \\
\hline
$\Omega_{\rm m}$                        & $0.2970\pm 0.0069$         & $0.2874\pm 0.0064$  \\
$\sigma_8$                              & $0.8129^{+0.0088}_{-0.0098}$  & $0.823\pm 0.010$  \\
$S_8$                                   & $0.8087\pm 0.0092$           & $0.8050\pm 0.0091$  \\
$r_s$ [Mpc]                             & $146.50^{+0.93}_{-0.67}$      & $145.6^{+1.1}_{-0.9}$  \\
%\hline
%$\Delta \chi^2$                         & $-6.0$ & $-2.7$  & $-16.8$  \\
\hline
\hline
\end{tabular}}
\caption{\label{tab:dig_S8} 
Constraints on the main and derived parameters (at 68\% CL if not otherwise stated) considering 
P18 + BAO and P18 + BAO + R19 combined to a Gaussian prior on $S_8$ for the IG model.}
\end{table*}

    \newpage
\begin{table*}[h!]
{\small
\centering
\begin{tabular}{l|ccc}
\hline
\hline
                                         & P18 & P18 + BAO & P18 + BAO + R19  \\
\hline
$\omega_{\rm b}$                         & $0.02242\pm 0.00016$             & $0.02242\pm 0.00013$     & $2.250\pm 0.013$  \\
$\omega_{\rm c}$                         & $0.1200\pm 0.0012$               & $0.11992\pm 0.00098$     & $0.1196\pm 0.0010$  \\
$H_0$ [km s$^{-1}$Mpc$^{-1}$]            & $68.34^{+0.71}_{-1.2}$           & $68.33^{+0.55}_{-0.72}$  & $69.45\pm 0.72$ \\
$\tau$                                   & $0.0563^{+0.0066}_{-0.0080}$     & $0.0543\pm 0.0072$       & $0.0580^{+0.0063}_{-0.0082}$  \\
$\ln \left(  10^{10} A_{\rm s} \right)$  & $3.050^{+0.013}_{-0.016}$        & $3.050^{+0.013}_{-0.015}$ & $3.055^{+0.013}_{-0.016}$  \\
$n_{\rm s}$                              & $0.9676^{+0.0045}_{-0.0055}$     & $0.9676\pm 0.0040$       & $0.9715\pm 0.0039$  \\
$10^{-5}\,\Delta$                         & $< 2.8$ (95\% CL)  & $< 2.3$ (95\% CL)  & $\left(2.01^{+0.86}_{-0.97}\right)$ \\
\hline
$\gamma_{\rm PN}$                                             & $> 0.999972$ (95\% CL)       & $> 0.999977$ (95\% CL)  & $0.99998^{+0.000009}_{-0.000010}$  \\
$\beta_{\rm PN}$                                              & $< 1.0000023$ (95\% CL)       & $< 1.0000019$ (95\% CL) & $1.0000017^{+0.0000007}_{-0.0000008}$  \\
$\delta G_\mathrm{N}/G_\mathrm{N}$ (z=0)                  & $> -0.026$ (95\% CL)  & $> -0.022$ (95\% CL)  & $-0.0191\pm 0.0083$  \\
$10^{13} \dot{G}_\mathrm{N}/G_{\rm N}$ (z=0) [yr$^{-1}$]  & $> -0.012$ (95\% CL)        & $> -0.0098$ (95\% CL)       & $-0.0085^{+0.0075}_{-0.0071}$  \\
$G_\mathrm{N}/G$ (z=0)                                    & $<  1.000041$ (95\% CL)  & $< 1.000034$  (95\% CL)        & $1.000030^{+0.000013}_{-0.000015}$  \\
$G_\mathrm{eff}/G$ (z=0)                                  & $< 1.000055$ (95\% CL)  & $< 1.000046$ (95\% CL)         & $1.000040^{+0.000017}_{-0.000019}$  \\
$\sigma_i$ [Mpl]    &   $0.223^{+0.097}_{-0.11}$  &   $0.221\pm 0.087$  &   $0.329^{+0.092}_{-0.061}$  \\
\hline
$\Omega_{\rm m}$                  & $0.305^{+0.011}_{-0.0091}$   & $0.3049\pm 0.0067$         & $0.2947\pm 0.0065$  \\
$\sigma_8$                        & $0.8205^{+0.0073}_{-0.010}$  & $0.8196^{+0.0073}_{-0.0098}$  & $0.828\pm 0.011$  \\
$S_8$                             & $0.827\pm 0.013$             & $0.826\pm 0.011$           & $0.821\pm 0.011$  \\
$r_s$ [Mpc]                       & $146.66^{+0.47}_{-0.33}$     & $146.71^{+0.46}_{-0.32}$      & $146.34\pm 0.50$  \\
%\hline
%$\Delta \chi^2$                   & $-5.0$ & $+0.1$  & $-13.3$  \\
\hline
\hline
\end{tabular}}
\caption{\label{tab:dcc} 
Constraints on the main and derived parameters (at 68\% CL if not otherwise stated) considering 
P18 in combination with BAO and BAO + R19 for the CC model.}
\end{table*}

    \newpage
\begin{table*}[h!]
{\small
\centering
\begin{tabular}{l|cc}
\hline
\hline
                                         & P18 + BAO & P18 + BAO + R19  \\
\hline
$\omega_{\rm b}$                         & $0.02236\pm 0.00013$     & $0.02242\pm 0.00013$  \\
$\omega_{\rm c}$                         & $0.1198\pm 0.0010$       & $0.1195\pm 0.0011$  \\
$H_0$ [km s$^{-1}$Mpc$^{-1}$]            & $68.38^{+0.54}_{-0.84}$  & $69.76\pm 0.80$ \\
$\tau$                                   & $0.0561\pm 0.0068$       & $0.0575^{+0.0063}_{-0.0084}$  \\
$\ln \left(  10^{10} A_{\rm s} \right)$  & $3.047\pm 0.014$         & $3.050^{+0.012}_{-0.016}$  \\
$n_{\rm s}$                              & $0.9660\pm 0.0038$       & $0.9687\pm 0.0036$  \\
$\xi$                                    & $< 0.089$ (95\% CL)     & $-$  \\
$\Delta$                                 & $> -0.018$ (95\% CL)    & $-0.0072^{+0.0053}_{-0.0020}$  \\
\hline
$\gamma_{\rm PN}$                                                             & $> 0.995$ (95\% CL)        & $> 0.991$  (95\% CL) \\
$\beta_{\rm PN}$                                                              & $> 0.9998$ (95\% CL)       & $> 0.9996$ (95\% CL) \\
$\delta G_\mathrm{N}/G_\mathrm{N}$ (z=0)                                  & $> -0.016$ (95\% CL)      & $-0.0150\pm 0.007$  \\
$10^{13} \dot{G}_\mathrm{N}/G_{\rm N}$ (z=0) [yr$^{-1}$]  & $> -1.3$ (95\% CL)        & $> -2.5$ (95\% CL)  \\
$G_\mathrm{N}/G$ (z=0)                                                    & $0.980^{+0.021}_{-0.0045}$ & $0.970^{+0.020}_{-0.0089}$  \\
$G_\mathrm{eff}/G$ (z=0)                                                  & $0.990^{+0.010}_{-0.0021}$ & $0.986^{+0.010}_{-0.0041}$  \\
$\sigma_i$ [Mpl]    &   $< 7.4$ (95\% CL)  &   $< 3.0$ (95\% CL)  \\
\hline
$\Omega_{\rm m}$                        & $0.3041^{+0.0080}_{-0.0068}$  & $0.2916^{+0.0065}_{-0.0072}$  \\
$\sigma_8$                              & $0.8202^{+0.0073}_{-0.0096}$  & $0.831\pm 0.011$  \\
$S_8$                                   & $0.826\pm 0.010$              & $0.819\pm 0.010$  \\
$r_s$ [Mpc]                             & $146.73^{+0.57}_{-0.37}$      & $146.14^{+0.75}_{-0.62}$  \\
%\hline
%$\Delta \chi^2$                         & $-1.1$ & $-15.6$ \\
\hline
\hline
\end{tabular}}
\caption{\label{tab:dnmc_xp} 
Constraints on the main and derived parameters (at 68\% CL if not otherwise stated) considering 
P18 in combination with BAO and BAO + R19 for the NMC+ model.}
\end{table*}

    \newpage
\begin{table*}[h!]
{\small
\centering
\begin{tabular}{l|cc}
\hline
\hline
                                         & P18 + BAO & P18 + BAO + R19  \\
\hline
$\omega_{\rm b}$                         & $0.02245\pm 0.00014$  & $2.254^{+0.014}_{-0.016}$         \\
$\omega_{\rm c}$                         & $0.1204\pm 0.0011$    & $0.1208^{+0.0012}_{-0.0014}$    \\
$H_0$ [km s$^{-1}$Mpc$^{-1}$]            & $68.38^{+0.56}_{-0.92}$ & $69.97^{+0.82}_{-0.97}$      \\
$\tau$                                   & $0.0560\pm 0.0069$    & $0.0566\pm 0.0071$   \\
$\ln \left(  10^{10} A_{\rm s} \right)$  & $3.052\pm 0.014$      & $3.059\pm 0.015$     \\
$n_{\rm s}$                              & $0.9671\pm 0.0038$    & $0.9702\pm 0.0039$   \\
$\xi$                                    & $> -0.041$ (95\% CL) & $-0.0164^{+0.013}_{-0.0042}$   \\
$\Delta$                                 & $< 0.021$ (95\% CL)  & $ < 0.030$ (95\% CL)  \\
\hline
$\gamma_{\rm PN}$                                                             & $> 0.998$ (95\% CL)        & $0.9979^{+0.0011}_{-0.00085}$  \\
$\beta_{\rm PN}$                                                              & $< 1.00002$ (95\% CL)      & $< 1.00003$ (95\% CL) \\
$\delta G_\mathrm{N}/G_\mathrm{N}$ (z=0)                                  & $> -0.020$ (95\% CL)     & $-0.0203\pm 0.0083$  \\
$10^{13} \dot{G}_\mathrm{N}/G_{\rm N}$ (z=0) [yr$^{-1}$]  & $> -0.69$ (95\% CL)        & $-0.65^{+0.33}_{-0.26}$ \\
$G_\mathrm{N}/G$ (z=0)                                                    & $1.028^{+0.010}_{-0.031}$ & $1.049^{+0.019}_{-0.046}$  \\
$G_\mathrm{eff}/G$ (z=0)                                                  & $1.014^{+0.005}_{-0.015}$ & $1.024^{+0.010}_{-0.022}$  \\
$\sigma_i$ [Mpl]    &   $< 5.9$ (95\% CL)  &   $2.35^{+0.74}_{-1.7}$  \\
\hline
$\Omega_{\rm m}$                        & $0.3056^{+0.0078}_{-0.0066}$ & $0.2929\pm 0.0069$        \\
$\sigma_8$                              & $0.8167^{+0.0072}_{-0.010}$  & $0.828^{+0.010}_{-0.012}$  \\
$S_8$                                   & $0.824\pm 0.011$             & $0.818\pm 0.011$      \\
$r_s$ [Mpc]                             & $146.39^{+0.77}_{-0.39}$     & $145.41\pm 0.86$    \\
%\hline
%$\Delta \chi^2$                         & $-0.1$ & $-16.5$  \\
\hline
\hline
\end{tabular}}
\caption{\label{tab:dnmc_xm} 
Constraints on the main and derived parameters (at 68\% CL if not otherwise stated) considering 
P18 in combination with BAO and BAO + R19 for the NMC$-$ model.}
\end{table*}

\newpage
    
\section{Triangle plots} \label{sec:appendix2}
\begin{figure}
\centering
\includegraphics[width=0.98\textwidth]{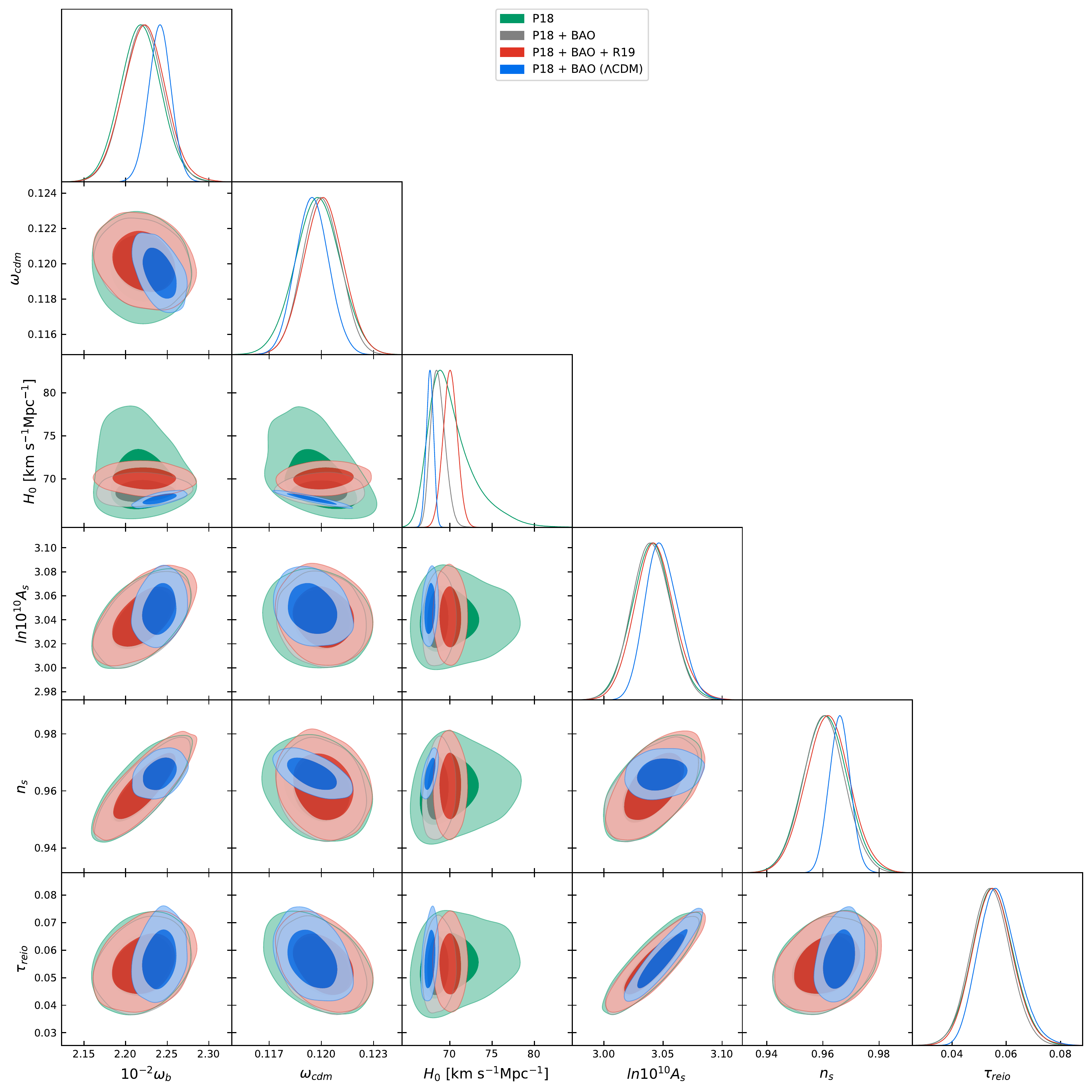}
\caption{Marginalized joint 68\% and 95\% CL regions 2D parameter space using 
the $Planck$ legacy data (green), its combination with BAO DR12, i.e. P18 + BAO (gray), 
and P18 + BAO + R19 (red) for the IG model. We include the contours for the $\Lambda$CDM 
in blue for P18 + BAO.}
\label{fig:dig_full}
\end{figure}

\newpage

\begin{figure}
\centering
\includegraphics[width=0.98\textwidth]{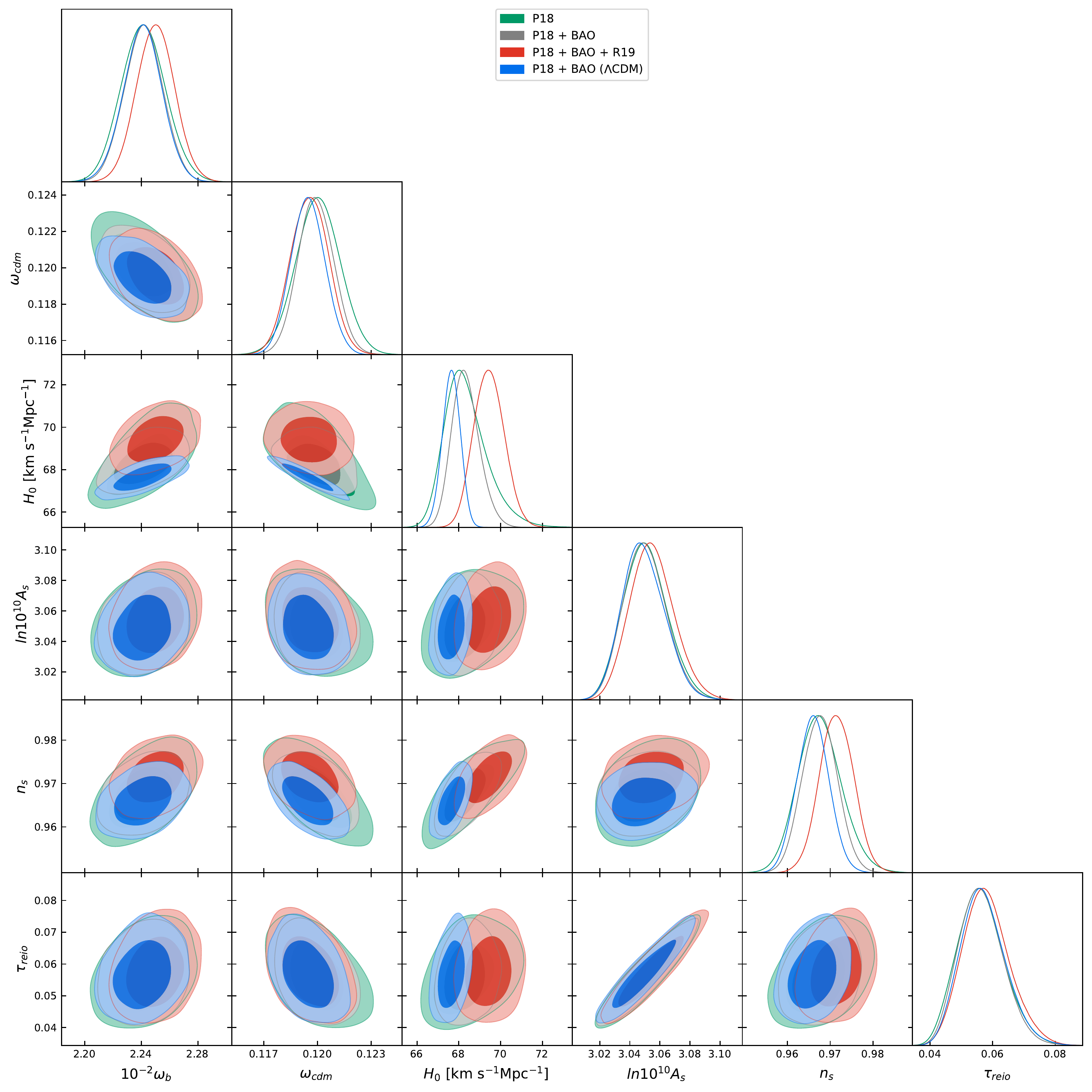}
\caption{Marginalized joint 68\% and 95\% CL regions 2D parameter space using 
the $Planck$ legacy data (green), its combination with BAO DR12, i.e. P18 + BAO (gray), 
and P18 + BAO + R19 (red) for the CC model. We include the contours for the $\Lambda$CDM 
in blue for P18 + BAO.}
\label{fig:dcc_full}
\end{figure}

\newpage

\begin{figure}
\centering
\includegraphics[width=0.98\textwidth]{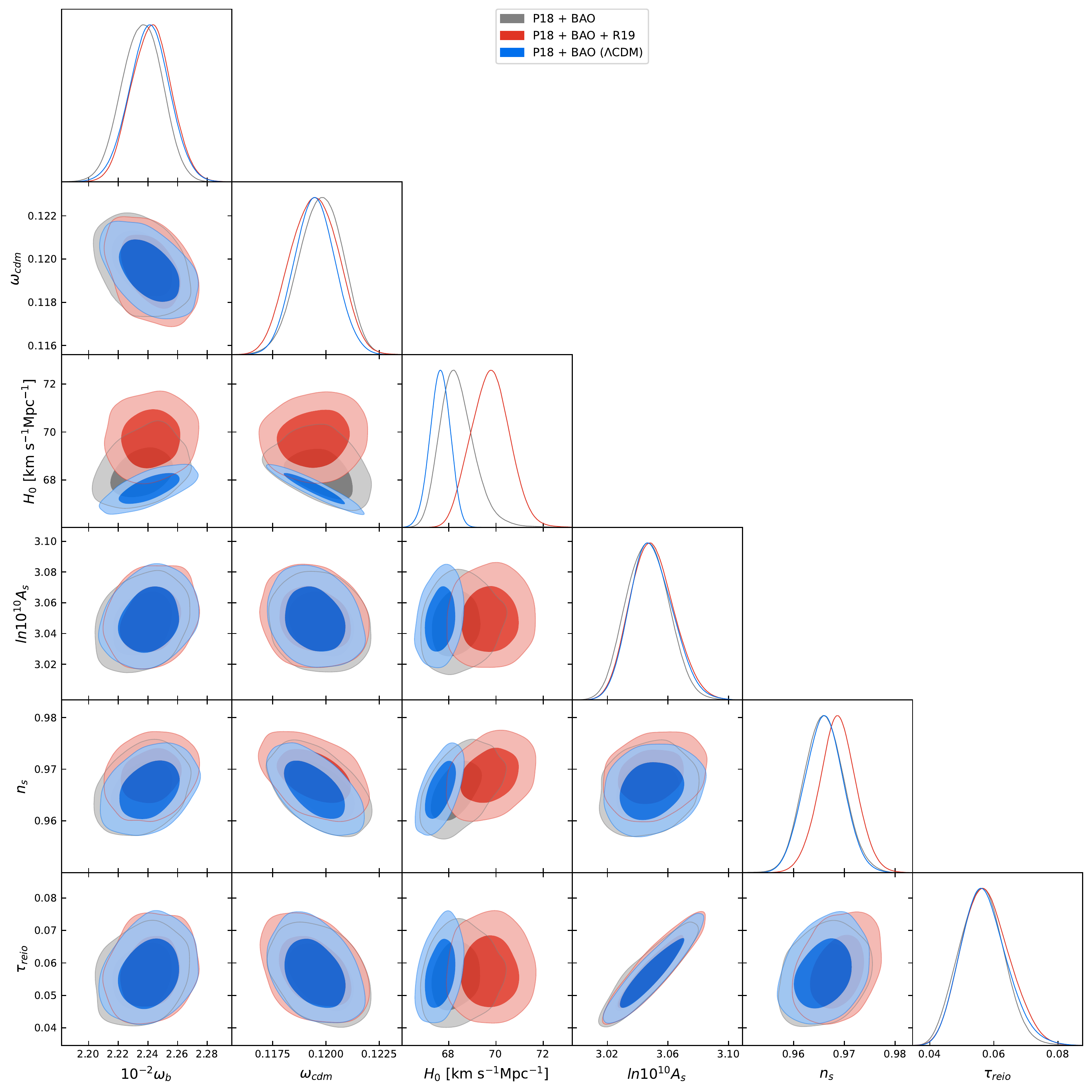}
\caption{Marginalized joint 68\% and 95\% CL regions 2D parameter space using 
the $Planck$ legacy data in combination with BAO DR12 (gray) and P18 + BAO + R19 (red) 
for the NMC+ model. We include the contours for the $\Lambda$CDM in blue for P18 + BAO.}
\label{fig:dnmc_xp_full}
\end{figure}

%\newpage

\begin{figure}
\centering
\includegraphics[width=0.98\textwidth]{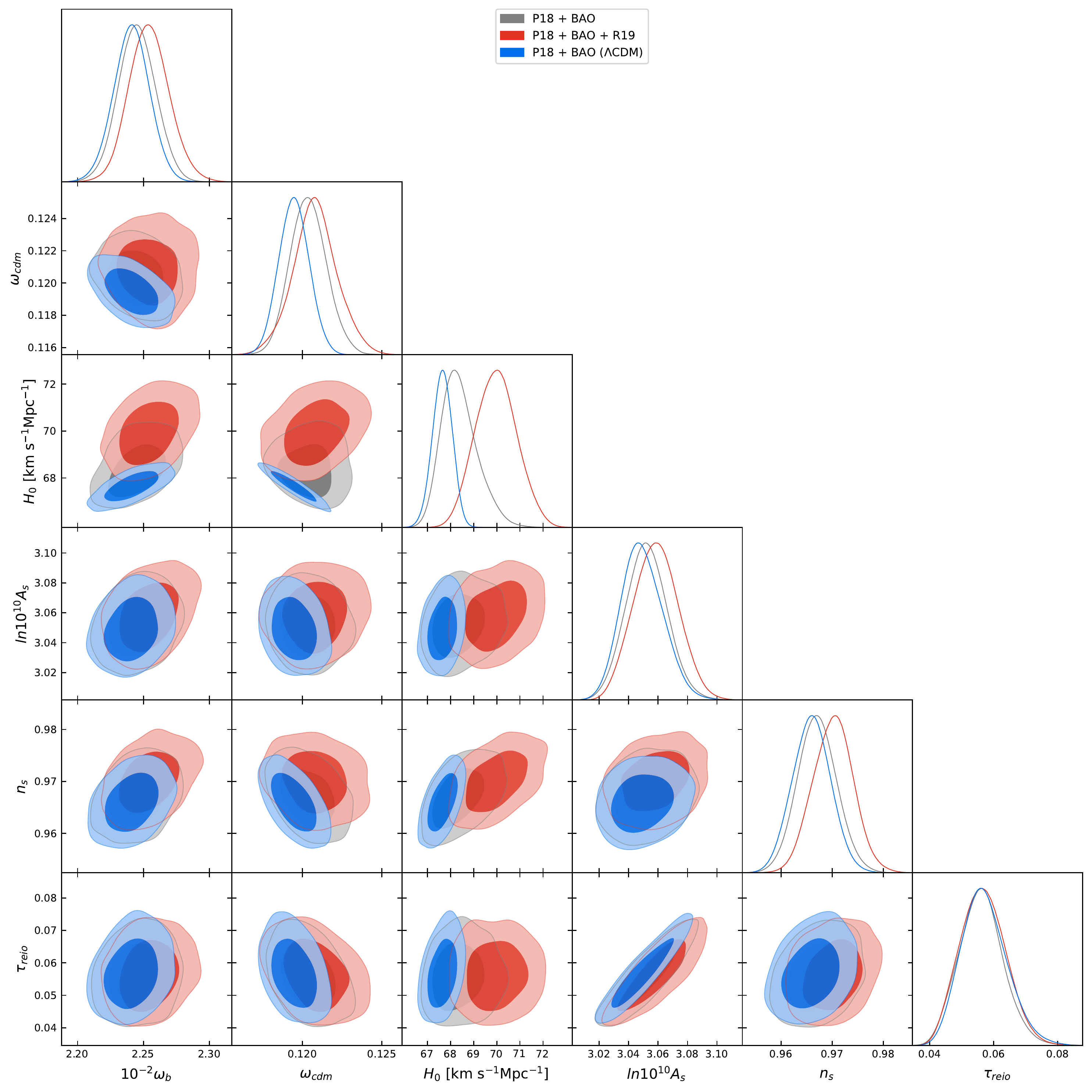}
\caption{Marginalized joint 68\% and 95\% CL regions 2D parameter space using 
the $Planck$ legacy data in combination with BAO DR12 (gray) and P18 + BAO + R19 (red) 
for the NMC- model. We include the contours for the $\Lambda$CDM in blue for P18 + BAO.}
\label{fig:dnmc_xm_full}
\end{figure}

\vspace{5cm}

\newpage
%%%%%%%%%%%%%%%%%
% Papers to add to the bib
% https://arxiv.org/abs/2102.06012
% https://arxiv.org/abs/2110.04336
%%%%%%%%%%%%%%%%%%

%%%%%%%%%%%%%%%%%%%%%%%%%%%%%%%%%%%%%%%%%%%%%%%%%%%%%%%%%%%%%%%%%%%%%%%%%%%%%%%
\bibliographystyle{unsrtnat}%unsrt}
\bibliography{Biblio}

\begin{thebibliography}{99}
\providecommand{\natexlab}[1]{#1}
\providecommand{\url}[1]{\texttt{#1}}
\expandafter\ifx\csname urlstyle\endcsname\relax
  \providecommand{\doi}[1]{doi: #1}\else
  \providecommand{\doi}{doi: \begingroup \urlstyle{rm}\Url}\fi

\bibitem[Uzan(2003)]{Uzan:2002vq}
Jean-Philippe Uzan.
\newblock {The Fundamental Constants and Their Variation: Observational Status
  and Theoretical Motivations}.
\newblock \emph{Rev. Mod. Phys.}, 75:\penalty0 403, 2003.
\newblock \doi{10.1103/RevModPhys.75.403}.

\bibitem[Dirac(1937)]{Dirac:1937ti}
Paul A.~M. Dirac.
\newblock {The Cosmological constants}.
\newblock \emph{Nature}, 139:\penalty0 323, 1937.
\newblock \doi{10.1038/139323a0}.

\bibitem[Will(2006)]{Will:2005va}
Clifford~M. Will.
\newblock {The Confrontation between general relativity and experiment}.
\newblock \emph{Living Rev. Rel.}, 9:\penalty0 3, 2006.
\newblock \doi{10.12942/lrr-2006-3}.

\bibitem[Bertotti et~al.(2003)Bertotti, Iess, and Tortora]{Bertotti:2003rm}
B.~Bertotti, L.~Iess, and P.~Tortora.
\newblock {A test of general relativity using radio links with the Cassini
  spacecraft}.
\newblock \emph{Nature}, 425:\penalty0 374--376, 2003.
\newblock \doi{10.1038/nature01997}.

\bibitem[Will(2014)]{Will:2014kxa}
Clifford~M. Will.
\newblock {The Confrontation between General Relativity and Experiment}.
\newblock \emph{Living Rev. Rel.}, 17:\penalty0 4, 2014.
\newblock \doi{10.12942/lrr-2014-4}.

\bibitem[Muller and Biskupek(2007)]{Muller:2007zzb}
Jurgen Muller and Liliane Biskupek.
\newblock {Variations of the gravitational constant from lunar laser ranging
  data}.
\newblock \emph{Class. Quant. Grav.}, 24:\penalty0 4533--4538, 2007.
\newblock \doi{10.1088/0264-9381/24/17/017}.

\bibitem[Garcia-Berro et~al.(2011)Garcia-Berro, Loren-Aguilar, Torres, Althaus,
  and Isern]{GarciaBerro:2011wc}
Enrique Garcia-Berro, Pablo Loren-Aguilar, Santiago Torres, Leandro~G. Althaus,
  and Jordi Isern.
\newblock {An upper limit to the secular variation of the gravitational
  constant from white dwarf stars}.
\newblock \emph{JCAP}, 05:\penalty0 021, 2011.
\newblock \doi{10.1088/1475-7516/2011/05/021}.

\bibitem[Mould and Uddin(2014)]{Mould:2014iga}
Jeremy Mould and Syed Uddin.
\newblock {Constraining a possible variation of G with Type Ia supernovae}.
\newblock \emph{Publ. Astron. Soc. Austral.}, 31:\penalty0 15, 2014.
\newblock \doi{10.1017/pasa.2014.9}.

\bibitem[Bellinger and Christensen-Dalsgaard(2019)]{Bellinger:2019lnl}
Earl~Patrick Bellinger and J\o{}rgen Christensen-Dalsgaard.
\newblock {Asteroseismic constraints on the cosmic-time variation of the
  gravitational constant from an ancient main-sequence star}.
\newblock \emph{Astrophys. J. Lett.}, 887\penalty0 (1):\penalty0 L1, 2019.
\newblock \doi{10.3847/2041-8213/ab43e7}.

\bibitem[Zhu et~al.(2019)]{Zhu:2018etc}
W.~W. Zhu et~al.
\newblock {Tests of Gravitational Symmetries with Pulsar Binary J1713+0747}.
\newblock \emph{Mon. Not. Roy. Astron. Soc.}, 482\penalty0 (3):\penalty0
  3249--3260, 2019.
\newblock \doi{10.1093/mnras/sty2905}.

\bibitem[Casas et~al.(1992{\natexlab{a}})Casas, Garcia-Bellido, and
  Quiros]{Casas:1990fz}
J.~A. Casas, J.~Garcia-Bellido, and M.~Quiros.
\newblock {Nucleosynthesis bounds on Jordan-Brans-Dicke theories of gravity}.
\newblock \emph{Mod. Phys. Lett. A}, 7:\penalty0 447--456, 1992{\natexlab{a}}.
\newblock \doi{10.1142/S0217732392000409}.

\bibitem[Casas et~al.(1992{\natexlab{b}})Casas, Garcia-Bellido, and
  Quiros]{Casas:1991ui}
J.~A. Casas, J.~Garcia-Bellido, and M.~Quiros.
\newblock {Updating nucleosynthesis bounds on Jordan-Brans-Dicke theories of
  gravity}.
\newblock \emph{Phys. Lett. B}, 278:\penalty0 94--96, 1992{\natexlab{b}}.
\newblock \doi{10.1016/0370-2693(92)90717-I}.

\bibitem[Santiago et~al.(1997)Santiago, Kalligas, and Wagoner]{Santiago:1997mu}
D.~I. Santiago, D.~Kalligas, and R.~V. Wagoner.
\newblock {Nucleosynthesis constraints on scalar - tensor theories of gravity}.
\newblock \emph{Phys. Rev. D}, 56:\penalty0 7627--7637, 1997.
\newblock \doi{10.1103/PhysRevD.56.7627}.

\bibitem[Clifton et~al.(2005)Clifton, Barrow, and Scherrer]{Clifton:2005xr}
Timothy Clifton, John~D. Barrow, and Robert~J. Scherrer.
\newblock {Constraints on the variation of G from primordial nucleosynthesis}.
\newblock \emph{Phys. Rev. D}, 71:\penalty0 123526, 2005.
\newblock \doi{10.1103/PhysRevD.71.123526}.

\bibitem[Bambi et~al.(2005)Bambi, Giannotti, and Villante]{Bambi:2005fi}
Cosimo Bambi, Maurizio Giannotti, and F.~L. Villante.
\newblock {The Response of primordial abundances to a general modification of
  G(N) and/or of the early Universe expansion rate}.
\newblock \emph{Phys. Rev. D}, 71:\penalty0 123524, 2005.
\newblock \doi{10.1103/PhysRevD.71.123524}.

\bibitem[Copi et~al.(2004)Copi, Davis, and Krauss]{Copi:2003xd}
Craig~J. Copi, Adam~N. Davis, and Lawrence~M. Krauss.
\newblock {A New nucleosynthesis constraint on the variation of G}.
\newblock \emph{Phys. Rev. Lett.}, 92:\penalty0 171301, 2004.
\newblock \doi{10.1103/PhysRevLett.92.171301}.

\bibitem[Alvey et~al.(2020)Alvey, Sabti, Escudero, and
  Fairbairn]{Alvey:2019ctk}
James Alvey, Nashwan Sabti, Miguel Escudero, and Malcolm Fairbairn.
\newblock {Improved BBN Constraints on the Variation of the Gravitational
  Constant}.
\newblock \emph{Eur. Phys. J. C}, 80\penalty0 (2):\penalty0 148, 2020.
\newblock \doi{10.1140/epjc/s10052-020-7727-y}.

\bibitem[Aghanim et~al.(2020{\natexlab{a}})]{Aghanim:2018eyx}
N.~Aghanim et~al.
\newblock {Planck 2018 results. VI. Cosmological parameters}.
\newblock \emph{Astron. Astrophys.}, 641:\penalty0 A6, 2020{\natexlab{a}}.
\newblock \doi{10.1051/0004-6361/201833910}.

\bibitem[Umezu et~al.(2005)Umezu, Ichiki, and Yahiro]{Umezu:2005ee}
Ken-ichi Umezu, Kiyotomo Ichiki, and Masanobu Yahiro.
\newblock {Cosmological constraints on Newton's constant}.
\newblock \emph{Phys. Rev. D}, 72:\penalty0 044010, 2005.
\newblock \doi{10.1103/PhysRevD.72.044010}.

\bibitem[Chang and Chu(2007)]{Chan:2007fe}
Kwan-Chuen Chang and M.~C. Chu.
\newblock {Constraining the Variation of G by Cosmic Microwave Background
  Anisotropies}.
\newblock \emph{Phys. Rev. D}, 75:\penalty0 083521, 2007.
\newblock \doi{10.1103/PhysRevD.75.083521}.

\bibitem[Perenon et~al.(2019)Perenon, Bel, Maartens, and de~la
  Cruz-Dombriz]{Perenon:2019dpc}
Louis Perenon, Julien Bel, Roy Maartens, and Alvaro de~la Cruz-Dombriz.
\newblock {Optimising growth of structure constraints on modified gravity}.
\newblock \emph{JCAP}, 06:\penalty0 020, 2019.
\newblock \doi{10.1088/1475-7516/2019/06/020}.

\bibitem[Han\i{}meli et~al.(2020)Han\i{}meli, Lamine, Blanchard, and
  Tutusaus]{Hanimeli:2019wrt}
Ekim~Taylan Han\i{}meli, Brahim Lamine, Alain Blanchard, and Isaac Tutusaus.
\newblock {Time-dependent $G$ in Einstein's equations as an alternative to the
  cosmological constant}.
\newblock \emph{Phys. Rev. D}, 101\penalty0 (6):\penalty0 063513, 2020.
\newblock \doi{10.1103/PhysRevD.101.063513}.

\bibitem[Sapone et~al.(2021)Sapone, Nesseris, and Bengaly]{Sapone:2020wwz}
Domenico Sapone, Savvas Nesseris, and Carlos A.~P. Bengaly.
\newblock {Is there any measurable redshift dependence on the SN Ia absolute
  magnitude?}
\newblock \emph{Phys. Dark Univ.}, 32:\penalty0 100814, 2021.
\newblock \doi{10.1016/j.dark.2021.100814}.

\bibitem[Sakr and Sapone(2021)]{Sakr:2021nja}
Ziad Sakr and Domenico Sapone.
\newblock {Can varying the gravitational constant alleviate the tensions?}
\newblock 12 2021.
\newblock arXiv:2112.14173.

\bibitem[Zahn and Zaldarriaga(2003)]{Zahn:2002rr}
Oliver Zahn and Matias Zaldarriaga.
\newblock {Probing the Friedmann equation during recombination with future CMB
  experiments}.
\newblock \emph{Phys. Rev. D}, 67:\penalty0 063002, 2003.
\newblock \doi{10.1103/PhysRevD.67.063002}.

\bibitem[Galli et~al.(2009)Galli, Melchiorri, Smoot, and Zahn]{Galli:2009pr}
Silvia Galli, Alessandro Melchiorri, George~F. Smoot, and Oliver Zahn.
\newblock {From Cavendish to PLANCK: Constraining Newton's Gravitational
  Constant with CMB Temperature and Polarization Anisotropy}.
\newblock \emph{Phys. Rev. D}, 80:\penalty0 023508, 2009.
\newblock \doi{10.1103/PhysRevD.80.023508}.

\bibitem[Martins et~al.(2010)Martins, Menegoni, Galli, Mangano, and
  Melchiorri]{Martins:2010gu}
C.~J. A.~P. Martins, Eloisa Menegoni, Silvia Galli, Gianpiero Mangano, and
  Alessandro Melchiorri.
\newblock {Varying couplings in the early universe: correlated variations of
  $\alpha$ and $G$}.
\newblock \emph{Phys. Rev. D}, 82:\penalty0 023532, 2010.
\newblock \doi{10.1103/PhysRevD.82.023532}.

\bibitem[Ballardini et~al.(2020)Ballardini, Braglia, Finelli, Paoletti,
  Starobinsky, and Umilt\`a]{Ballardini:2020iws}
Mario Ballardini, Matteo Braglia, Fabio Finelli, Daniela Paoletti, Alexei~A.
  Starobinsky, and Caterina Umilt\`a.
\newblock {Scalar-tensor theories of gravity, neutrino physics, and the $H_0$
  tension}.
\newblock \emph{JCAP}, 10:\penalty0 044, 2020.
\newblock \doi{10.1088/1475-7516/2020/10/044}.

\bibitem[Paoletti et~al.(2019)Paoletti, Braglia, Finelli, Ballardini, and
  Umilt\`a]{Paoletti:2018xet}
D.~Paoletti, M.~Braglia, F.~Finelli, M.~Ballardini, and C.~Umilt\`a.
\newblock {Isocurvature fluctuations in the effective Newton\textquoteright{}s
  constant}.
\newblock \emph{Phys. Dark Univ.}, 25:\penalty0 100307, 2019.
\newblock \doi{10.1016/j.dark.2019.100307}.

\bibitem[Umilt\`a et~al.(2015)Umilt\`a, Ballardini, Finelli, and
  Paoletti]{Umilta:2015cta}
C.~Umilt\`a, M.~Ballardini, F.~Finelli, and D.~Paoletti.
\newblock {CMB and BAO constraints for an induced gravity dark energy model
  with a quartic potential}.
\newblock \emph{JCAP}, 08:\penalty0 017, 2015.
\newblock \doi{10.1088/1475-7516/2015/08/017}.

\bibitem[Ballardini et~al.(2016{\natexlab{a}})Ballardini, Finelli, Umilt\`a,
  and Paoletti]{Ballardini:2016cvy}
Mario Ballardini, Fabio Finelli, Caterina Umilt\`a, and Daniela Paoletti.
\newblock {Cosmological constraints on induced gravity dark energy models}.
\newblock \emph{JCAP}, 05:\penalty0 067, 2016{\natexlab{a}}.
\newblock \doi{10.1088/1475-7516/2016/05/067}.

\bibitem[Rossi et~al.(2019)Rossi, Ballardini, Braglia, Finelli, Paoletti,
  Starobinsky, and Umilt\`a]{Rossi:2019lgt}
Massimo Rossi, Mario Ballardini, Matteo Braglia, Fabio Finelli, Daniela
  Paoletti, Alexei~A. Starobinsky, and Caterina Umilt\`a.
\newblock {Cosmological constraints on post-Newtonian parameters in effectively
  massless scalar-tensor theories of gravity}.
\newblock \emph{Phys. Rev. D}, 100\penalty0 (10):\penalty0 103524, 2019.
\newblock \doi{10.1103/PhysRevD.100.103524}.

\bibitem[Sch\"oneberg et~al.(2021)Sch\"oneberg, Franco~Abell\'an,
  P\'erez~S\'anchez, Witte, Poulin, and Lesgourgues]{Schoneberg:2021qvd}
Nils Sch\"oneberg, Guillermo Franco~Abell\'an, Andrea P\'erez~S\'anchez,
  Samuel~J. Witte, Vivian Poulin, and Julien Lesgourgues.
\newblock {The $H_0$ Olympics: A fair ranking of proposed models}.
\newblock 7 2021.
\newblock arXiv: 2107.10291.

\bibitem[Di~Valentino et~al.(2021)]{DiValentino:2020zio}
Eleonora Di~Valentino et~al.
\newblock {Snowmass2021 - Letter of interest cosmology intertwined II: The
  hubble constant tension}.
\newblock \emph{Astropart. Phys.}, 131:\penalty0 102605, 2021.
\newblock \doi{10.1016/j.astropartphys.2021.102605}.

\bibitem[Marra and Perivolaropoulos(2021)]{Marra:2021fvf}
Valerio Marra and Leandros Perivolaropoulos.
\newblock {Rapid transition of Geff at zt\ensuremath{\simeq}0.01 as a possible
  solution of the Hubble and growth tensions}.
\newblock \emph{Phys. Rev. D}, 104\penalty0 (2):\penalty0 L021303, 2021.
\newblock \doi{10.1103/PhysRevD.104.L021303}.

\bibitem[Alestas et~al.(2021)Alestas, Camarena, Di~Valentino, Kazantzidis,
  Marra, Nesseris, and Perivolaropoulos]{Alestas:2021luu}
George Alestas, David Camarena, Eleonora Di~Valentino, Lavrentios Kazantzidis,
  Valerio Marra, Savvas Nesseris, and Leandros Perivolaropoulos.
\newblock {Late-transition vs smooth $H(z)$ deformation models for the
  resolution of the Hubble crisis}.
\newblock 10 2021.
\newblock arXiv: 2110.04336.

\bibitem[Jordan(1949)]{Jordan:1949zz}
Pascual Jordan.
\newblock {Formation of the Stars and Development of the Universe}.
\newblock \emph{Nature}, 164:\penalty0 637--640, 1949.
\newblock \doi{10.1038/164637a0}.

\bibitem[Brans and Dicke(1961)]{Brans:1961sx}
C.~Brans and R.~H. Dicke.
\newblock {Mach's principle and a relativistic theory of gravitation}.
\newblock \emph{Phys. Rev.}, 124:\penalty0 925--935, 1961.
\newblock \doi{10.1103/PhysRev.124.925}.

\bibitem[Chen and Kamionkowski(1999)]{Chen:1999qh}
Xue-lei Chen and Marc Kamionkowski.
\newblock {Cosmic microwave background temperature and polarization anisotropy
  in Brans-Dicke cosmology}.
\newblock \emph{Phys. Rev. D}, 60:\penalty0 104036, 1999.
\newblock \doi{10.1103/PhysRevD.60.104036}.

\bibitem[Nagata et~al.(2004)Nagata, Chiba, and Sugiyama]{Nagata:2003qn}
Ryo Nagata, Takeshi Chiba, and Naoshi Sugiyama.
\newblock {WMAP constraints on scalar- tensor cosmology and the variation of
  the gravitational constant}.
\newblock \emph{Phys. Rev. D}, 69:\penalty0 083512, 2004.
\newblock \doi{10.1103/PhysRevD.69.083512}.

\bibitem[Acquaviva et~al.(2005)Acquaviva, Baccigalupi, Leach, Liddle, and
  Perrotta]{Acquaviva:2004ti}
Viviana Acquaviva, Carlo Baccigalupi, Samuel~M. Leach, Andrew~R. Liddle, and
  Francesca Perrotta.
\newblock {Structure formation constraints on the Jordan-Brans-Dicke theory}.
\newblock \emph{Phys. Rev. D}, 71:\penalty0 104025, 2005.
\newblock \doi{10.1103/PhysRevD.71.104025}.

\bibitem[Li et~al.(2013)Li, Wu, and Chen]{Li:2013nwa}
Yi-Chao Li, Feng-Quan Wu, and Xuelei Chen.
\newblock {Constraints on the Brans-Dicke gravity theory with the Planck data}.
\newblock \emph{Phys. Rev. D}, 88:\penalty0 084053, 2013.
\newblock \doi{10.1103/PhysRevD.88.084053}.

\bibitem[Avilez and Skordis(2014)]{Avilez:2013dxa}
A.~Avilez and C.~Skordis.
\newblock {Cosmological constraints on Brans-Dicke theory}.
\newblock \emph{Phys. Rev. Lett.}, 113\penalty0 (1):\penalty0 011101, 2014.
\newblock \doi{10.1103/PhysRevLett.113.011101}.

\bibitem[Ooba et~al.(2016)Ooba, Ichiki, Chiba, and Sugiyama]{Ooba:2016slp}
Junpei Ooba, Kiyotomo Ichiki, Takeshi Chiba, and Naoshi Sugiyama.
\newblock {Planck constraints on scalar-tensor cosmology and the variation of
  the gravitational constant}.
\newblock \emph{Phys. Rev. D}, 93\penalty0 (12):\penalty0 122002, 2016.
\newblock \doi{10.1103/PhysRevD.93.122002}.

\bibitem[Ooba et~al.(2017)Ooba, Ichiki, Chiba, and Sugiyama]{Ooba:2017gyn}
Junpei Ooba, Kiyotomo Ichiki, Takeshi Chiba, and Naoshi Sugiyama.
\newblock {Cosmological constraints on scalar\textendash{}tensor gravity and
  the variation of the gravitational constant}.
\newblock \emph{PTEP}, 2017\penalty0 (4):\penalty0 043E03, 2017.
\newblock \doi{10.1093/ptep/ptx046}.

\bibitem[Sol\`a~Peracaula et~al.(2019)Sol\`a~Peracaula, Gomez-Valent,
  de~Cruz~P\'erez, and Moreno-Pulido]{SolaPeracaula:2019zsl}
Joan Sol\`a~Peracaula, Adria Gomez-Valent, Javier de~Cruz~P\'erez, and Cristian
  Moreno-Pulido.
\newblock {Brans\textendash{}Dicke Gravity with a Cosmological Constant
  Smoothes Out $\Lambda$CDM Tensions}.
\newblock \emph{Astrophys. J. Lett.}, 886\penalty0 (1):\penalty0 L6, 2019.
\newblock \doi{10.3847/2041-8213/ab53e9}.

\bibitem[Ballesteros et~al.(2020)Ballesteros, Notari, and
  Rompineve]{Ballesteros:2020sik}
Guillermo Ballesteros, Alessio Notari, and Fabrizio Rompineve.
\newblock {The $H_0$ tension: $\Delta G_N$ vs. $\Delta N_{\rm eff}$}.
\newblock \emph{JCAP}, 11:\penalty0 024, 2020.
\newblock \doi{10.1088/1475-7516/2020/11/024}.

\bibitem[Braglia et~al.(2020)Braglia, Ballardini, Emond, Finelli, Gumrukcuoglu,
  Koyama, and Paoletti]{Braglia:2020iik}
Matteo Braglia, Mario Ballardini, William~T. Emond, Fabio Finelli, A.~Emir
  Gumrukcuoglu, Kazuya Koyama, and Daniela Paoletti.
\newblock {Larger value for $H_0$ by an evolving gravitational constant}.
\newblock \emph{Phys. Rev. D}, 102\penalty0 (2):\penalty0 023529, 2020.
\newblock \doi{10.1103/PhysRevD.102.023529}.

\bibitem[Braglia et~al.(2021)Braglia, Ballardini, Finelli, and
  Koyama]{Braglia:2020auw}
Matteo Braglia, Mario Ballardini, Fabio Finelli, and Kazuya Koyama.
\newblock {Early modified gravity in light of the $H_0$ tension and LSS data}.
\newblock \emph{Phys. Rev. D}, 103\penalty0 (4):\penalty0 043528, 2021.
\newblock \doi{10.1103/PhysRevD.103.043528}.

\bibitem[Cheng et~al.(2021)Cheng, Wu, and Chen]{Cheng:2021yvh}
Gong Cheng, Fengquan Wu, and Xuelei Chen.
\newblock {Cosmological test of an extended quintessence model}.
\newblock \emph{Phys. Rev. D}, 103\penalty0 (10):\penalty0 103527, 2021.
\newblock \doi{10.1103/PhysRevD.103.103527}.

\bibitem[Joudaki et~al.(2020)Joudaki, Ferreira, Lima, and
  Winther]{Joudaki:2020shz}
Shahab Joudaki, Pedro~G. Ferreira, Nelson~A. Lima, and Hans~A. Winther.
\newblock {Testing Gravity on Cosmic Scales: A Case Study of Jordan-Brans-Dicke
  Theory}.
\newblock 10 2020.
\newblock arXiv: 2010.15278.

\bibitem[Amendola(1999)]{Amendola:1999qq}
Luca Amendola.
\newblock {Scaling solutions in general nonminimal coupling theories}.
\newblock \emph{Phys. Rev. D}, 60:\penalty0 043501, 1999.
\newblock \doi{10.1103/PhysRevD.60.043501}.

\bibitem[Finelli et~al.(2008)Finelli, Tronconi, and Venturi]{Finelli:2007wb}
F.~Finelli, A.~Tronconi, and Giovanni Venturi.
\newblock {Dark Energy, Induced Gravity and Broken Scale Invariance}.
\newblock \emph{Phys. Lett. B}, 659:\penalty0 466--470, 2008.
\newblock \doi{10.1016/j.physletb.2007.11.053}.

\bibitem[Boisseau et~al.(2000)Boisseau, Esposito-Farese, Polarski, and
  Starobinsky]{Boisseau:2000pr}
B.~Boisseau, Gilles Esposito-Farese, D.~Polarski, and Alexei~A. Starobinsky.
\newblock {Reconstruction of a scalar tensor theory of gravity in an
  accelerating universe}.
\newblock \emph{Phys. Rev. Lett.}, 85:\penalty0 2236, 2000.
\newblock \doi{10.1103/PhysRevLett.85.2236}.

\bibitem[Gannouji et~al.(2006)Gannouji, Polarski, Ranquet, and
  Starobinsky]{Gannouji:2006jm}
Radouane Gannouji, David Polarski, Andre Ranquet, and Alexei~A. Starobinsky.
\newblock {Scalar-Tensor Models of Normal and Phantom Dark Energy}.
\newblock \emph{JCAP}, 09:\penalty0 016, 2006.
\newblock \doi{10.1088/1475-7516/2006/09/016}.

\bibitem[Audren et~al.(2013)Audren, Lesgourgues, Benabed, and
  Prunet]{Audren:2012wb}
Benjamin Audren, Julien Lesgourgues, Karim Benabed, and Simon Prunet.
\newblock {Conservative Constraints on Early Cosmology: an illustration of the
  Monte Python cosmological parameter inference code}.
\newblock \emph{JCAP}, 02:\penalty0 001, 2013.
\newblock \doi{10.1088/1475-7516/2013/02/001}.

\bibitem[Brinckmann and Lesgourgues(2019)]{Brinckmann:2018cvx}
Thejs Brinckmann and Julien Lesgourgues.
\newblock {MontePython 3: boosted MCMC sampler and other features}.
\newblock \emph{Phys. Dark Univ.}, 24:\penalty0 100260, 2019.
\newblock \doi{10.1016/j.dark.2018.100260}.

\bibitem[Lesgourgues(2011)]{Lesgourgues:2011re}
Julien Lesgourgues.
\newblock {The Cosmic Linear Anisotropy Solving System (CLASS) I: Overview}.
\newblock 4 2011.
\newblock arXiv: 1104.2932.

\bibitem[Blas et~al.(2011)Blas, Lesgourgues, and Tram]{Blas:2011rf}
Diego Blas, Julien Lesgourgues, and Thomas Tram.
\newblock {The Cosmic Linear Anisotropy Solving System (CLASS) II:
  Approximation schemes}.
\newblock \emph{JCAP}, 07:\penalty0 034, 2011.
\newblock \doi{10.1088/1475-7516/2011/07/034}.

\bibitem[Lewis(2019)]{Lewis:2019xzd}
Antony Lewis.
\newblock {GetDist: a Python package for analysing Monte Carlo samples}.
\newblock 10 2019.

\bibitem[Pisanti et~al.(2008)Pisanti, Cirillo, Esposito, Iocco, Mangano, Miele,
  and Serpico]{Pisanti:2007hk}
O.~Pisanti, A.~Cirillo, S.~Esposito, F.~Iocco, G.~Mangano, G.~Miele, and P.~D.
  Serpico.
\newblock {PArthENoPE: Public Algorithm Evaluating the Nucleosynthesis of
  Primordial Elements}.
\newblock \emph{Comput. Phys. Commun.}, 178:\penalty0 956--971, 2008.
\newblock \doi{10.1016/j.cpc.2008.02.015}.

\bibitem[Consiglio et~al.(2018)Consiglio, de~Salas, Mangano, Miele, Pastor, and
  Pisanti]{Consiglio:2017pot}
R.~Consiglio, P.~F. de~Salas, G.~Mangano, G.~Miele, S.~Pastor, and O.~Pisanti.
\newblock {PArthENoPE reloaded}.
\newblock \emph{Comput. Phys. Commun.}, 233:\penalty0 237--242, 2018.
\newblock \doi{10.1016/j.cpc.2018.06.022}.

\bibitem[Hamann et~al.(2008)Hamann, Lesgourgues, and Mangano]{Hamann:2007sb}
Jan Hamann, Julien Lesgourgues, and Gianpiero Mangano.
\newblock {Using BBN in cosmological parameter extraction from CMB: A Forecast
  for PLANCK}.
\newblock \emph{JCAP}, 03:\penalty0 004, 2008.
\newblock \doi{10.1088/1475-7516/2008/03/004}.

\bibitem[Reid et~al.(2019)Reid, Pesce, and Riess]{Reid:2019tiq}
M.~J. Reid, D.~W. Pesce, and A.~G. Riess.
\newblock {An Improved Distance to NGC 4258 and its Implications for the Hubble
  Constant}.
\newblock \emph{Astrophys. J. Lett.}, 886\penalty0 (2):\penalty0 L27, 2019.
\newblock \doi{10.3847/2041-8213/ab552d}.

\bibitem[Aghanim et~al.(2020{\natexlab{b}})]{Aghanim:2019ame}
N.~Aghanim et~al.
\newblock {Planck 2018 results. V. CMB power spectra and likelihoods}.
\newblock \emph{Astron. Astrophys.}, 641:\penalty0 A5, 2020{\natexlab{b}}.
\newblock \doi{10.1051/0004-6361/201936386}.

\bibitem[Aghanim et~al.(2020{\natexlab{c}})]{Aghanim:2018oex}
N.~Aghanim et~al.
\newblock {Planck 2018 results. VIII. Gravitational lensing}.
\newblock \emph{Astron. Astrophys.}, 641:\penalty0 A8, 2020{\natexlab{c}}.
\newblock \doi{10.1051/0004-6361/201833886}.

\bibitem[Alam et~al.(2017{\natexlab{a}})]{Alam:2016hwk}
Shadab Alam et~al.
\newblock {The clustering of galaxies in the completed SDSS-III Baryon
  Oscillation Spectroscopic Survey: cosmological analysis of the DR12 galaxy
  sample}.
\newblock \emph{Mon. Not. Roy. Astron. Soc.}, 470\penalty0 (3):\penalty0
  2617--2652, 2017{\natexlab{a}}.
\newblock \doi{10.1093/mnras/stx721}.

\bibitem[Beutler et~al.(2011)Beutler, Blake, Colless, Jones, Staveley-Smith,
  Campbell, Parker, Saunders, and Watson]{Beutler:2011hx}
Florian Beutler, Chris Blake, Matthew Colless, D.~Heath Jones, Lister
  Staveley-Smith, Lachlan Campbell, Quentin Parker, Will Saunders, and Fred
  Watson.
\newblock {The 6dF Galaxy Survey: Baryon Acoustic Oscillations and the Local
  Hubble Constant}.
\newblock \emph{Mon. Not. Roy. Astron. Soc.}, 416:\penalty0 3017--3032, 2011.
\newblock \doi{10.1111/j.1365-2966.2011.19250.x}.

\bibitem[Ross et~al.(2015)Ross, Samushia, Howlett, Percival, Burden, and
  Manera]{Ross:2014qpa}
Ashley~J. Ross, Lado Samushia, Cullan Howlett, Will~J. Percival, Angela Burden,
  and Marc Manera.
\newblock {The clustering of the SDSS DR7 main Galaxy sample \textendash{} I. A
  4 per cent distance measure at $z = 0.15$}.
\newblock \emph{Mon. Not. Roy. Astron. Soc.}, 449\penalty0 (1):\penalty0
  835--847, 2015.
\newblock \doi{10.1093/mnras/stv154}.

\bibitem[Abbott et~al.(2018)]{DES:2017myr}
T.~M.~C. Abbott et~al.
\newblock {Dark Energy Survey year 1 results: Cosmological constraints from
  galaxy clustering and weak lensing}.
\newblock \emph{Phys. Rev. D}, 98\penalty0 (4):\penalty0 043526, 2018.
\newblock \doi{10.1103/PhysRevD.98.043526}.

\bibitem[Hildebrandt et~al.(2017)]{Hildebrandt:2016iqg}
H.~Hildebrandt et~al.
\newblock {KiDS-450: Cosmological parameter constraints from tomographic weak
  gravitational lensing}.
\newblock \emph{Mon. Not. Roy. Astron. Soc.}, 465:\penalty0 1454, 2017.
\newblock \doi{10.1093/mnras/stw2805}.

\bibitem[Hildebrandt et~al.(2020)]{Hildebrandt:2018yau}
H.~Hildebrandt et~al.
\newblock {KiDS+VIKING-450: Cosmic shear tomography with optical and infrared
  data}.
\newblock \emph{Astron. Astrophys.}, 633:\penalty0 A69, 2020.
\newblock \doi{10.1051/0004-6361/201834878}.

\bibitem[Hikage et~al.(2019)]{HSC:2018mrq}
Chiaki Hikage et~al.
\newblock {Cosmology from cosmic shear power spectra with Subaru Hyper
  Suprime-Cam first-year data}.
\newblock \emph{Publ. Astron. Soc. Jap.}, 71\penalty0 (2):\penalty0 43, 2019.
\newblock \doi{10.1093/pasj/psz010}.

\bibitem[Hazumi et~al.(2020)]{LiteBIRD:2020khw}
M.~Hazumi et~al.
\newblock {LiteBIRD: JAXA's new strategic L-class mission for all-sky surveys
  of cosmic microwave background polarization}.
\newblock \emph{Proc. SPIE Int. Soc. Opt. Eng.}, 11443:\penalty0 114432F, 2020.
\newblock \doi{10.1117/12.2563050}.

\bibitem[Abazajian et~al.(2019)]{Abazajian:2019eic}
Kevork Abazajian et~al.
\newblock {CMB-S4 Science Case, Reference Design, and Project Plan}.
\newblock 7 2019.
\newblock arXiv: 1907.04473.

\bibitem[Benson et~al.(2014)]{SPT-3G:2014dbx}
B.~A. Benson et~al.
\newblock {SPT-3G: A Next-Generation Cosmic Microwave Background Polarization
  Experiment on the South Pole Telescope}.
\newblock \emph{Proc. SPIE Int. Soc. Opt. Eng.}, 9153:\penalty0 91531P, 2014.
\newblock \doi{10.1117/12.2057305}.

\bibitem[Ade et~al.(2019)]{SimonsObservatory:2018koc}
Peter Ade et~al.
\newblock {The Simons Observatory: Science goals and forecasts}.
\newblock \emph{JCAP}, 02:\penalty0 056, 2019.
\newblock \doi{10.1088/1475-7516/2019/02/056}.

\bibitem[Ballardini et~al.(2019)Ballardini, Sapone, Umilt\`a, Finelli, and
  Paoletti]{Ballardini:2019tho}
M.~Ballardini, D.~Sapone, C.~Umilt\`a, F.~Finelli, and D.~Paoletti.
\newblock {Testing extended Jordan-Brans-Dicke theories with future
  cosmological observations}.
\newblock \emph{JCAP}, 05:\penalty0 049, 2019.
\newblock \doi{10.1088/1475-7516/2019/05/049}.

\bibitem[Hu and Okamoto(2002)]{Hu:2001kj}
Wayne Hu and Takemi Okamoto.
\newblock {Mass reconstruction with cmb polarization}.
\newblock \emph{Astrophys. J.}, 574:\penalty0 566--574, 2002.
\newblock \doi{10.1086/341110}.

\bibitem[Hirata and Seljak(2003)]{Hirata:2003ka}
Christopher~M. Hirata and Uros Seljak.
\newblock {Reconstruction of lensing from the cosmic microwave background
  polarization}.
\newblock \emph{Phys. Rev. D}, 68:\penalty0 083002, 2003.
\newblock \doi{10.1103/PhysRevD.68.083002}.

\bibitem[Smith et~al.(2012)Smith, Hanson, LoVerde, Hirata, and
  Zahn]{Smith:2010gu}
Kendrick~M. Smith, Duncan Hanson, Marilena LoVerde, Christopher~M. Hirata, and
  Oliver Zahn.
\newblock {Delensing CMB Polarization with External Datasets}.
\newblock \emph{JCAP}, 06:\penalty0 014, 2012.
\newblock \doi{10.1088/1475-7516/2012/06/014}.

\bibitem[Seo and Eisenstein(2003)]{Seo:2003pu}
Hee-Jong Seo and Daniel~J. Eisenstein.
\newblock {Probing dark energy with baryonic acoustic oscillations from future
  large galaxy redshift surveys}.
\newblock \emph{Astrophys. J.}, 598:\penalty0 720--740, 2003.
\newblock \doi{10.1086/379122}.

\bibitem[Song and Percival(2009)]{Song:2008qt}
Yong-Seon Song and Will~J. Percival.
\newblock {Reconstructing the history of structure formation using Redshift
  Distortions}.
\newblock \emph{JCAP}, 10:\penalty0 004, 2009.
\newblock \doi{10.1088/1475-7516/2009/10/004}.

\bibitem[Wang et~al.(2013)Wang, Chuang, and Hirata]{Wang:2012bx}
Yun Wang, Chia-Hsun Chuang, and Christopher~M. Hirata.
\newblock {Toward More Realistic Forecasting of Dark Energy Constraints from
  Galaxy Redshift Surveys}.
\newblock \emph{Mon. Not. Roy. Astron. Soc.}, 430:\penalty0 2446, 2013.
\newblock \doi{10.1093/mnras/stt068}.

\bibitem[Blanchard et~al.(2020)]{Euclid:2019clj}
A.~Blanchard et~al.
\newblock {Euclid preparation: VII. Forecast validation for Euclid cosmological
  probes}.
\newblock \emph{Astron. Astrophys.}, 642:\penalty0 A191, 2020.
\newblock \doi{10.1051/0004-6361/202038071}.

\bibitem[Alcock and Paczynski(1979)]{Alcock:1979mp}
C.~Alcock and B.~Paczynski.
\newblock {An evolution free test for non-zero cosmological constant}.
\newblock \emph{Nature}, 281:\penalty0 358--359, 1979.
\newblock \doi{10.1038/281358a0}.

\bibitem[Kaiser(1987)]{Kaiser:1987qv}
N.~Kaiser.
\newblock {Clustering in real space and in redshift space}.
\newblock \emph{Mon. Not. Roy. Astron. Soc.}, 227:\penalty0 1--27, 1987.

\bibitem[Boyle and Komatsu(2018)]{Boyle:2017lzt}
Aoife Boyle and Eiichiro Komatsu.
\newblock {Deconstructing the neutrino mass constraint from galaxy redshift
  surveys}.
\newblock \emph{JCAP}, 03:\penalty0 035, 2018.
\newblock \doi{10.1088/1475-7516/2018/03/035}.

\bibitem[Aghamousa et~al.(2016)]{DESI:2016fyo}
Amir Aghamousa et~al.
\newblock {The DESI Experiment Part I: Science,Targeting, and Survey Design}.
\newblock 10 2016.
\newblock arXiv: 1611.00036.

\bibitem[Ballardini et~al.(2016{\natexlab{b}})Ballardini, Finelli, Fedeli, and
  Moscardini]{Ballardini:2016hpi}
Mario Ballardini, Fabio Finelli, Cosimo Fedeli, and Lauro Moscardini.
\newblock {Probing primordial features with future galaxy surveys}.
\newblock \emph{JCAP}, 10:\penalty0 041, 2016{\natexlab{b}}.
\newblock \doi{10.1088/1475-7516/2016/10/041}.
\newblock [Erratum: JCAP 04, E01 (2018)].

\bibitem[Alam et~al.(2017{\natexlab{b}})]{BOSS:2016wmc}
Shadab Alam et~al.
\newblock {The clustering of galaxies in the completed SDSS-III Baryon
  Oscillation Spectroscopic Survey: cosmological analysis of the DR12 galaxy
  sample}.
\newblock \emph{Mon. Not. Roy. Astron. Soc.}, 470\penalty0 (3):\penalty0
  2617--2652, 2017{\natexlab{b}}.
\newblock \doi{10.1093/mnras/stx721}.

\bibitem[Amendola et~al.(2008)Amendola, Kunz, and Sapone]{Amendola:2007rr}
Luca Amendola, Martin Kunz, and Domenico Sapone.
\newblock {Measuring the dark side (with weak lensing)}.
\newblock \emph{JCAP}, 04:\penalty0 013, 2008.
\newblock \doi{10.1088/1475-7516/2008/04/013}.

\bibitem[Majerotto et~al.(2016)Majerotto, Sapone, and
  Sch\"afer]{Majerotto:2015bra}
Elisabetta Majerotto, Domenico Sapone, and Bj\"orn~Malte Sch\"afer.
\newblock {Combined constraints on deviations of dark energy from an ideal
  fluid from Euclid and Planck}.
\newblock \emph{Mon. Not. Roy. Astron. Soc.}, 456\penalty0 (1):\penalty0
  109--118, 2016.
\newblock \doi{10.1093/mnras/stv2640}.

\bibitem[Palma et~al.(2018)Palma, Sapone, and Sypsas]{Palma:2017wxu}
Gonzalo~A. Palma, Domenico Sapone, and Spyros Sypsas.
\newblock {Constraints on inflation with LSS surveys: features in the
  primordial power spectrum}.
\newblock \emph{JCAP}, 06:\penalty0 004, 2018.
\newblock \doi{10.1088/1475-7516/2018/06/004}.

\bibitem[Schaan et~al.(2017)Schaan, Krause, Eifler, Dor\'e, Miyatake, Rhodes,
  and Spergel]{Schaan:2016ois}
Emmanuel Schaan, Elisabeth Krause, Tim Eifler, Olivier Dor\'e, Hironao
  Miyatake, Jason Rhodes, and David~N. Spergel.
\newblock {Looking through the same lens: Shear calibration for LSST, Euclid,
  and WFIRST with stage 4 CMB lensing}.
\newblock \emph{Phys. Rev. D}, 95\penalty0 (12):\penalty0 123512, 2017.
\newblock \doi{10.1103/PhysRevD.95.123512}.

\bibitem[Mandelbaum et~al.(2018)]{LSSTDarkEnergyScience:2018jkl}
Rachel Mandelbaum et~al.
\newblock {The LSST Dark Energy Science Collaboration (DESC) Science
  Requirements Document}.
\newblock 9 2018.
\newblock arXiv: 1809.01669.

\bibitem[Tassev et~al.(2013)Tassev, Zaldarriaga, and Eisenstein]{Tassev:2013pn}
Svetlin Tassev, Matias Zaldarriaga, and Daniel Eisenstein.
\newblock {Solving Large Scale Structure in Ten Easy Steps with COLA}.
\newblock \emph{JCAP}, 06:\penalty0 036, 2013.
\newblock \doi{10.1088/1475-7516/2013/06/036}.

\bibitem[Winther et~al.(2017)Winther, Koyama, Manera, Wright, and
  Zhao]{Winther:2017jof}
Hans~A. Winther, Kazuya Koyama, Marc Manera, Bill~S. Wright, and Gong-Bo Zhao.
\newblock {COLA with scale-dependent growth: applications to screened modified
  gravity models}.
\newblock \emph{JCAP}, 08:\penalty0 006, 2017.
\newblock \doi{10.1088/1475-7516/2017/08/006}.

\bibitem[Alonso et~al.(2017)Alonso, Bellini, Ferreira, and
  Zumalac\'arregui]{Alonso:2016suf}
David Alonso, Emilio Bellini, Pedro~G. Ferreira, and Miguel Zumalac\'arregui.
\newblock {Observational future of cosmological scalar-tensor theories}.
\newblock \emph{Phys. Rev. D}, 95\penalty0 (6):\penalty0 063502, 2017.
\newblock \doi{10.1103/PhysRevD.95.063502}.

\end{thebibliography}
%%%%%%%%%%%%%%%%%%%%%%%%%%%%%%%%%%%%%%%%%%%%%%%%%%%%%%%%%%%%%%%%%%%%%%%%%%%%%%%

\end{document}